\documentclass[fleqn,usenatbib]{mnras}

\usepackage{etoolbox,xparse}
\usepackage[T2A,T1]{fontenc}
\usepackage[utf8]{inputenc}
\usepackage[bulgarian,british]{babel}
\usepackage{datetime2}
\usepackage[usenames,dvipsnames]{xcolor}
\usepackage{rotating}

\usepackage[final,defaultcolor=magenta,authormarkup=none,commentmarkup=uwave]{changes}
\definechangesauthor[color=black]{s}
\definechangesauthor[color=orange]{t}
\definechangesauthor[color=magenta]{r}

\usepackage{multirow}
\usepackage{booktabs}  %
\usepackage{graphicx}
\usepackage{amsmath}
\usepackage{newtxtext}
\usepackage{newtxmath}
\usepackage[style=english,english=british]{csquotes}  %
\usepackage{siunitx}  %
\newcommand*{\hourang}[2][]{%
    \ang[
        angle-symbol-degree=\textsuperscript{h},
        angle-symbol-minute=\textsuperscript{m},
        angle-symbol-second=\textsuperscript{s},
        #1]{#2}%
}

\newcommand{\ut}{\textsc{ut}}
\newcommand{\tablenotemark}[1]{\textsuperscript{\textit{#1}}}

\DeclareSIPostPower{\nominal}{N}
\DeclareSIQualifier{\earth}{\ensuremath{\oplus}}
\DeclareSIQualifier{\jupiter}{J}
\DeclareSIQualifier{\planet}{p}
\DeclareSIQualifier{\etoile}{\ensuremath{\star}}
\DeclareSIQualifier{\sun}{\ensuremath{\odot}}
\DeclareSIUnit\angstrom{\AA}
\DeclareSIUnit{\au}{au}
\DeclareSIUnit{\density}{\ensuremath{\mathnormal{\rho}}}
\DeclareSIUnit{\erg}{erg}
\DeclareSIUnit{\luminosity}{\ensuremath{\mathrm{L}}}
\DeclareSIUnit{\magnitude}{mag}
\DeclareSIUnit{\mass}{\ensuremath{\mathrm{M}}}
\DeclareSIUnit{\mas}{\milliarcsecond}
\DeclareSIUnit{\milliarcsecond}{mas}
\DeclareSIUnit{\parsec}{pc}
\DeclareSIUnit{\radius}{\ensuremath{\mathrm{R}}}
\DeclareSIUnit{\year}{yr}

\newcommand{\bjdtdb}{\ensuremath{\mathrm{BJD}_\text{TDB}}}
\newcommand{\feh}{\ensuremath{[\text{Fe}/\text{H}]}}
\newcommand{\logg}{\ensuremath{\log g}}
\newcommand{\tempeff}{\ensuremath{T_{\text{eff}}}}
\newcommand{\tempeq}{\ensuremath{T_{\text{eq}}}}
\newcommand{\vsini}{\ensuremath{v \sin i}}

\newcommand{\normaldist}{\mathcal{N}}
\newcommand{\uniformdist}{\mathcal{U}}

\newcommand{\gaiaG}{\ensuremath{G}}
\newcommand{\gaiaBP}{\ensuremath{G_\mathrm{BP}}}
\newcommand{\gaiaRP}{\ensuremath{G_\mathrm{RP}}}
\newcommand{\filterB}{\ensuremath{B}}
\newcommand{\filterIc}{\ensuremath{I_\mathrm{c}}}
\newcommand{\filterJ}{\ensuremath{J}}
\newcommand{\filterH}{\ensuremath{H}}
\newcommand{\filterK}{\ensuremath{K_\mathrm{s}}}

\newcommand{\filterRc}{\ensuremath{R_\mathrm{c}}}
\newcommand{\filterV}{\ensuremath{V}}
\newcommand{\filterWone}{\ensuremath{W1}}

\newcommand{\filterWfour}{\ensuremath{W4}}
\newcommand{\filtergp}{\ensuremath{g'}}
\newcommand{\filterip}{\ensuremath{i'}}
\newcommand{\filterrp}{\ensuremath{r'}}
\newcommand{\filterzs}{\ensuremath{z_\mathrm{s}}}

\newcommand{\ares}{\textsc{ares}}
\newcommand{\arviz}{\textsc{arviz}}

\newcommand{\celeritetwo}{\textsc{celerite2}}

\newcommand{\exofast}{\textsc{exofast}}
\newcommand{\exoplanetpy}{\textsc{exoplanet}}
\newcommand{\isochrones}{\textsc{isochrones}}
\newcommand{\mist}{MIST}
\newcommand{\molusc}{\textsc{molusc}}
\newcommand{\moog}{\textsc{moog}}
\newcommand{\pymc}{\textsc{PyMC}}
\newcommand{\qlp}{\textsc{qlp}}

\newcommand{\triceratops}{\textsc{triceratops}}

\newcommand{\planet}[2]{#1\thinspace #2}
\newcommand{\toitwothousand}{TOI-2000}
\newcommand{\toitwothousandtic}{371188886}
\newcommand{\planetinner}{\planet{\toitwothousand}{b}}
\newcommand{\planetouter}{\planet{\toitwothousand}{c}}
\newcommand{\waspfortyseven}{WASP-47}
\newcommand{\waspfortysevenhj}{\planet{\waspfortyseven}{b}}
\newcommand{\waspfortyseveninner}{\planet{\waspfortyseven}{e}}
\newcommand{\waspfortysevenneptune}{\planet{\waspfortyseven}{d}}

\newcommand{\keplerseventhirtyhj}{\planet{Kepler-730}{b}}
\newcommand{\keplerseventhirtyinner}{\planet{Kepler-730}{c}}
\newcommand{\toieleventhirty}{TOI-1130}
\newcommand{\toieleventhirtyinner}{\planet{\toieleventhirty}{b}}
\newcommand{\toieleventhirtyhj}{\planet{\toieleventhirty}{c}}
\newcommand{\keplerthirty}{Kepler-30}
\newcommand{\keplerthirtyb}{\planet{\keplerthirty}{b}}
\newcommand{\keplereightynine}{KOI-94}
\newcommand{\keplereightynineb}{\planet{\keplereightynine}{b}}

\newcommand{\wasponethirtytwohj}{\planet{WASP-132}{b}}
\newcommand{\wasponethirtytwoinner}{\planet{WASP-132}{c}}

\newcommand{\waspeightyfourhj}{\planet{WASP-84}{b}}
\newcommand{\waspeightyfourinner}{\planet{WASP-84}{c}}

\newcommand{\astep}{ASTEP}
\newcommand{\chiron}{CHI\-RON}
\newcommand{\edrthree}{EDR3}
\newcommand{\feros}{FEROS}
\newcommand{\gaia}{\emph{Gaia}}
\newcommand{\geminisouth}{Gemini South}
\newcommand{\harps}{HARPS}
\newcommand{\jwst}{\emph{JWST}}
\newcommand{\kepler}{\emph{Kepler}}

\newcommand{\lcogt}{LCOGT}
\newcommand{\pest}{PEST}
\newcommand{\soar}{SOAR}
\newcommand{\tess}{\emph{TESS}}
\newcommand{\twomass}{2MASS}
\newcommand{\wise}{\emph{WISE}}

\newcommand{\sysParamPeriodSubZero}{$9.1270550$}
\newcommand{\sysParamPeriodSubZeroUnc}{$_{-0.0000072}^{+0.0000073}$}
\newcommand{\sysParamPeriodSubOne}{$3.098330$}
\newcommand{\sysParamPeriodSubOneUnc}{$_{-0.000019}^{+0.000021}$}
\newcommand{\sysParamTzeroSubZero}{$2459110.06588$}
\newcommand{\sysParamTzeroSubZeroUnc}{$_{-0.00028}^{+0.00027}$}
\newcommand{\sysParamTzeroSubOne}{$2458855.2442$}
\newcommand{\sysParamTzeroSubOneUnc}{$_{-0.0021}^{+0.0022}$}
\newcommand{\sysParamRpSubZero}{$0.06581$}
\newcommand{\sysParamRpSubZeroUnc}{$\pm 0.00068$}
\newcommand{\sysParamRpSubOne}{$0.02182$}
\newcommand{\sysParamRpSubOneUnc}{$_{-0.00097}^{+0.00089}$}
\newcommand{\sysParamRPlanetJupiterSubZero}{$0.727$}
\newcommand{\sysParamRPlanetJupiterSubZeroUnc}{$_{-0.027}^{+0.028}$}
\newcommand{\sysParamRPlanetJupiterSubOne}{$0.241$}
\newcommand{\sysParamRPlanetJupiterSubOneUnc}{$\pm 0.014$}
\newcommand{\sysParamBSubZero}{$0.631$}
\newcommand{\sysParamBSubZeroUnc}{$_{-0.047}^{+0.039}$}
\newcommand{\sysParamBSubOne}{$0.770$}
\newcommand{\sysParamBSubOneUnc}{$_{-0.071}^{+0.038}$}
\newcommand{\sysParamTdurSubZero}{$3.654$}
\newcommand{\sysParamTdurSubZeroUnc}{$_{-0.028}^{+0.030}$}
\newcommand{\sysParamTdurSubOne}{$1.959$}
\newcommand{\sysParamTdurSubOneUnc}{$_{-0.098}^{+0.209}$}
\newcommand{\sysParamSqrtEccVecZeroSubZero}{$-0.215$}
\newcommand{\sysParamSqrtEccVecZeroSubZeroUnc}{$_{-0.046}^{+0.061}$}
\newcommand{\sysParamSqrtEccVecZeroSubOne}{$-0.06$}
\newcommand{\sysParamSqrtEccVecZeroSubOneUnc}{$_{-0.12}^{+0.14}$}
\newcommand{\sysParamEccSubZero}{$0.063$}
\newcommand{\sysParamEccSubZeroUnc}{$_{-0.022}^{+0.023}$}

\newcommand{\sysParamRPlanetEarthSubZero}{$8.14$}
\newcommand{\sysParamRPlanetEarthSubZeroUnc}{$_{-0.30}^{+0.31}$}
\newcommand{\sysParamRPlanetEarthSubOne}{$2.70$}
\newcommand{\sysParamRPlanetEarthSubOneUnc}{$\pm 0.15$}
\newcommand{\sysParamMPlanetJupiterSubZero}{$0.257$}
\newcommand{\sysParamMPlanetJupiterSubZeroUnc}{$_{-0.014}^{+0.015}$}
\newcommand{\sysParamMPlanetJupiterSubOne}{$0.0347$}
\newcommand{\sysParamMPlanetJupiterSubOneUnc}{$_{-0.0075}^{+0.0077}$}
\newcommand{\sysParamMPlanetEarthSubZero}{$81.7$}
\newcommand{\sysParamMPlanetEarthSubZeroUnc}{$_{-4.6}^{+4.7}$}
\newcommand{\sysParamMPlanetEarthSubOne}{$11.0$}
\newcommand{\sysParamMPlanetEarthSubOneUnc}{$\pm 2.4$}
\newcommand{\sysParamRhoPlanetSubZero}{$0.829$}
\newcommand{\sysParamRhoPlanetSubZeroUnc}{$_{-0.096}^{+0.111}$}
\newcommand{\sysParamRhoPlanetSubOne}{$3.07$}
\newcommand{\sysParamRhoPlanetSubOneUnc}{$_{-0.78}^{+0.94}$}
\newcommand{\sysParamKSubZero}{$23.7$}

\newcommand{\sysParamKSubOne}{$4.59$}
\newcommand{\sysParamKSubOneUnc}{$_{-0.99}^{+1.00}$}
\newcommand{\sysParamMStarZero}{$1.083$}
\newcommand{\sysParamMStarZeroUnc}{$_{-0.050}^{+0.059}$}
\newcommand{\sysParamFehZero}{$0.417$}
\newcommand{\sysParamFehZeroUnc}{$_{-0.041}^{+0.039}$}
\newcommand{\sysParamEep}{$385$}
\newcommand{\sysParamEepUnc}{$_{-38}^{+24}$}
\newcommand{\sysParamMStar}{$1.082$}
\newcommand{\sysParamMStarUnc}{$_{-0.050}^{+0.059}$}
\newcommand{\sysParamRStar}{$1.134$}
\newcommand{\sysParamRStarUnc}{$_{-0.036}^{+0.037}$}
\newcommand{\sysParamRhoStar}{$1.047$}
\newcommand{\sysParamRhoStarUnc}{$_{-0.100}^{+0.114}$}
\newcommand{\sysParamLoggStar}{$4.363$}
\newcommand{\sysParamLoggStarUnc}{$_{-0.032}^{+0.034}$}
\newcommand{\sysParamTeff}{$5611$}
\newcommand{\sysParamTeffUnc}{$_{-82}^{+85}$}
\newcommand{\sysParamFeh}{$0.439$}
\newcommand{\sysParamFehUnc}{$_{-0.043}^{+0.041}$}
\newcommand{\sysParamAge}{$5.3$}
\newcommand{\sysParamAgeUnc}{$\pm 2.7$}
\newcommand{\sysParamParallax}{$5.773$}
\newcommand{\sysParamParallaxUnc}{$\pm 0.010$}
\newcommand{\sysParamAv}{$0.22$}
\newcommand{\sysParamAvUnc}{$\pm 0.11$}
\newcommand{\sysParamLStar}{$1.147$}
\newcommand{\sysParamLStarUnc}{$_{-0.084}^{+0.094}$}

\newcommand{\sysParamRStarSed}{$1.103$}
\newcommand{\sysParamRStarSedUnc}{$\pm 0.014$}
\newcommand{\sysParamTeffSed}{$5687$}
\newcommand{\sysParamTeffSedUnc}{$_{-133}^{+142}$}

\newcommand{\sysParamSedUncScale}{$1.27$}
\newcommand{\sysParamSedUncScaleUnc}{$_{-0.20}^{+0.39}$}
\newcommand{\sysParamUTessSubZero}{$0.130$}
\newcommand{\sysParamUTessSubZeroUnc}{$_{-0.095}^{+0.170}$}
\newcommand{\sysParamUTessSubOne}{$0.32$}
\newcommand{\sysParamUTessSubOneUnc}{$_{-0.24}^{+0.16}$}
\newcommand{\sysParamURcSubZero}{$0.89$}
\newcommand{\sysParamURcSubZeroUnc}{$_{-0.28}^{+0.21}$}
\newcommand{\sysParamURcSubOne}{$-0.29$}
\newcommand{\sysParamURcSubOneUnc}{$_{-0.20}^{+0.33}$}
\newcommand{\sysParamUZsSubZero}{$0.41$}
\newcommand{\sysParamUZsSubZeroUnc}{$_{-0.28}^{+0.35}$}
\newcommand{\sysParamUZsSubOne}{$0.01$}
\newcommand{\sysParamUZsSubOneUnc}{$_{-0.27}^{+0.32}$}
\newcommand{\sysParamMeanFluxZero}{$(-1.3$}
\newcommand{\sysParamMeanFluxZeroUnc}{$\pm 1.2) \times 10^{-5}$}
\newcommand{\sysParamMeanFluxOne}{$(1.5$}
\newcommand{\sysParamMeanFluxOneUnc}{$\pm 1.2) \times 10^{-5}$}

\newcommand{\sysParamLcJitterSubZero}{$(9.2$}
\newcommand{\sysParamLcJitterSubZeroUnc}{$_{-4.2}^{+3.1}) \times 10^{-5}$}
\newcommand{\sysParamLcJitterSubOne}{$(2.06$}
\newcommand{\sysParamLcJitterSubOneUnc}{$_{-1.15}^{+0.87}) \times 10^{-4}$}
\newcommand{\sysParamLcJitterSubTwo}{$(1.25$}
\newcommand{\sysParamLcJitterSubTwoUnc}{$_{-0.87}^{+1.23}) \times 10^{-4}$}
\newcommand{\sysParamLcJitterSubThree}{$0.00123$}
\newcommand{\sysParamLcJitterSubThreeUnc}{$\pm 0.00014$}
\newcommand{\sysParamRvGammaSubZero}{$6647.7$}
\newcommand{\sysParamRvGammaSubZeroUnc}{$_{-12.8}^{+8.8}$}
\newcommand{\sysParamRvGammaSubOne}{$8116$}
\newcommand{\sysParamRvGammaSubOneUnc}{$_{-14}^{+11}$}
\newcommand{\sysParamRvGammaSubTwo}{$8118.1$}
\newcommand{\sysParamRvGammaSubTwoUnc}{$_{-12.2}^{+8.2}$}
\newcommand{\sysParamRvJitterSubZero}{$3.4$}
\newcommand{\sysParamRvJitterSubZeroUnc}{$_{-2.3}^{+3.2}$}
\newcommand{\sysParamRvJitterSubOne}{$14.0$}
\newcommand{\sysParamRvJitterSubOneUnc}{$_{-3.3}^{+4.5}$}
\newcommand{\sysParamRvJitterSubTwo}{$3.14$}
\newcommand{\sysParamRvJitterSubTwoUnc}{$_{-0.78}^{+0.82}$}
\newcommand{\sysParamGpSigma}{$21.4$}
\newcommand{\sysParamGpSigmaUnc}{$_{-7.6}^{+14.6}$}
\newcommand{\sysParamGpRho}{$111$}
\newcommand{\sysParamGpRhoUnc}{$_{-37}^{+57}$}
\newcommand{\sysParamGpTau}{$37$}
\newcommand{\sysParamGpTauUnc}{$_{-31}^{+95}$}
\newcommand{\sysParamASubZero}{$0.0878$}
\newcommand{\sysParamASubZeroUnc}{$_{-0.0014}^{+0.0016}$}
\newcommand{\sysParamASubOne}{$0.04271$}
\newcommand{\sysParamASubOneUnc}{$_{-0.00067}^{+0.00076}$}
\newcommand{\sysParamAorSubZero}{$16.64$}
\newcommand{\sysParamAorSubZeroUnc}{$_{-0.55}^{+0.59}$}
\newcommand{\sysParamAorSubOne}{$8.10$}
\newcommand{\sysParamAorSubOneUnc}{$_{-0.27}^{+0.28}$}
\newcommand{\sysParamInclSubZero}{$87.86$}
\newcommand{\sysParamInclSubZeroUnc}{$_{-0.18}^{+0.19}$}
\newcommand{\sysParamInclSubOne}{$84.59$}
\newcommand{\sysParamInclSubOneUnc}{$_{-0.43}^{+0.54}$}
\newcommand{\sysParamDistance}{$173.22$}
\newcommand{\sysParamDistanceUnc}{$\pm 0.31$}
\newcommand{\sysParamMPlanetSubZero}{$(2.26$}
\newcommand{\sysParamMPlanetSubZeroUnc}{$_{-0.12}^{+0.11}) \times 10^{-4}$}
\newcommand{\sysParamMPlanetSubOne}{$(3.06$}
\newcommand{\sysParamMPlanetSubOneUnc}{$\pm 0.66) \times 10^{-5}$}
\newcommand{\sysParamIrradiationSubZero}{$(2.03$}
\newcommand{\sysParamIrradiationSubZeroUnc}{$_{-0.14}^{+0.16}) \times 10^{8}$}
\newcommand{\sysParamIrradiationSubOne}{$(8.55$}
\newcommand{\sysParamIrradiationSubOneUnc}{$_{-0.59}^{+0.66}) \times 10^{8}$}
\newcommand{\sysParamTempEqSubZero}{$1038$}
\newcommand{\sysParamTempEqSubZeroUnc}{$_{-111}^{+84}$}
\newcommand{\sysParamTempEqSubOne}{$1488$}
\newcommand{\sysParamTempEqSubOneUnc}{$_{-160}^{+122}$}
\newcommand{\sysParamOmegaFoldSubZero}{$196$}
\newcommand{\sysParamOmegaFoldSubZeroUnc}{$_{-34}^{+29}$}

\date{Received 2023 May 29; in original form 2022 September 23.}
\pubyear{2023}

\graphicspath{{./}{figures/}}

\title[TESS discovery of the TOI-2000 system]{\textit{TESS} spots a mini-neptune interior to a hot saturn in the TOI-2000 system}

\author[L. Sha et al.]{%
Lizhou~Sha%
\textsuperscript{\href{https://orcid.org/0000-0001-5401-8079}{\includegraphics[width=2.5mm]{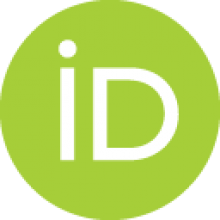}}}%
,\textsuperscript{1}%
\thanks{Email: \href{mailto:lsha@wisc.edu}{lsha@wisc.edu}}
Andrew~M.~Vanderburg%
\textsuperscript{\href{https://orcid.org/0000-0001-7246-5438}{\includegraphics[width=2.5mm]{orcid-ID.png}}}%
,\textsuperscript{2}
Chelsea~X.~Huang%
\textsuperscript{\href{https://orcid.org/0000-0003-0918-7484}{\includegraphics[width=2.5mm]{orcid-ID.png}}}%
,\textsuperscript{3}%
\thanks{ARC DECRA Fellow}
David~J.~Armstrong%
\textsuperscript{\href{https://orcid.org/0000-0002-5080-4117}{\includegraphics[width=2.5mm]{orcid-ID.png}}}%
,\textsuperscript{4,5}
\newauthor
Rafael~Brahm%
\textsuperscript{\href{https://orcid.org/0000-0002-9158-7315}{\includegraphics[width=2.5mm]{orcid-ID.png}}}%
,\textsuperscript{6,7,8}
Steven~Giacalone%
\textsuperscript{\href{https://orcid.org/0000-0002-8965-3969}{\includegraphics[width=2.5mm]{orcid-ID.png}}}%
,\textsuperscript{9}
Mackenna~L.~Wood%
\textsuperscript{\href{https://orcid.org/0000-0001-7336-7725}{\includegraphics[width=2.5mm]{orcid-ID.png}}}%
,\textsuperscript{10}
Karen~A.~Collins%
\textsuperscript{\href{https://orcid.org/0000-0001-6588-9574}{\includegraphics[width=2.5mm]{orcid-ID.png}}}%
,\textsuperscript{11}
\newauthor
Louise~D.~Nielsen%
\textsuperscript{\href{https://orcid.org/0000-0002-5254-2499}{\includegraphics[width=2.5mm]{orcid-ID.png}}}%
,\textsuperscript{12}
Melissa~J.~Hobson%
\textsuperscript{\href{https://orcid.org/0000-0002-5945-7975}{\includegraphics[width=2.5mm]{orcid-ID.png}}}%
,\textsuperscript{13,14}
Carl~Ziegler%
\textsuperscript{\href{https://orcid.org/0000-0002-0619-7639}{\includegraphics[width=2.5mm]{orcid-ID.png}}}%
,\textsuperscript{15}
Steve~B.~Howell%
\textsuperscript{\href{https://orcid.org/0000-0002-2532-2853}{\includegraphics[width=2.5mm]{orcid-ID.png}}}%
,\textsuperscript{16}
\newauthor
Pascal~Torres-Miranda,\textsuperscript{7,17}
Andrew~W.~Mann%
\textsuperscript{\href{https://orcid.org/0000-0003-3654-1602}{\includegraphics[width=2.5mm]{orcid-ID.png}}}%
,\textsuperscript{10}
George~Zhou%
\textsuperscript{\href{https://orcid.org/0000-0002-4891-3517}{\includegraphics[width=2.5mm]{orcid-ID.png}}}%
,\textsuperscript{3}%
\footnotemark[\value{footnote}]
Elisa~Delgado-Mena%
\textsuperscript{\href{https://orcid.org/0000-0003-4434-2195}{\includegraphics[width=2.5mm]{orcid-ID.png}}}%
,\textsuperscript{18}
\newauthor
Felipe~I.~Rojas%
\textsuperscript{\href{https://orcid.org/0000-0003-3047-6272}{\includegraphics[width=2.5mm]{orcid-ID.png}}}%
,\textsuperscript{17}
Lyu~Abe%
\textsuperscript{\href{https://orcid.org/0000-0002-0856-4527}{\includegraphics[width=2.5mm]{orcid-ID.png}}}%
,\textsuperscript{19}
Trifon~Trifonov%
\textsuperscript{\href{https://orcid.org/0000-0002-0236-775X}{\includegraphics[width=2.5mm]{orcid-ID.png}}}%
,\textsuperscript{13,20}
Vardan~Adibekyan%
\textsuperscript{\href{https://orcid.org/0000-0002-0601-6199}{\includegraphics[width=2.5mm]{orcid-ID.png}}}%
,\textsuperscript{18}
Sérgio~G.~Sousa%
\textsuperscript{\href{https://orcid.org/0000-0001-9047-2965}{\includegraphics[width=2.5mm]{orcid-ID.png}}}%
,\textsuperscript{18}
\newauthor
Sergio~B.~Fajardo-Acosta%
\textsuperscript{\href{https://orcid.org/0000-0001-9309-0102}{\includegraphics[width=2.5mm]{orcid-ID.png}}}%
,\textsuperscript{21}
Tristan~Guillot%
\textsuperscript{\href{https://orcid.org/0000-0002-7188-8428}{\includegraphics[width=2.5mm]{orcid-ID.png}}}%
,\textsuperscript{19}
Saburo~Howard%
\textsuperscript{\href{https://orcid.org/0000-0003-4894-7271}{\includegraphics[width=2.5mm]{orcid-ID.png}}}%
,\textsuperscript{19}
Colin~Littlefield%
\textsuperscript{\href{https://orcid.org/0000-0001-7746-5795}{\includegraphics[width=2.5mm]{orcid-ID.png}}}%
,\textsuperscript{16}
\newauthor
Faith~Hawthorn%
\textsuperscript{\href{https://orcid.org/0000-0002-8675-182X}{\includegraphics[width=2.5mm]{orcid-ID.png}}}%
,\textsuperscript{4,5}
François-Xavier~Schmider%
\textsuperscript{\href{https://orcid.org/0000-0003-3914-3546}{\includegraphics[width=2.5mm]{orcid-ID.png}}}%
,\textsuperscript{19}
Jan~Eberhardt%
\textsuperscript{\href{https://orcid.org/0000-0003-3130-2768}{\includegraphics[width=2.5mm]{orcid-ID.png}}}%
,\textsuperscript{13}
Thiam-Guan~Tan%
\textsuperscript{\href{https://orcid.org/0000-0001-5603-6895}{\includegraphics[width=2.5mm]{orcid-ID.png}}}%
,\textsuperscript{22}
\newauthor
Ares~Osborn%
\textsuperscript{\href{https://orcid.org/0000-0002-5899-7750}{\includegraphics[width=2.5mm]{orcid-ID.png}}}%
,\textsuperscript{4,5}
Richard~P.~Schwarz%
\textsuperscript{\href{https://orcid.org/0000-0001-8227-1020}{\includegraphics[width=2.5mm]{orcid-ID.png}}}%
,\textsuperscript{11}
Paul~Strøm%
\textsuperscript{\href{https://orcid.org/0000-0002-7823-1090}{\includegraphics[width=2.5mm]{orcid-ID.png}}}%
,\textsuperscript{4}
Andrés~Jordán%
\textsuperscript{\href{https://orcid.org/0000-0002-5389-3944}{\includegraphics[width=2.5mm]{orcid-ID.png}}}%
,\textsuperscript{6,7,8}
Gavin~Wang%
\textsuperscript{\href{https://orcid.org/0000-0003-3092-4418}{\includegraphics[width=2.5mm]{orcid-ID.png}}}%
,\textsuperscript{23}
\newauthor
Thomas~Henning%
\textsuperscript{\href{https://orcid.org/0000-0002-1493-300X}{\includegraphics[width=2.5mm]{orcid-ID.png}}}%
,\textsuperscript{13}
Bob~Massey%
\textsuperscript{\href{https://orcid.org/0000-0001-8879-7138}{\includegraphics[width=2.5mm]{orcid-ID.png}}}%
,\textsuperscript{24}
Nicholas~Law%
\textsuperscript{\href{https://orcid.org/0000-0001-9380-6457}{\includegraphics[width=2.5mm]{orcid-ID.png}}}%
,\textsuperscript{10}
Chris~Stockdale%
\textsuperscript{\href{https://orcid.org/0000-0003-2163-1437}{\includegraphics[width=2.5mm]{orcid-ID.png}}}%
,\textsuperscript{25}
Elise~Furlan%
\textsuperscript{\href{https://orcid.org/0000-0001-9800-6248}{\includegraphics[width=2.5mm]{orcid-ID.png}}}%
,\textsuperscript{26}
\newauthor
Gregor~Srdoc,\textsuperscript{27}
Peter~J.~Wheatley%
\textsuperscript{\href{https://orcid.org/0000-0003-1452-2240}{\includegraphics[width=2.5mm]{orcid-ID.png}}}%
,\textsuperscript{4,5}
David~Barrado~Navascués%
\textsuperscript{\href{https://orcid.org/0000-0002-5971-9242}{\includegraphics[width=2.5mm]{orcid-ID.png}}}%
,\textsuperscript{28}
Jack~J.~Lissauer%
\textsuperscript{\href{https://orcid.org/0000-0001-6513-1659}{\includegraphics[width=2.5mm]{orcid-ID.png}}}%
,\textsuperscript{16}
\newauthor
Keivan~G.~Stassun%
\textsuperscript{\href{https://orcid.org/0000-0002-3481-9052}{\includegraphics[width=2.5mm]{orcid-ID.png}}}%
,\textsuperscript{29}
George~R.~Ricker%
\textsuperscript{\href{https://orcid.org/0000-0003-2058-6662}{\includegraphics[width=2.5mm]{orcid-ID.png}}}%
,\textsuperscript{2}
Roland~K.~Vanderspek%
\textsuperscript{\href{https://orcid.org/0000-0001-6763-6562}{\includegraphics[width=2.5mm]{orcid-ID.png}}}%
,\textsuperscript{2}
David~W.~Latham%
\textsuperscript{\href{https://orcid.org/0000-0001-9911-7388}{\includegraphics[width=2.5mm]{orcid-ID.png}}}%
,\textsuperscript{11}
\newauthor
Joshua~N.~Winn%
\textsuperscript{\href{https://orcid.org/0000-0002-4265-047X}{\includegraphics[width=2.5mm]{orcid-ID.png}}}%
,\textsuperscript{30}
Sara~Seager%
\textsuperscript{\href{https://orcid.org/0000-0002-6892-6948}{\includegraphics[width=2.5mm]{orcid-ID.png}}}%
,\textsuperscript{2,31,32}
Jon~M.~Jenkins%
\textsuperscript{\href{https://orcid.org/0000-0002-4715-9460}{\includegraphics[width=2.5mm]{orcid-ID.png}}}%
,\textsuperscript{16}
Thomas~Barclay%
\textsuperscript{\href{https://orcid.org/0000-0001-7139-2724}{\includegraphics[width=2.5mm]{orcid-ID.png}}}%
,\textsuperscript{33,34}
\newauthor
Luke~G.~Bouma%
\textsuperscript{\href{https://orcid.org/0000-0002-0514-5538}{\includegraphics[width=2.5mm]{orcid-ID.png}}}%
,\textsuperscript{35}%
\thanks{51 Pegasi b Postdoctoral Fellow}
Jessie~L.~Christiansen%
\textsuperscript{\href{https://orcid.org/0000-0002-8035-4778}{\includegraphics[width=2.5mm]{orcid-ID.png}}}%
,\textsuperscript{26}
Natalia~Guerrero%
\textsuperscript{\href{https://orcid.org/0000-0002-5169-9427}{\includegraphics[width=2.5mm]{orcid-ID.png}}}%
\thinspace\textsuperscript{36}
and
Mark~E.~Rose%
\textsuperscript{\href{https://orcid.org/0000-0003-4724-745X}{\includegraphics[width=2.5mm]{orcid-ID.png}}}%
\thinspace\textsuperscript{16}
\\
\\
Affiliations are listed at the end of the paper
}

\begin{document}
\label{firstpage}
\pagerange{\pageref{firstpage}--\pageref{lastpage}}

\maketitle

\begin{abstract}
Hot jupiters ($P < 10\,\text{d}, M > 60\, \mathrm{M}_\oplus$)
are almost always found alone around their stars,
but four out of hundreds known have inner companion planets.
These rare companions allow us to constrain the hot jupiter's formation history
by ruling out high-eccentricity tidal migration.
Less is known about inner companions to hot Saturn-mass planets.
We report here the discovery of the TOI-2000 system, which features a hot Saturn-mass planet with a smaller inner
companion.
The mini-neptune TOI-2000\thinspace b
(\sysParamRPlanetEarthSubOne\sysParamRPlanetEarthSubOneUnc$\, \mathrm{R}_\oplus$,
\sysParamMPlanetEarthSubOne\sysParamMPlanetEarthSubOneUnc$\, \mathrm{M}_\oplus$)
is in a
3.10-day orbit,
and the hot saturn TOI-2000\thinspace c
(\sysParamRPlanetEarthSubZero\sysParamRPlanetEarthSubZeroUnc$\, \mathrm{R}_\oplus$,
\sysParamMPlanetEarthSubZero\sysParamMPlanetEarthSubZeroUnc$\, \mathrm{M}_\oplus$)
is in a
9.13-day orbit.
Both planets transit their host star TOI-2000
(TIC\thinspace 371188886, $V = 10.98$, \textit{TESS}~magnitude~$= 10.36$),
a metal-rich ($[\text{Fe}/\text{H}] =$ \sysParamFeh\sysParamFehUnc)
G~dwarf
174\thinspace pc away.
\emph{TESS} observed the two planets in sectors~9–11 and 36–38,
and we followed up with ground-based photometry, spectroscopy, and speckle imaging.
Radial velocities from \added[id=s]{CHIRON, FEROS, and} HARPS allowed us to confirm both planets by direct mass measurement.
In addition, we demonstrate constraining
planetary and stellar parameters
with MIST stellar evolutionary tracks
through Hamiltonian Monte Carlo under the PyMC framework,
achieving higher sampling efficiency and shorter run time
compared to traditional Markov chain Monte Carlo.
Having the brightest host star in the $V$ band among similar systems,
TOI-2000\thinspace b and c are superb candidates for atmospheric characterization by the \emph{JWST},
which can potentially distinguish whether they formed together or
TOI-2000\thinspace c swept along material during migration to form TOI-2000\thinspace b.
\end{abstract}


\section{Introduction}

Hot gas giant planets
($P < 10$\thinspace d,
$M > \SI{60}{\mass\earth}$; also known as hot jupiters)
are rarely observed with an inner companion planet
\citep{2016ApJ...825...98H,2021AJ....162..263H}.
This relative scarcity is consistent with the hypothesis that
high-eccentricity migration (HEM) may be responsible for the formation
of many, if not most, hot gas giants
(see \citealt{2018ARA&A..56..175D} for a review).
Under HEM, a hot gas giant first forms \enquote{cold} at an orbital separation of several astronomical units
and then enters into an orbit of high eccentricity by interacting with other planets in the system
or another star
\citep{1996Sci...274..954R,1996Natur.384..619W,2003ApJ...589..605W}.
Later, tidal interaction with the host star dissipates the hot gas giant's orbital energy
and circularizes its orbit.
This dynamically disruptive process would likely have eliminated any inner companions in the system
\citep{2015ApJ...808...14M}.

Two alternative mechanisms may explain how hot gas giants with inner companions formed.
Under disc migration,
tidal interactions with the protoplanetary disc move the initially cold gas giant to its present location
(\citealt{1996Natur.380..606L};
see \citealt{2014prpl.conf..667B} for a review),
sweeping material along its mean-motion resonances (MMRs) to form inner companions
\citep{2006Sci...313.1413R}.
The other alternative has the giant planet forming \emph{in situ},
near its present location
\citep{2014ApJ...797...95L,2016ApJ...829..114B,2016ApJ...817L..17B,2016ApJ...817...90L,2021MNRAS.505.2500P}.
Thus, measuring the occurrence rate of inner companions
may quantify the fraction of hot gas giants that formed under these two mechanisms
as opposed to under HEM.

Until recently, efforts to measure the occurrence rate of hot gas giants' inner companions
were hampered by the lack of detections.
Early studies looked for transit timing variations (TTVs) of known hot gas giants
induced by possible companions near MMRs
but found no significant TTVs of their targets
\citep{2005MNRAS.364L..96S,2008ApJ...682..586M,2008ApJ...682..593M,2009ApJ...700.1078G}.
After the launch of \kepler{},
\citet{2011ApJ...732L..24L} analysed 117 transiting multiplanet systems
that have at least one hot planet candidate ($P < \SI{10}{\day}$) and concluded that
only $4^{+9}_{-2}$~per cent have giant planets bigger than Neptune,
a rate considerably lower than the $31^{+4}_{-3}$~per cent
of 405 hot planet candidates analysed in total that are bigger than Neptune.
\citet{2012PNAS..109.7982S} additionally
searched for TTV signals in \kepler{} light curves of hot jupiters
and ruled out the existence of all but the least massive
($M \lesssim \SI{1}{\mass\earth}$)
companions near MMRs.
Using data from the full \kepler{} mission,
\citet{2016ApJ...825...98H} found
no inner companions to 45 hot jupiters in their sample,
but found that half or more of 27 warm jupiters
($\SI{10}{\day} \leq P < \SI{200}{\day}$)
searched had small coplanar nearby companions.
More recently,
\citet{2021AJ....162..263H} found no additional transiting planets
in the light curves of 184 confirmed hot jupiters
from the first year of observations
by the \emph{Transiting Exoplanet Survey Satellite}
\citep[\tess{};][]{tess_mission_paper}.

Nevertheless, \kepler{} and \tess{} have detected a handful of inner companions to hot gas giants.
Out of the almost 500 transiting hot gas giants published in the literature,
four are known to have inner companions:
\waspfortysevenhj{}
\citep{2015ApJ...812L..18B,2017AJ....154..237V,2022AJ....163..197B,2012MNRAS.426..739H},
\keplerseventhirtyhj{} \citep{2018RNAAS...2..160Z,2019ApJ...870L..17C},
\toieleventhirtyhj{} (\citealt{2020ApJ...892L...7H,2023arXiv230515565K})
and \wasponethirtytwohj{} \citep{2022AJ....164...13H,2017MNRAS.465.3693H}.
These exceptional planets may be important for understanding the formation of hot gas giant systems,
so it is natural to ask if they share a formation history
that is distinct from hot gas giants without inner companions.
Currently, however, the sample of systems with companions
is too small and relatively uncharacterized to answer that question.

Adding to this growing, but still small, family,
we report the discovery of the \toitwothousand{} system,
which hosts the smallest hot gas giant known to have an interior planet.
We found the system through a systematic multi-planet search of the full-frame image (FFI) light curves of
all hot gas giants observed by \tess{}
during its two-year prime mission.
We conducted a series of ground-based follow-up
photometry, spectroscopy, and speckle imaging (\autoref{sec:obs}).
By constructing a joint model of transit light curves, radial velocities, and broadband photometry,
we derive the planetary masses and other physical parameters of the \toitwothousand{} system
and demonstrate a way to interpolate MIST stellar evolutionary tracks
when using Hamiltonian Monte Carlo (HMC) for parameter fitting
(\autoref{sec:analysis}).
We then argue for the confirmation of
the inner mini-neptune (\planetinner{})
and the outer hot saturn (\planetouter{})
by carefully considering and rejecting alternative explanations for the two planets' signals
(\autoref{sec:confirm}).
Finally, we compare the two planets to those in similar planetary systems,
point out they might show TTVs,
and
explore future observations that may constrain how they formed,
ending with a note on why interpolating MIST tracks under HMC is generally preferred over traditional Markov chain Monte Carlo
(\autoref{sec:discussion}).

\section{Observations and Data Reduction} \label{sec:obs}

\subsection{Photometry}

\begin{figure*}
    \centering
    \includegraphics[width=\textwidth]{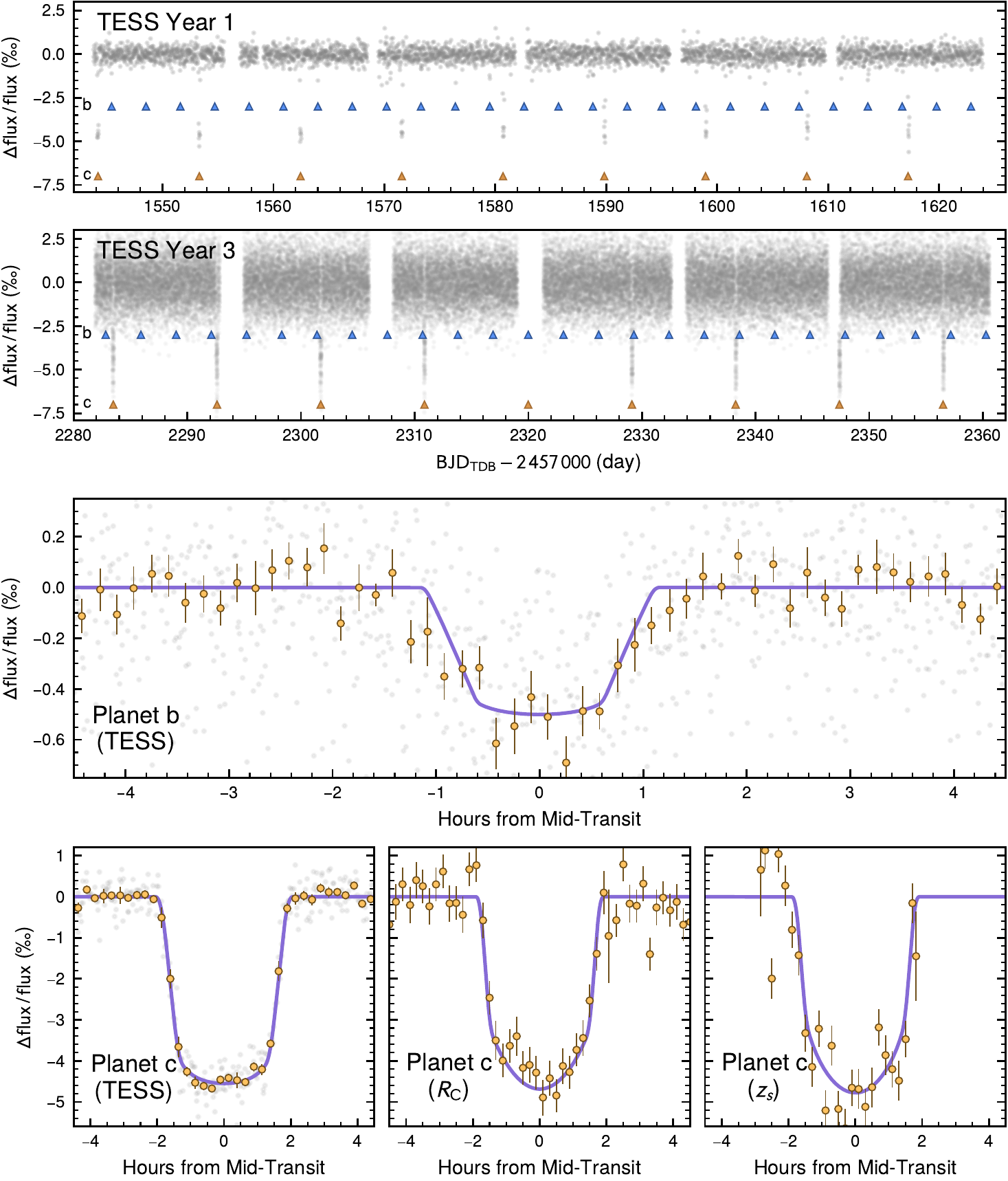}
    \caption{Light curves of \toitwothousand{}.
    Top two rows: Detrended light curves from \tess{} years~1 and 3.
    Year~1 points are from the 30-min full-frame images,
    whilst year~3 points are binned to 2~min from the 20-s time series,
    which results in higher scatter per cadence but lower scatter at 30~min.
    The upright triangles indicate transits of planet~b (blue) and planet~c (orange).
    Bottom two rows:
    Phase-folded light curves zoomed in on the transits of planets~b (third row)
    and c (last row).
    The purple line is the transit model,
    \added[id=s]{and the label in each panel indicates the passband of the limb darkening parameters.}
    For the \tess{} light curves,
    the faint grey marks are observations both years,
    with year~3 observations binned to 30~min to be consistent with those from year~1.
    \added[id=s]{The orange marks of the \tess{} light curves are binned means (15-min bins for planet~b and 10~min for planet~c),}
    with the error bars representing the standard error of the mean.
    \added[id=s]{The middle and right panels of the bottom row show planet~c transit light curves from ground observations
    (\astep{} in the \filterRc{} band, \lcogt{}~SSO in \filterzs{})}.
    For the ground observations,
    the orange marks are 12-min binned means weighted by inverse variance
    and the error bars are the standard error of the weighted mean.
    }
    \label{fig:lc}
\end{figure*}

We present space and ground-based photometry of \toitwothousand{} in this section.
\autoref{tab:phot} summarizes the observations,
and \autoref{fig:lc} shows the light curves.

\begin{table*}
    \caption{Summary of photometric observations.}
    \label{tab:phot}
    \begin{tabular}{llccccccc}
        \toprule
        Facility \& Telescope
        & Date(s)
        & Planet
        & Transit
        & Transit(s)
        & No.\ Images
        & Exp.\ Time
        & Filter
        & Used in
        \\
        & (UT) & & Coverage & Detected? & & (s) & & Joint Model? \\
        \midrule
        \tess{} camera 3    & 2019 Feb 28 -- May 21 & All   & Full  & Y & 1612  & 1800  & \textit{TESS} & Y \\
        \tess{} camera 3    & 2021 Mar 7 -- May 26  & All   & Full  & Y &111410 & 20    & \textit{TESS} & Y \\
        \midrule
        \lcogt{} SSO 1\,m   & 2021 Feb 10   & c & Full      & Y & 129   & 30    & \filterB              & N \\
        \lcogt{} SSO 1\,m   & 2021 Feb 10   & c & Full      & Y & 128   & 30    &           \filterzs   & Y \\
        \pest{}             & 2021 Apr 24   & c & Full      & Y & 168   & 60    & \filterB              & N \\
        \pest{}             & 2021 Apr 24   & c & Full      & Y & 167   & 60    &           \filterIc   & N \\
        \astep{} 0.4\,m     & 2021 May 3    & c & Full      & Y & 427   & 70    & \filterRc             & Y \\
        \midrule
        \lcogt{} SSO 1\,m   & 2021 Jan 18               & b & Full & N & 467 & 20 & \filterip & N \\
        \lcogt{} CTIO 1\,m  & 2021 Feb 6                & b & Full & N & 515 & 20 & \filterip & N \\
        \lcogt{} SAAO 1\,m  & 2021 Apr 27               & b & Ingress & N & 190 & 15 & \filterip & N \\
        \lcogt{} CTIO 1\,m  &         2021  Dec 16      & b & Full & N & 305 & 15 & \filterip & N \\
        \lcogt{} SAAO 1\,m  &                2022 Jan 7 & b & Full & N & 324 & 15 & \filterip & N \\
        \lcogt{} CTIO 0.4\,m&         2022  May 16      & b & Full & N & 133 &100 & \filterip & N \\
        \bottomrule
    \end{tabular}
\end{table*}

\begin{table}
    \centering
    \caption{\tess{} light curve of \toitwothousand{} from year 1 at 30-minute cadence.
    (The entire table is available electronically in machine-readable format.)}
    \label{tab:tess30m}
    \begin{tabular}{ccc}
        \toprule
        \bjdtdb & Flux & Detrended Flux \\
        ${} - \num{2457000}$ & (normalized) & (normalized)\\
        \midrule
        1543.783216 & 1.00716 & 1.00027 \\
        1543.804029 & 1.00661 & 0.99986 \\
        1543.824903 & 1.00622 & 0.99960 \\
        1543.845716 & 1.00698 & 1.00049 \\
        1543.866530 & 1.00644 & 1.00008 \\
        \ldots      & \ldots  & \ldots \\
        \bottomrule
    \end{tabular}
\end{table}

\begin{table}
    \centering
    \caption{\tess{} light curve of \toitwothousand{} from year 3 at 20-second cadence.
    (The entire table is available electronically in machine-readable format.)}
    \label{tab:tess20s}
    \begin{tabular}{ccc}
        \toprule
        \bjdtdb{} & Detrended Flux & Uncertainty \\
        ${} - \num{2457000}$ & (normalized) \\
        \midrule
        2281.876943 & 0.9993 & 0.0027 \\
        2281.877174 & 0.9989 & 0.0027 \\
        2281.877406 & 0.9986 & 0.0027 \\
        2281.877637 & 0.9989 & 0.0027 \\
        2281.877869 & 1.0034 & 0.0027 \\
        \ldots      & \ldots  & \ldots \\
        \bottomrule
    \end{tabular}
\end{table}

\begin{table*}
    \centering
    \caption{\lcogt{} (Siding Spring Observatory) light curve of \toitwothousand{}.
    (The entire table is available electronically in machine-readable format.)}
    \label{tab:lcogt}
    \begin{tabular}{cccccccc}
        \toprule
        \bjdtdb & Flux & Unc. & Width & Sky & Airmass & Exp.\ Time & Filter \\
        ${} - \num{2457000}$ & (normalized) & & (pixel) & (count/pixel) & & (s) \\
        \midrule
        2255.980346 & 0.9950 & 0.0011 & 15.323563 & 19.373182 & 1.392324 & 29.976 & $B$ \\
        2255.981878 & 1.0028 & 0.0010 & 11.535994 & 15.302150 & 1.387933 & 29.976 & $B$ \\
        2255.983397 & 1.0044 & 0.0010 & 12.790320 & 15.547044 & 1.383667 & 29.972 & $B$ \\
        2255.984914 & 1.0060 & 0.0010 & 12.572493 & 15.866800 & 1.379476 & 29.972 & $B$ \\
        2255.986431 & 1.0042 & 0.0010 & 11.599984 & 15.431446 & 1.375335 & 29.972 & $B$ \\
        \ldots & \ldots & \ldots & \ldots & \ldots & \ldots & \ldots & \ldots\\
        \bottomrule
    \end{tabular}
\end{table*}

\begin{table}
    \centering
    \caption{\pest{} light curve of \toitwothousand.
    (The entire table is available electronically in machine-readable format.)}
    \label{tab:pest}
    \begin{tabular}{ccccc}
        \toprule
        \bjdtdb & Flux & Unc. & Airmass & Filter\\
        ${} - \num{2457000}$ & (normalized) \\
        \midrule
        2328.9512276 & 1.0023 & 0.0047 & 1.2571 & $B$ \\ 
        2328.9527553 & 1.0050 & 0.0047 & 1.2556 & $B$ \\ 
        2328.9542715 & 1.0014 & 0.0047 & 1.2541 & $B$ \\ 
        2328.9557993 & 1.0022 & 0.0046 & 1.2527 & $B$ \\ 
        2328.9573271 & 1.0000 & 0.0046 & 1.2513 & $B$ \\ 
        \ldots    & \ldots  & \ldots & \ldots & \ldots \\
        \bottomrule
    \end{tabular}
\end{table}

\begin{table}
    \centering
    \caption{\astep{} light curve of \toitwothousand{}.
    (The entire table is available electronically in machine-readable format.)}
    \label{tab:astep}
    \begin{tabular}{ccccc}
        \toprule
        \bjdtdb & Flux & Unc. & Airmass & Sky \\
        ${} - \num{2457000}$ & (normalized) & & & (count) \\
        \midrule
        2338.006101 & 0.9992 & 0.0012 & 1.017 & 201 \\
        2338.007462 & 1.0007 & 0.0012 & 1.017 & 203 \\
        2338.009370 & 0.9985 & 0.0012 & 1.017 & 203 \\
        2338.010466 & 1.0016 & 0.0012 & 1.018 & 203 \\
        2338.011563 & 0.9982 & 0.0012 & 1.018 & 209 \\
        \ldots    & \ldots  & \ldots & \ldots & \ldots \\
        \bottomrule
    \end{tabular}
\end{table}

\subsubsection[TESS photometry]{\tess{} Photometry} \label{sec:obs_tess}

\tess{} observed \toitwothousand{} in camera~3 during years~1 and 3 of its mission.
In year~1, \toitwothousand{} was observed in the FFIs at a 30-min cadence during
sectors~9--11 (\ut{} 2019 February 28 -- May 21).
The MIT Quick Look Pipeline \citep[\qlp{};][]{qlp_rnaas_1,qlp_rnaas_2}
detected the outer planet as a 9.13-day transit signal with a depth of 0.43 per cent
at a signal--to--pink noise ratio (S/PN) of 30.62,  %
and it was released as \tess{} Object of Interest (TOI) 2000.01,
having passed all vetting criteria \citep{2021ApJS..254...39G}.
We renamed TOI-2000.01 to \planetouter{} following its confirmation in this paper.

After removing the points where \planetouter{} was in transit,
we detected the 3.10-day signal of the inner planet
with a depth of $474_{-42}^{+39}$~parts per million (ppm)
at a S/PN of $10.56$
through a boxed least-squares (BLS) search \citep{kovacs_bls_2002}
of the \qlp{} light curve.
This search was part of a systematic effort to identify possible inner companions
to all confirmed and candidate hot gas giants
observed by \tess{} during its two-year prime mission.
Later in year~3, \toitwothousand{} was selected by the \tess{} mission for 20-s fast cadence observation
during sectors~36--38 in camera~3 (\ut{} 2021 March 7 -- May 26).
Subsequently, the \tess{} Science Processing Operations Center pipeline
\citep[SPOC pipeline;][]{Jenkins:2016,2010SPIE.7740E..0DJ,2020TPSkdph,2002ApJ...575..493J}
independently detected the signal of the inner planet
in a search%
\footnote{We note that whilst the SPOC search nominally included data from years~1 and 3,
since \toitwothousand{} was only observed in targeted fast cadence in year~3,
the SPOC search only included these sectors.
We performed our own search of the full six sectors of light curves
and identified no additional planetary signals.}
of sectors~1--39 in 2021 August
with a multi-event statistic of $9.5$ and
a signal-to-noise ratio (S/N) of $10.3$.
The transit signature of the inner planet passed all the diagnostic tests
\citep{Twicken:DVdiagnostics2018,Li:DVmodelFit2019}
and the \tess{} Science Office issued an alert on 2022 March 24 for the planet
as TOI-2000.02, which we renamed to \planetinner{} in this paper.
In addition,
the difference image centroiding of the SPOC pipeline
located the source of the transit signatures to within
$\ang{;;0.17} \pm \ang{;;2.4}$
and
$\ang{;;4.2} \pm \ang{;;3.9}$
of \toitwothousand{}
for planets~c and b, respectively.

To produce the year~1 light curve, we used the SPOC-calibrated FFIs
obtained from the TESSCut service \citep{tesscut}.
We performed photometry with a series of 20 different apertures
and corrected these light curves for dilution from the light of other nearby stars.
To make this correction, we first determined the fraction of light from other stars in each aperture
by simulating the \tess{} image with and without contaminating sources
using the location and brightness of nearby stars from the \tess{} Input Catalog
\citep[TIC;][]{tess_tic7,tess_tic8}
and the measured instrument pixel response function%
\footnote{Available online from MAST at \url{https://archive.stsci.edu/missions/tess/models/prf_fitsfiles/}.},
which we determined at the position of the star on the detector using bilinear interpolation.
We then subtracted the contaminating flux and re-normalized the resulting light curves.

After correcting for dilution in the year~1 light curves of each aperture,
we removed instrumental systematics from the light curves
by decorrelating the light curves against the mean and standard deviation
of the spacecraft pointing quaternion time series within each exposure
\citep[processed similarly to][]{vanderburg2019}
and the \tess{} 2-min cadence pre-search data conditioning (PDC) band~3 (fast-timescale) cotrending basis vectors (CBVs) binned to 30 min.
We modeled low-frequency light curve variability with a B-spline
and excluded the points during transits from the systematics correction.
We performed the fit using least squares while iteratively identifying and removing outliers.
After correcting systematics in each aperture, we selected the aperture in each sector that minimized the scatter in the light curve.
The best apertures chosen for the final light curve were all roughly circular and included a total of 14, 8, and 9 pixels
in sectors~9, 10, and 11, respectively.
The processed light curve with and without detrending by B-spline can be found in \autoref{tab:tess30m}.

For the year~3 \tess{} data,
we used the SPOC pipeline's simple aperture photometry flux (\texttt{SAP\_FLUX}) from the 20-s target pixel file
and performed our own systematics correction using a process similar to the 30-min cadence year~1 data.
As before, we exclude points during transits from the systematics correction
and decorrelated against the spacecraft pointing quaternion time series
and the 20-s PDC band~3 CBVs
while modelling the low-frequency variability as a B-spline.
We then corrected for the contaminating flux from nearby stars using the value in the target pixel file's \texttt{CROWDSAP} header.
The processed light curve can be found in \autoref{tab:tess20s}.

\added[id=s]{%
To look for additional planets in the \tess{} light curve,
we performed BLS searches
after masking out the portions containing the two known planets' transits,
following procedures described by \citet{vanderburg_planetary_2016}.
No additional transiting planets were found above our detection threshold ($\mathrm{S}/\mathrm{N} > 9$).
}

\subsubsection{Ground-based photometry of TOI-2000~c} \label{sec:groundphotc}

We observed multiple transits of \planetouter{} using ground-based seeing-limited photometry
to confirm that the transit signal originated from the expected host star.
We observed three full transits with high S/N at different facilities and in various passbands during the first half of 2021.
We detail the three full-transit observations below.
The lower right panel in \autoref{fig:lc} shows the phase-folded light curve of
all ground-based observations in bins weighted by inverse variance.

The Las Cumbres Observatory Global Telescope \citep[LCOGT;][]{Brown:2013}
1-m network node at Siding Spring Observatory, Australia
observed a full transit of \planetouter{}
on \ut{} 2021 February 10
in the \filterB{} and \filterzs{} bands.
Differential photometric data were extracted using \textsc{AstroImageJ} \citep{Collins:2017}
and circular photometric apertures with radii \ang{;;5.8} and \ang{;;6.2}.
The apertures exclude flux from all nearby \gaia{} \edrthree{} stars
that are bright enough to cause the event in the \tess{} aperture.
The event arrived on time.
The light curves in both filters can be found in \autoref{tab:lcogt}.

The Perth Exoplanet Survey Telescope (\pest{}, a $12''$ Meade LX200 SCT Schmidt--Cassegrain telescope in Perth, Australia) observed a full transit of \planetouter{} in alternating \filterB{} and \filterIc{} bands at \SI{60}{\second} cadence
on \ut{} 2021 April 24.
The photometric apertures were uncontaminated and sized \ang{;;9.2} and \ang{;;7.8}, respectively.
The transit was detected on time in the \filterIc{} band
but was more marginal in the \filterB{} band.
Detrending by airmass improved the \filterB{} band signal.
\autoref{tab:pest} contains the light curve in both filters.

The 0.4-m telescope of the Antarctica Search for Transiting Exoplanets
\citep[\astep{};][]{Guillot+2015}
programme at Dome C, Antarctica
observed a full transit of \planetouter{} on \ut{} 2021 May 3 with an uncontaminated \ang{;;9.3} aperture in the \filterRc{} band.
The event arrived on time.
The light curve,
extracted through procedures described by \citet{Mekarnia+2016},
can be found in \autoref{tab:astep}.

\subsubsection{Ground-based photometry of TOI-2000~b} \label{sec:groundphotb}

We observed \toitwothousand{} using \lcogt{} telescopes on six nights%
\footnote{These light curves are available from the ExoFOP-TESS site at
\url{https://exofop.ipac.caltech.edu/tess/target.php?id=371188886}.}
in 2021 and 2022.
The transit of TOI-2000~b is too shallow ($\approx 470 \, \text{ppm}$)
to detect on target in standard ground-based observations,
so instead the goal of our observations was to rule out
that the periodic transit signal of \planetinner{} detected in \tess{} light curves
actually originated from nearby sources in the sky.
As detailed in \autoref{tab:phot},
we mostly used the 1-m telescopes on the \lcogt{} sites at
Siding Spring Observatory in Australia,
Cerro Tololo Inter-American Observatory in Chile,
and South African Astronomical Observatory in South Africa.
Using the observations on \ut{} 2021 February 6 at Cerro Tololo Inter-American Observatory, Chile,
which covered nearly 6 hours before the transit ingress, the full transit ingress, and more than 70 per cent of the transit window,
we were able to rule out that the transit event occurred on all nearby \gaia{} \edrthree{} targets within $2'$,
except for a
$\Delta \mathrm{mag}= 8.7$ (\textit{TESS} band)
neighbour \ang{;;5.3} north of the target, TIC\thinspace 845089585.

As expected,
we could not confidently detect the shallow transit signal due to \planetinner{} during those observations.
Nevertheless, the ground-based photometry rules out virtually all potential nearby sources of contamination,
which allows us to rule out the false-positive scenario due to resolved nearby eclipsing binaries in \autoref{sec:analyzeb}.

\begin{figure*}
    \centering
    \includegraphics[width=\textwidth]{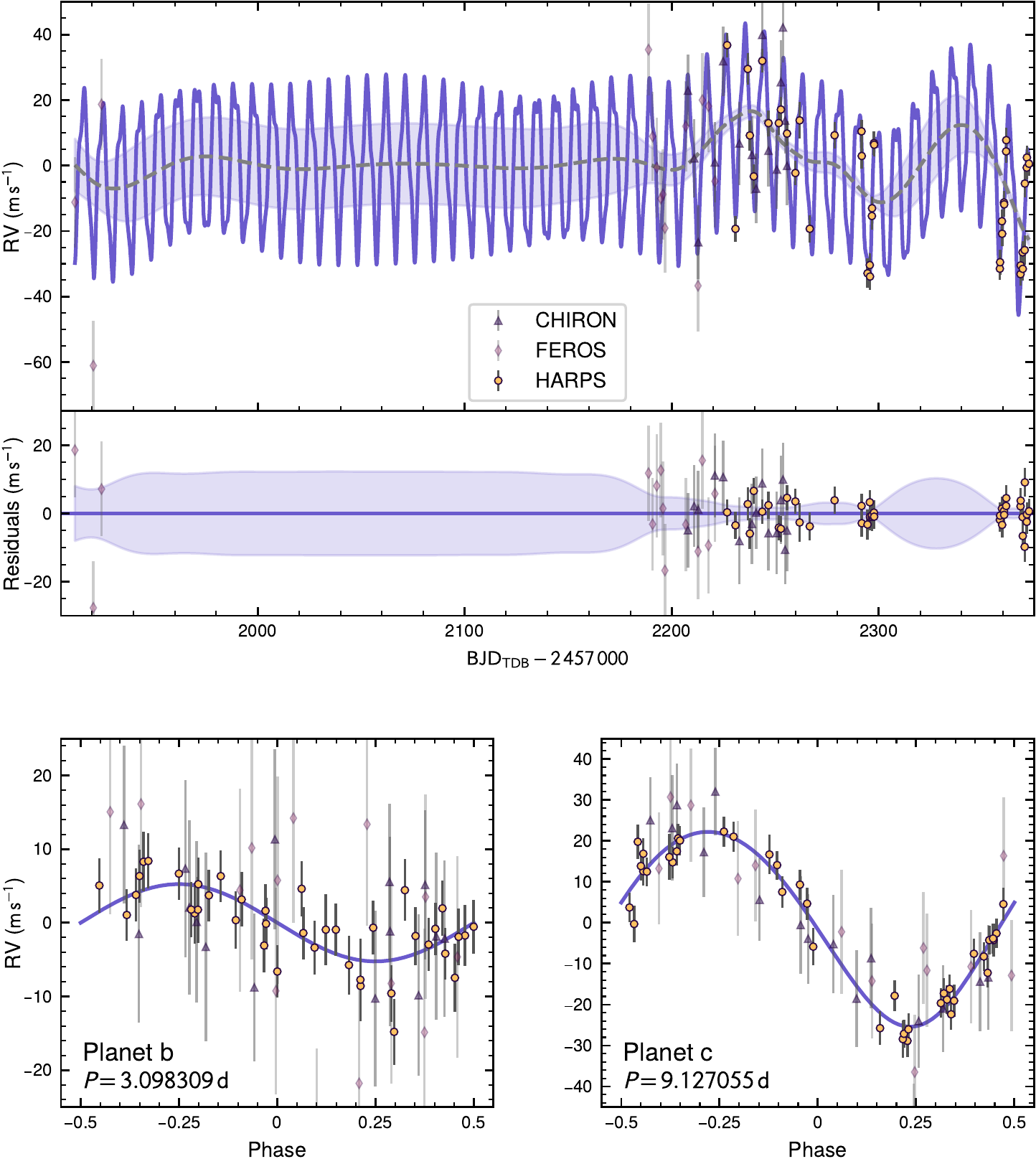}
    \caption{RV measurements of \toitwothousand{}.
    The shapes and colours of each mark indicate which spectrograph it corresponds to,
    and its error bar is the quadrature sum of uncertainty and an instrumental jitter term (\autoref{sec:globalmodel}).
    Solid purple lines represent the best-fitting RV model.
    Top: The best-fitting RV and Gaussian process (GP) model.
    The dashed grey line represents the inferred GP model of residuals in excess of known planet-induced variations.
    The purple-shaded interval represent the $1\sigma$ uncertainty of the GP model.
    Middle: Residual of the RV measurements with respect to the best-fitting RV and GP model.
    Bottom left: Phase-folded RV variations due to planet~b only.
    Bottom right: Phase-folded RV variations due to planet~c only.}
    \label{fig:rv}
\end{figure*}

\begin{table*}
    \caption{Summary of spectroscopic observations.}
    \label{tab:spec}
    \renewcommand{\arraystretch}{1.25}
    \begin{tabular}{lcccccr@{}lr@{}l}
        \toprule
        Instrument
        & Date(s)
        & No. Spectra
        & Resolution
        & Wavelengths
        & S/N
        & \multicolumn2c{Jitter}
        & \multicolumn2c{$\gamma$}
        \\
        & (UT) &
        & $(\lambda / \Delta\lambda / 1000)$
        & (nm)
        & (at \SI{500}{\nano\meter})
        & \multicolumn2c{(\si{\meter\per\second})}
        & \multicolumn2c{(\si{\meter\per\second})}
        \\
        \midrule
        \chiron{}   & 2020 Dec 24 -- 2021 Feb 10& 15    &  80   & 410--870  & 70 &
        \sysParamRvJitterSubZero & \sysParamRvJitterSubZeroUnc & \sysParamRvGammaSubZero & \sysParamRvGammaSubZeroUnc \\
        \feros{}    & 2020 Mar 3 -- 2021 Jan 6  & 14    &  48   & 350--920  & 56--97 &
        \sysParamRvJitterSubOne & \sysParamRvJitterSubOneUnc & \sysParamRvGammaSubOne & \sysParamRvGammaSubOneUnc \\
        \harps{}    & 2021 Jan 12 -- Jun 7      & 41    & 115   & 378--691  & 27.9--56.4 &
        \sysParamRvJitterSubTwo & \sysParamRvJitterSubTwoUnc & \sysParamRvGammaSubTwo & \sysParamRvGammaSubTwoUnc \\
        \bottomrule
    \end{tabular}
\end{table*}

\begin{table}
    \caption{Radial velocity measurements of \toitwothousand{}.
    The RVs \added[id=s]{from \feros{} and \harps{}} are \added[id=s]{measured} relative to the Solar System barycentre.
    (The entire table is available electronically in machine-readable format.)}
    \label{tab:rv}
    \begin{tabular}{@{}SS@{}Sr}
    \toprule
    {\bjdtdb{}}
        & {RV}
        & {Uncertainty}
        & {Instrument} \\
    {${} - \num{2457000}$}
        & {(\si{\meter\per\second})}
        & {(\si{\meter\per\second})}
        & \\
    \midrule
2207.81850 &	6671.1 &	8.2 &	CHIRON \\
2210.82757 &	6650.5 &	9.6 &	CHIRON \\
2212.80571 &	6624.8 &	8.7 &	CHIRON \\
{\ldots} & {\ldots} & {\ldots} & {\ldots} \\
1911.72374 &	8106.2 &	6.3 &	FEROS \\
1920.62204 &	8056.3 &	6.1 &	FEROS \\
1924.57801 &	8136.1 &	6.6 &	FEROS \\
{\ldots} & {\ldots} & {\ldots} & {\ldots} \\
2226.762805 &	8155.81 &	2.10 &	HARPS \\
2230.798279 &	8099.73 &	2.46 &	HARPS \\
2236.721230 &	8148.55 &	3.74 &	HARPS \\
{\ldots} & {\ldots} & {\ldots} & {\ldots} \\
    \bottomrule
    \end{tabular}
\end{table}

\subsection{Spectroscopy}  \label{sec:rv}

We observed \toitwothousand{} with the \chiron{}, \feros{}, and \harps{} spectrographs,
and the details are summarized in \autoref{tab:spec}.
\autoref{tab:rv} lists all the radial velocity (RV) measurements and their uncertainties,
which are also shown in \autoref{fig:rv}.

We obtained 15 spectra of \toitwothousand{} with the \chiron{} spectrograph
\citep{2013PASP..125.1336T}
on the \SI{1.5}{\meter} SMARTS telescope
located at Cerro Tololo Inter-American Observatory, Chile.
\chiron{} is a high-resolution echelle spectrograph, fed via a fibre bundle,
with a spectral resolving power of $\lambda / \Delta \lambda \equiv R \approx \num{80000}$
from \SIrange{4100}{8700}{\angstrom} for slicer mode observations.
The spectra were extracted by the standard \chiron{} pipeline
\citep{2021AJ....162..176P}.
The RVs were derived for each spectrum by cross-correlating against a median-combined template spectrum.
The template spectrum was a median combination of all \chiron{} spectra,
each shifted to rest after an approximate velocity measurement
via cross correlation against a synthetic template.
The measured velocity of each spectrum is that of the mean velocity from each spectral order, weighted by their cross correlation function heights.
The velocity uncertainties were estimated from the scatter of the per-order velocities.
We find a mean internal uncertainty of \SI{9}{\meter\per\second}
and S/N $\approx 70$ per spectral resolution element.

\added[id=s]{We acquired 14 spectra of \toitwothousand{}
at $R \approx \num{48000}$ with}
the \feros{} spectrograph \citep{1999Msngr..95....8K}
mounted on the MPG/ESO 2.2-m telescope
at La Silla observatory, Chile
between \ut{} 2020 March 3 and 2021 January 6.
Most of the spectra had an exposure time of \SI{600}{\second},
but two had an exposure time of \SI{900}{\second} due to poor weather.
The mean and median S/N per spectral resolution element were 82.4 and 83, respectively,
ranging 56--97 in total.
The instrumental drift was calibrated by simultaneously observing a fibre illuminated with a ThAr+Ne lamp.
The data were processed with the CERES suite of echelle pipelines
\citep{2017PASP..129c4002B},
which produced RVs and bisector spans in addition to reduced spectra.

\added[id=s]{We acquired 41 spectra of \toitwothousand{}
at $R \approx \num{115000}$ with}
the High Accuracy Radial velocity Planet Searcher spectrograph
\citep[\harps{};][]{2003Msngr.114...20M}
between \ut{} 2021 January 12 and June 7.%
\footnote{\harps{} programme IDs 1102.C-0249 (PI: D.~Armstrong) and 106.21ER.001 (PI: R.~Brahm).}
\harps{} is fibre-fed by the Cassegrain focus of the 3.6-m telescope
at La Silla Observatory, Chile.
The spectra were taken with exposure times between \SIrange{900}{1800}{\second}
in high-accuracy mode (HAM),
resulting in S/N of 27.1--56.5 \added[id=s]{at spectral order 60 for individual spectra},
except for a poor quality one at \bjdtdb{} \num{2459282.76},
which we excluded from the joint model (\autoref{sec:globalmodel}).
The spectra were reduced
with the standard Data Reduction Software
\citep[DRS;][]{2002A&A...388..632P,1996A&AS..119..373B},
\added[id=s]{%
using a K0 template to correct the flux balance over spectral orders
before applying a G2 binary cross-correlation function (CCF) mask.
Such colour correction allowed us to minimize the impact of the CCF mask mismatch
between the observed spectral type and the CCF binary mask.
The K0 flux template was selected as the closest one to a G5--G6 star like \toitwothousand{},
and the G2 binary mask was selected to minimize photon noise uncertainty.
}

\subsection{Speckle imaging}

\begin{figure}
    \centering
    \includegraphics[width=\columnwidth]{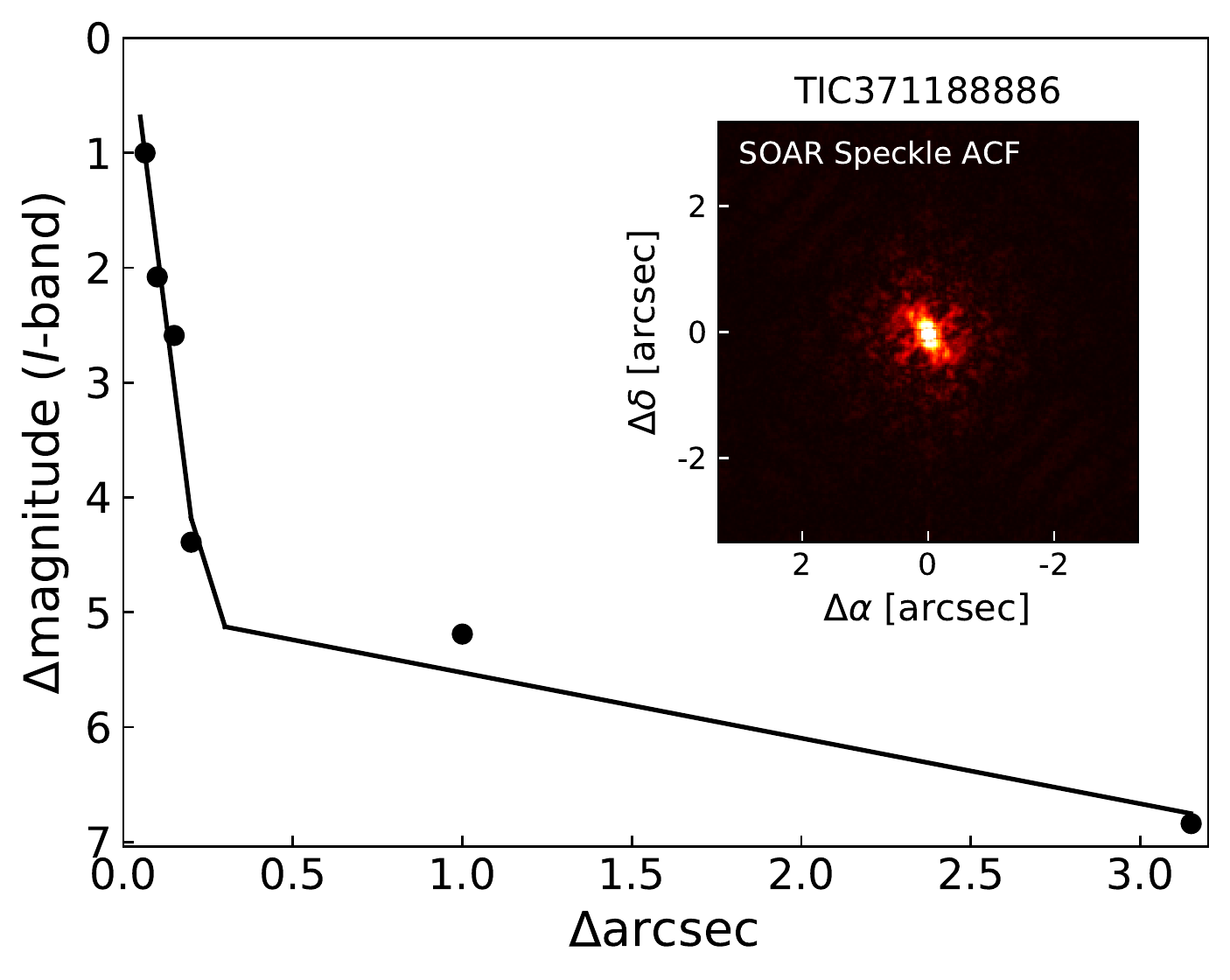}
    \caption{SOAR speckle observations of \toitwothousand{}.
    The curve is the $5\sigma$ detection sensitivity,
    and the inset is the speckle imaging autocorrelation function (ACF).
    No companion is detected within the contrast limit.}
    \label{fig:soar}
\end{figure}

\begin{figure}
    \centering
    \includegraphics[width=\columnwidth]{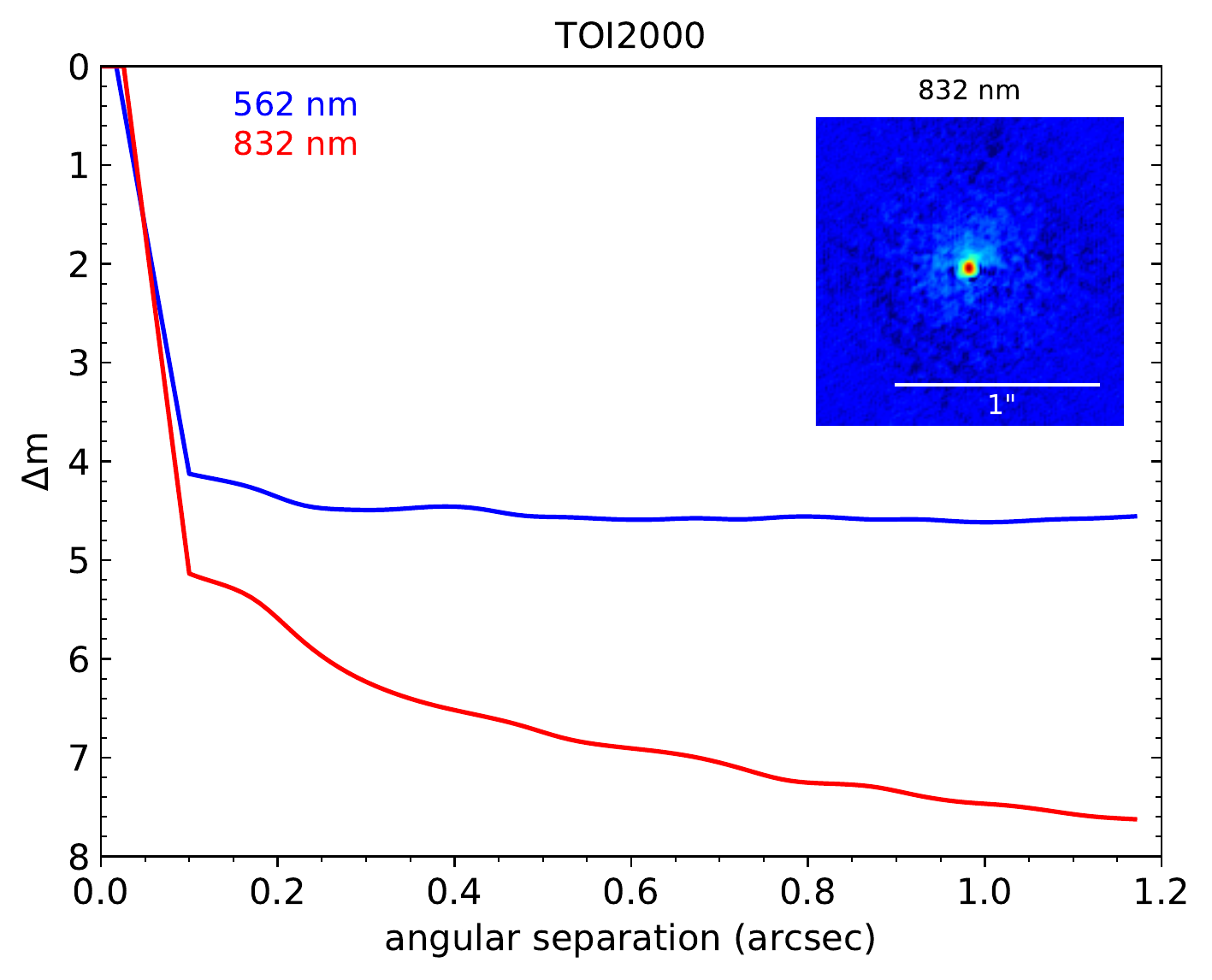}
    \caption{Gemini speckle observations of \toitwothousand{}.
    The curves are the $5\sigma$ detection sensitivity in the
    \SI{562}{\nano\meter} (\emph{blue}) and
    \SI{832}{\nano\meter} (\emph{red}) bands,
    and the inset is the reconstructed speckle image in \SI{832}{\nano\meter}.
    No companion is detected within the contrast limit.}
    \label{fig:gemini}
\end{figure}

We searched for unresolved companions of \toitwothousand{}
with speckle imaging from two telescopes on Cerro Pachón, Chile.
The first set of data was acquired by
the HRCam instrument \citep{2018PASP..130c5002T}
on the 4.1-m Southern Astrophysical Research (\soar{}) telescope
on \ut{} 2020 October 31.
The observations were in a passband similar to \tess{}'s
and were reduced with procedures described by \citet{2020AJ....159...19Z}.
No companion was found
with a contrast of 6.8 magnitudes at $1''$.
The $5\sigma$ sensitivity
and the speckle autocorrelation function (ACF)
from the observations are plotted in
\autoref{fig:soar}.

The second set of speckle imaging data was acquired by the Zorro speckle instrument
on the 8-m \geminisouth{} telescope%
\footnote{\url{https://www.gemini.edu/sciops/instruments/alopeke-zorro/}}
\citep{2021FrASS...8..138S}
on \ut{} 2022 March 17.
Zorro collected 20 sets of 1000 speckle imaging observations simultaneously in two bands
(\SI{562}{\nano\meter} and \SI{832}{\nano\meter})
with an integration time of \SI{60}{\milli\second} per frame.
These thousands of observations were reduced using the method described by \citet{2011AJ....142...19H},
yielding a high-resolution view of the sky near \toitwothousand{}.
\autoref{fig:gemini} shows the two $5\sigma$ contrast curves and the reconstructed speckle image in \SI{562}{\nano\meter} and \SI{832}{\nano\meter}.
Again, we found no companion to \toitwothousand{} in the 832-nm band within a contrast of 5--8 magnitudes
from \ang{;;0.1} to \ang{;;1.2} separation,
which corresponds to a projected separation of \SIrange{17.4}{208}{\au}
at the distance of \toitwothousand{} (\SI{173.6}{\parsec}).

\section{Analysis} \label{sec:analysis}

\subsection{Stellar parameters} \label{sec:stellarparams}

\subsubsection{Spectral energy distribution} \label{sec:sed}

\begin{figure}
    \centering
    \includegraphics[width=\columnwidth]{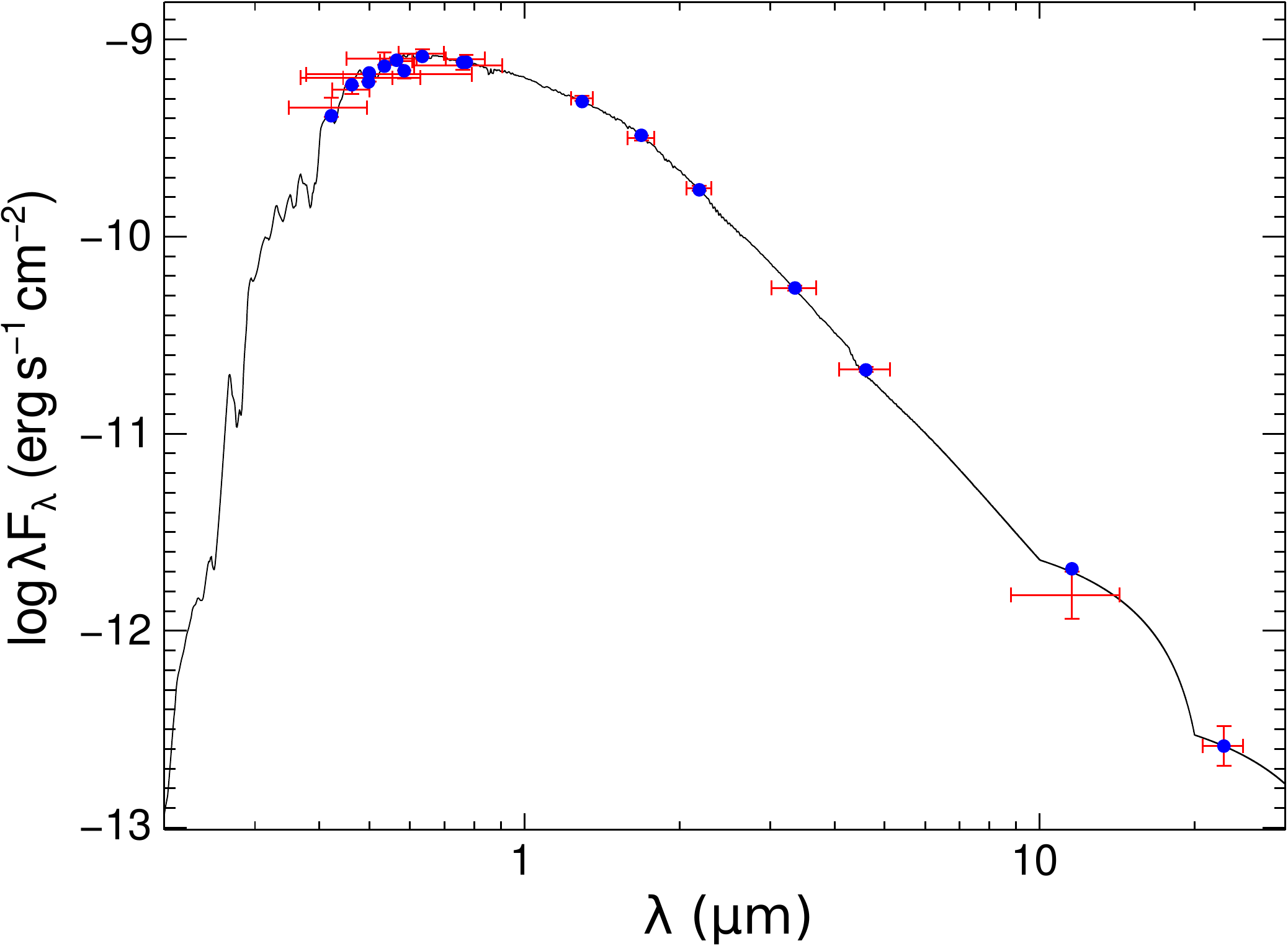}
    \caption{Spectral energy distribution of \toitwothousand{}.
    Red symbols represent the observed photometric measurements,
    where the horizontal bars represent the effective width of the passband.
    Blue symbols are the model fluxes from the best-fitting Kurucz atmosphere model (black).  \label{fig:sed}}
    \label{fig:my_label}
\end{figure}

Independently of the obtained spectra in \autoref{sec:rv},
we analysed the broadband spectral energy distribution (SED) of the star
following the procedures described by \citet{Stassun:2016} and \citet{Stassun:2017,Stassun:2018}.
We used the the \filterB{}, \filterV{}, \filtergp{}, \filterrp{}, \filterip{} magnitudes from APASS,
the \filterJ{}, \filterH{}, \filterK{} magnitudes from \twomass{},
the \filterWone{}--\filterWfour{} magnitudes from \wise{},
and the \gaiaG{}, \gaiaBP{}, \gaiaRP{} magnitudes from \gaia{}.
Together, the available photometry spans the wavelength range \SIrange[range-phrase=--]{0.4}{10}{\micro\meter}
(\autoref{fig:sed}).

We fit the SED using \citet{Kurucz-93} stellar atmosphere models,
with effective temperature \tempeff{},
metallicity \feh{},
and extinction $A_V$
as free parameters.
We limited $A_V$ to the maximum line-of-sight value from the Galactic dust maps of \citet{Schlegel:1998}.
The resulting best-fitting (reduced $\chi^2 = 1.2$) values are
$\tempeff{} = \SI{5550 \pm 75}{\kelvin}$,
$\feh{} = 0.3 \pm 0.2$,
and $A_V = 0.15 \pm 0.04$.
Integrating the unreddened SED model gives the bolometric flux at Earth
$F_{\rm bol} = \SI{1.169 \pm 0.014e-9}{\erg\per\second\per\centi\meter\squared}$.
Together with the \gaia{} \edrthree{} parallax,
the $F_{\rm bol}$ and \tempeff{} values
imply a stellar radius $R_\star = \SI{1.135 \pm 0.031}{\radius\sun}$,
consistent with the value from joint modelling (\autoref{sec:globalmodel}).

\subsubsection{Spectroscopy and chemical abundances} \label{sec:specanalysis}

\begin{table}
    \centering
    \caption{Chemical abundances of \toitwothousand{} with respect to the Sun.}
    \label{tab:abudances}
    \begin{tabular}{ccc}
        \toprule
        Element (X) & [X/H] & Uncertainty \\
        \midrule
        \ion{C}{I}   & 0.26  &  0.02 \\
        \ion{O}{I}   & 0.30  &  0.15 \\
        \ion{Mg}{I}  & 0.44  &  0.06 \\
        \ion{Al}{I}  & 0.52  &  0.04 \\
        \ion{Si}{I}  & 0.43  &  0.03 \\
        \ion{Ca}{I}  & 0.42  &  0.07 \\
        \ion{Ti}{I}  & 0.47  &  0.06 \\
        \ion{Cr}{I}  & 0.45  &  0.05 \\
        \ion{Ni}{I}  & 0.50  &  0.05 \\
        \ion{Cu}{I}  & 0.66  &  0.05 \\
        \ion{Zn}{I}  & 0.45  &  0.03 \\
        \ion{Sr}{I}  & 0.46  &  0.08 \\
        \ion{Y}{II}  & 0.34  &  0.07 \\
        \ion{Zr}{II} & 0.28  &  0.04 \\
        \ion{Ba}{II} & 0.25  &  0.03 \\
        \ion{Ce}{II} & 0.37  &  0.06 \\
        \ion{Nd}{II} & 0.40  &  0.03 \\
        \bottomrule
    \end{tabular}
\end{table}

We used \ares{}+\moog{} \citep{Sousa-14,Santos-13}
to obtain more precise stellar atmospheric parameters
(\tempeff{}, surface gravity \logg{}, microturbulence, \feh{})
from the combined \harps{} spectrum of \toitwothousand{}.
\added[id=s]{The combined spectrum achieved an S/N of 275 at spectral order 60.}
We measured the equivalent widths of iron lines using the
Automatic Routine for line Equivalent widths in stellar Spectra
(\ares{}) v2 code%
\footnote{The latest version of \ares{}~v2 can be downloaded from \url{http://www.astro.up.pt/~sousasag/ares}}
\citep{Sousa-15}.
Then, we used a minimization process
where we assumed ionization and excitation equilibrium to converge
to the best set of spectroscopic parameters.
This process made use of a grid of \citet{Kurucz-93} model atmospheres
and the radiative transfer code \moog{} \citep{Sneden-73},
yielding the values
$\tempeff{} = \SI{5568 \pm 66}{\kelvin}$,
$\feh{} = 0.438 \pm 0.044$.
\added[id=s]{%
These \tempeff{} and \feh{} values were then used to constrain the \mist{} evolutionary track portion
of the joint model (\autoref{sec:jointmodelmist}).
We also used the \gaia{} \edrthree{} parallax measurement to derive a stellar surface gravity
$\logg{} = 4.38 \pm 0.03$,
consistent with the value independently derived by the joint model (\autoref{sec:globalmodel}).
The spectroscopic \logg{} was \emph{not} used to constrain the joint model.
}

The combined \harps{} spectrum also gave the abundances of various chemical elements in \toitwothousand{}
(see \autoref{tab:abudances}).
These abundances were derived
from the same code and models as the stellar parameters,
using classical curve of growth analysis
assuming local thermodynamic equilibrium.
For the derivation of abundances of refractory elements,
we closely followed the methods described by
\citet{Adibekyan-12, Adibekyan-15} and \citet{Delgado-17}.
Abundances of the volatile elements C and O were derived following the methods of
\citet{Delgado-21} and \citet{Bertrandelis-15}.
All the [X/H] ratios were calculated by differential analysis with respect to a
high S/N Solar (Vesta) spectrum from \harps{}.

These detailed abundances allowed us to measure the age of \toitwothousand{} through chemical clocks,
or abundance ratios strongly correlated with age.
We found the [X/Fe] ratios of TOI-2000 to be typical for a thin-disc star.
Considering the variation in age due to \tempeff{} and \feh{},
we applied the 3D formulas described by \citet{Delgado-19} in their Table~10
to calculate ages associated with the chemical clocks
[Y/Mg], [Y/Zn], [Y/Ti], [Y/Si], [Y/Al], [Sr/Ti], [Sr/Mg], and [Sr/Si].
Their weighted average
gave an independently measured age of \SI{4.1 \pm 1.6}{\giga\year},
within the uncertainty of the value from joint modelling
(\autoref{sec:globalmodel}).

\added[id=s]{%
In addition to chemical abundances,
we also measured the projected rotational velocity
$\vsini$ to be \SI{2.99\pm 0.20}{\km\per\s}
by performing spectral synthesis with \moog{} on 36 isolated iron lines
and by fixing the stellar parameters and limb-darkening coefficient \citep{CostaSilva-20}.
The limb-darkening coefficient ($\approx 0.6$) was determined using the stellar parameters
as described by \citet{Espinoza-15} assuming a linear limb darkening law.}

\subsection{Radial velocity variations} \label{sec:rv_analysis}

\added[id=t]{
Since transits are sensitive only to planets with inclinations precisely aligned to our line of sight,
it is possible additional planets may orbit in the \toitwothousand{} system but not be detected in the \tess{} observations.
Therefore, we proceed to investigate whether additional planetary signals are present
by computing and examining periodograms of our RV measurements.
}

\subsubsection{Frequency analysis of RVs} 

\begin{figure*}
    \centering
    \includegraphics[width=\textwidth]{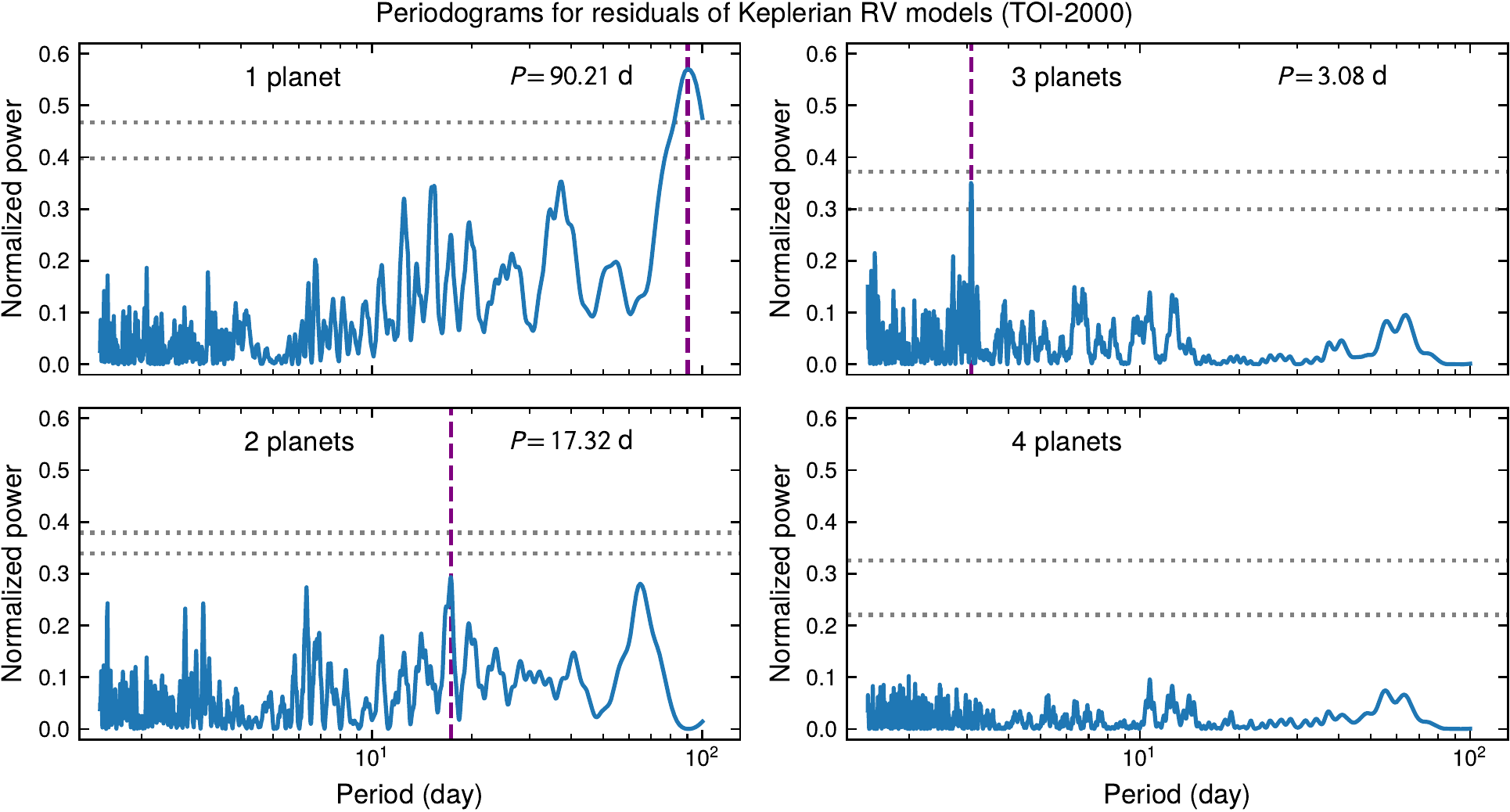}
    \caption{Periodograms of RV residuals after subtracting Keplerian signals.
    At each step, a Keplerian signal corresponding to the period peak in the previous periodogram is added.
    The vertical dashed purple line marks the most significant peak in the period interval from \SIrange{1.5}{100}{\day},
    and the horizontal dotted grey lines indicate the 1 and 0.1 per cent false-alarm levels computed with bootstrap resampling.}
    \label{fig:rv_period}
\end{figure*}

\added[id=t]{%
We investigate the possibility that there are signals for additional planets in the system
in the RV measurements (\autoref{sec:rv}).
Starting from the first 9.1-d signal of \planetouter{},
we iteratively remove new Keplerian signals at periodogram peaks.
At each step, we calculate the Lomb--Scargle \citep[LS;][]{1976Ap&SS..39..447L,1982ApJ...263..835S} periodogram,
fit for a new circular Keplerian (i.e. sinusoidal) signal at the next significant peak,
remove the signal due to the new planet,
and then recalculate the LS periodogram.
\autoref{fig:rv_period} shows this process.}%
\footnote{\added[id=t]{This procedure is performed using the Data and Analysis Center for Exoplanets
(\href{https://dace.unige.ch}{DACE}) facility
from the University of Geneva.}}
\added[id=t]{%
We found two additional periodogram peaks:
a second one at \SI{90.2}{\day}, which is highly formally significant based on the false-alarm probability (FAP) calculated by bootstrap resampling
\citep[see e.g.][]{2020sdmm.book.....I},
and a third one at \SI{17.3}{\day} (FAP $\approx 5$ per cent).
After removing these two signals, we finally identify a fourth peak at \SI{3.1}{\day}, which corresponds to the period of \planetinner{}, with an FAP $\approx 1$ per cent.
}

\added[id=t]{%
Given these detections, we assess whether the two additional RV signals are viable planet candidates.
Although both signals were detected before that of the transiting mini-neptune \planetinner{},
there are reasons to caution that they may not be planetary.
In particular, the 17.3-d signal has a relatively high FAP ($> 1$ per cent).
Although the 90-d signal is formally statistically significant,
our observations cover fewer than two full phase cycles
and consequently have yet to establish that the signal is repeating.
Out of these considerations, we proceed to investigate alternative explanations for the additional
two peaks at \SIlist{17.3;90}{\day},
namely stellar activity and rotation, in the following subsections.
}

\subsubsection{Correlation with stellar activity}

\begin{figure*}
    \centering
    \includegraphics[width=\textwidth]{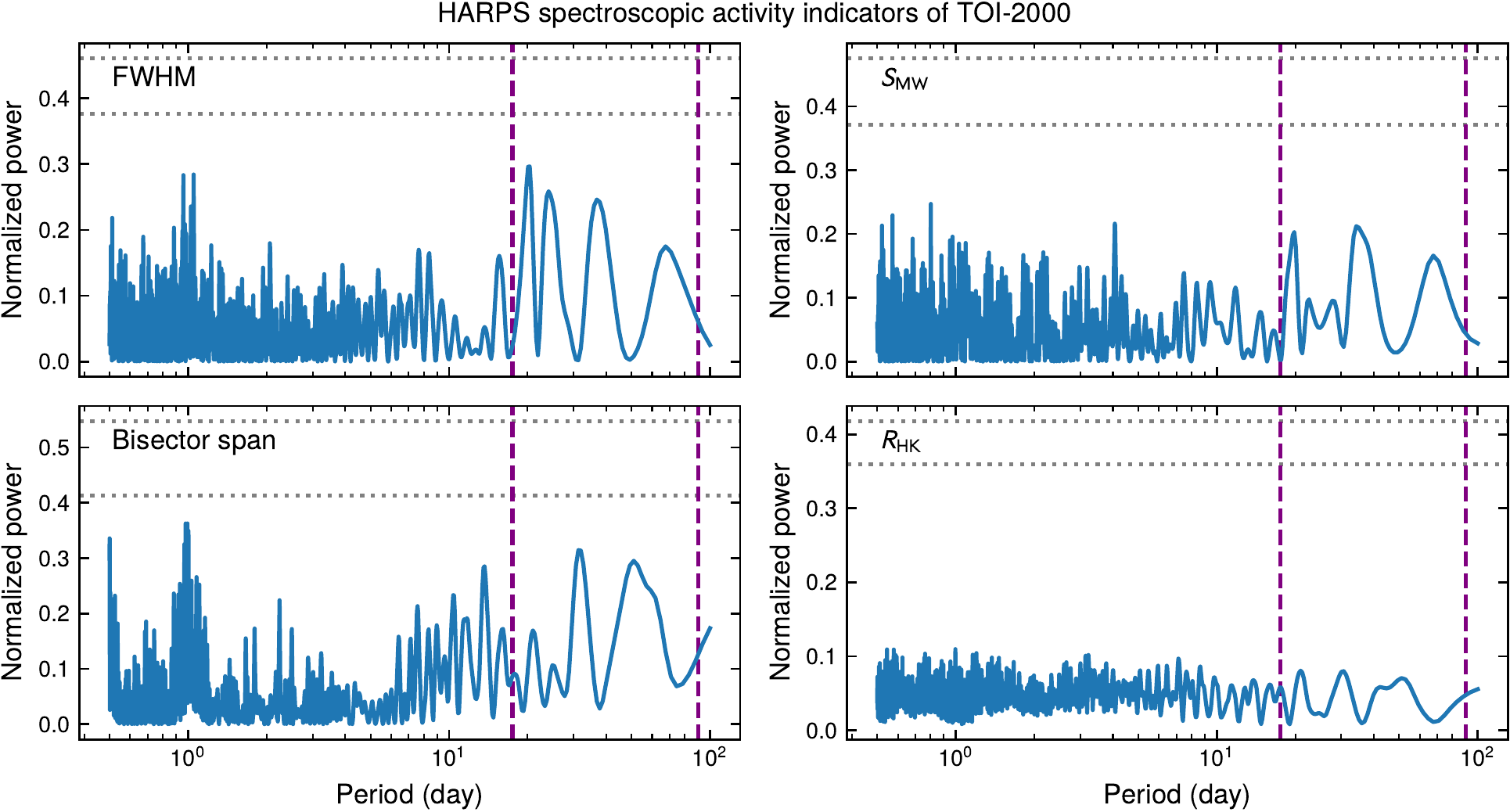}
    \caption{Lomb--Scargle periodograms of \harps{} spectroscopic activity signals for \toitwothousand{}.
    The activity signals are
    full-width half maximum (FWHM, \emph{Top Left}),
    bisector span (\emph{Bottom Left}),
    Mount Wilson $S$-index (\emph{Top Right}),
    and spectral index of the \ion{Ca}{II} H and K lines $R_\text{HK}$ (\emph{Bottom Right}).
    The two vertical dashed purple lines mark the periods of the RV signals at \SIlist{17.3;90}{\day},
    and the three horizontal dotted grey lines indicate the 1 and 0.1 per cent
    false-alarm levels computed from bootstrap resampling.
    There are no significant peaks at the periods of the candidate planets.
    The activity signals are tabulated in \autoref{tab:rv_harps}.}
    \label{fig:rv_activity}
\end{figure*}

\added[id=t]{%
We check the standard stellar activity indicators from the \harps{} RV measurements
to test for the possibility that the two additional RV signals are due to stellar activity.
\autoref{fig:rv_activity}
shows LS periodograms of the HARPS full-width half maximum, bisector span, Mount Wilson $S$-index,
and the $R_\text{HK}$ spectral index of \ion{Ca}{II} H and K lines.
None of the four indicators show significant peaks corresponding to the RV signals at \SIlist{17.3;90}{\day}.
}

\subsubsection{Correlation with stellar rotation}

\begin{figure*}
    \centering
    \includegraphics[width=\textwidth]{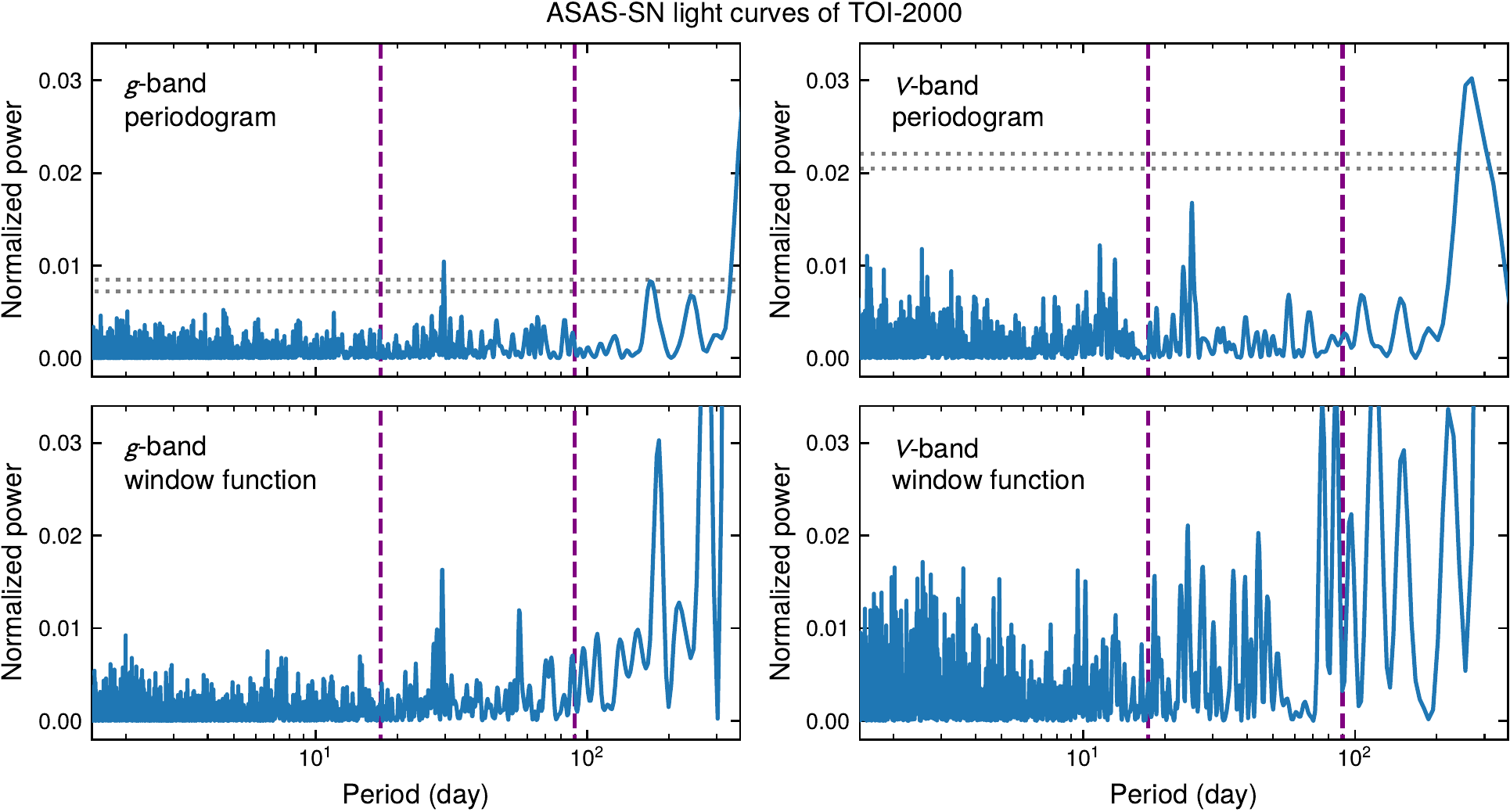}
    \caption{Lomb--Scargle periodogram of the ASAS-SN light curve of \toitwothousand{} in the $g$ (\emph{Left})
    and \filterV{} (\emph{Right}) bands,
    with associated window functions (\emph{Bottom}).
    The $x$-axis ranges from \SI{1.5}{\day} to \SI{1}{\year}.
    The two vertical dashed purple lines mark the period of the candidate non-transiting planets at \SIlist{17.3;90}{\day},
    and the two horizontal dotted grey lines indicate the 1 and 0.1 per cent false-alarm levels
    computed from bootstrap resampling.
    There is no significant peak that does not correspond to a peak in the window function,
    and there is no peak that corresponds to the 17.3 or 90-d RV signal.
    }
    \label{fig:assasn}
\end{figure*}

\added[id=t]{%
As the \tess{} light curve of \toitwothousand{} does not show significant periodic variations,
we check Wide Angle Search for Planets (WASP) and
the All-Sky Automated Survey for Supernovae
\citep[ASAS-SN;][]{2014ApJ...788...48S}
archival data for hints of the stellar rotation signal.
There is no WASP light curve for \toitwothousand{},
but ASAS-SN has light curves in the $g$ and $V$ bands
spanning \ut{} 2016 January 29 to 2023 February 14
\citep{2017PASP..129j4502K}.
\autoref{fig:assasn} shows LS periodograms of the ASAS-SN light curves
and their window functions.
Again, no stellar rotation is detected,
as there is no significant peak that does not correspond to a peak in the window function.
In addition, the periodograms do not show peaks at the periods of
the 17.3 and 90-d RV signals.
Thus, we cannot determine whether the additional RV signals are associated with stellar rotation.
}

\subsubsection{Two-planet vs. four-planet solutions}

\added[id=t]{%
In the remainder of \autoref{sec:analysis},
we present models assuming two or more planets in the \toitwothousand{} system.
In \autoref{sec:globalmodel}, we describe a two-planet solution
that jointly models transits, RVs, and stellar parameters.
In \autoref{sec:extra_planets}, we describe an alternative four-planet solution
based on an RV-only model,
compare it to the two-planet solution,
and justify our present preference for the two-planet solution.
}

\subsection{Joint modelling} \label{sec:globalmodel}

\begin{table*}
\caption{\toitwothousand{} stellar parameters.}
\label{tab:stellar}
\renewcommand{\arraystretch}{1.25}
\begin{tabular}{lr@{}lcl}
\toprule
Parameter \dotfill Unit
& \multicolumn2c{Value}
& Prior
& Source
\\
\midrule
\addlinespace[4pt]
Identifying Information \\
\addlinespace[1pt]
$\alpha$, right ascension (J2016) \dotfill h:m:s     & \multicolumn2c{\hourang{09;45;35.289}}    & -- & \gaia{} \edrthree{} \\
$\delta$, declination (J2016) \dotfill d:m:s    & \multicolumn2c{\ang{-66;41;11.86}}        & -- & \gaia{} \edrthree{} \\
$\mu_\alpha$, R.A. proper motion \dotfill       \si{\mas\per\year}      & $-73.219$ & $\pm 0.012$  & -- & \gaia{} \edrthree{} \\
$\mu_\delta$, decl.\ proper motion \dotfill     \si{\mas\per\year}      & $12.623$  & $\pm 0.012$  & -- & \gaia{} \edrthree{} \\
TIC ID \dotfill & \multicolumn2c{\toitwothousandtic} & -- & TIC\thinspace v8.2 \\
\gaia{} \edrthree{} ID \dotfill & \multicolumn2c{5244434756689177088} & -- & \gaia{} \edrthree{} \\
\addlinespace[4pt]
Photometric Properties \\
\addlinespace[1pt]
\tess{} \dotfill mag    & $10.358$ & $\pm 0.006$ & -- & TIC\thinspace v8.2 \\
$V$ \dotfill mag        & $10.984$ & $\pm 0.012$ & -- & TIC\thinspace v8.2 \\
\gaiaG{} \dotfill mag   & $10.85420$ & $\pm 0.00016$ & -- & \gaia{} \edrthree{} \\
\addlinespace[4pt]
Spectroscopic Properties \\
$v \sin i$, rotational speed\tablenotemark{a}   \dotfill \si{\km\per\s} & $2.99$ & $\pm 0.20$ & -- & \harps{} \\
\addlinespace[1pt]
\addlinespace[4pt]
Sampled Properties \\
\addlinespace[1pt]
$M_{\star, 0}$, initial mass \dotfill           \si{\mass\sun}  & \sysParamMStarZero & \sysParamMStarZeroUnc & $\sim \uniformdist(0.1, 4)\,\normaldist(1.05, 0.2)$ & -- \\
$\feh_0$, initial metallicity \dotfill          dex                     & \sysParamFehZero & \sysParamFehZeroUnc & $\uniformdist(0, 0.5)$ & -- \\
EEP, MESA equivalent evolutionary point \ldots & \sysParamEep & \sysParamEepUnc & $\uniformdist(202, 454)$\tablenotemark{b} & -- \\
Parallax \dotfill                               mas                     & \sysParamParallax  & \sysParamParallaxUnc & $\normaldist(5.773, 0.010)$\tablenotemark{c} & \gaia{} \edrthree{} \\
$A_V$, extinction \dotfill                      mag                     & \sysParamAv & \sysParamAvUnc & $\uniformdist(0, 0.70)$ & Schlegel 1998 \\
\addlinespace[4pt]
Derived Properties \\
\addlinespace[1pt]
\tempeff{}, effective temperature \dotfill      K                       & \sysParamTeff & \sysParamTeffUnc & $\normaldist(5568, 100)\,\normaldist(\text{MIST},3\%)$\tablenotemark{d} & \harps{} \\
\feh{}, metallicity \dotfill                    dex                     & \sysParamFeh & \sysParamFehUnc & $\normaldist(0.438, 0.044)\,\normaldist(\text{MIST},3\%)$\tablenotemark{d} & \harps{} \\
\logg{}, surface gravity \dotfill           dex(\si{\cm\per\s\squared}) & \sysParamLoggStar & \sysParamLoggStarUnc & -- & -- \\
$M_\star$, mass \dotfill                        \si{\mass\sun}          & \sysParamMStar & \sysParamMStarUnc & -- & -- \\
$R_\star$, radius \dotfill                      \si{\radius\sun}        & \sysParamRStar & \sysParamRStarUnc & $\normaldist(\text{MIST},3\%)$\tablenotemark{d} & -- \\
$\rho_\star$, density \dotfill                  \si{\gram\per\cm\cubed} & \sysParamRhoStar & \sysParamRhoStarUnc & -- & -- \\
$L_\star$, bolometric luminosity \dotfill $\si{\luminosity\sun}$& \sysParamLStar & \sysParamLStarUnc & -- & -- \\
Age \dotfill                                    Gyr                     & \sysParamAge & \sysParamAgeUnc & $\uniformdist(0, 10)$\tablenotemark{e} & -- \\
distance                                        \dotfill pc             & \sysParamDistance & \sysParamDistanceUnc & -- & -- \\
\bottomrule
\addlinespace[4pt]
\end{tabular}

\raggedright

\tablenotemark{a}%
Not part of the joint model, measured directly from HARPS spectra.

\tablenotemark{b}%
Constrains the star to be main sequence, which we know from spectroscopic analysis.
The effective prior is uniform in age rather than EEP because the model likelihood is multiplied by 
$ \partial(\mathrm{EEP}) / \partial(\mathrm{age}) $.

\tablenotemark{c}%
Corrected for the \gaia{} \edrthree{} parallax zero-point as a function of magnitude, colour, and position
using the prescription of \citet{2021A&A...649A...4L}.

\tablenotemark{d}%
At each step of the HMC chain,
the sampled values of $(M_{\star,0}, \feh_0, \mathrm{EEP})$
are used as the input to the \mist{} grid,
which output $M_{\star,\text{MIST}}$,
$\logg{}_{\star,\text{MIST}}$,
$\tempeff{}_{,\text{MIST}}$,
$\feh_{\text{MIST}}$,
and stellar age.
These interpolated values are used to compute
$R_{\star,\text{MIST}}$.
The shorthand $\normaldist(\text{MIST}, 3\%)$ means that
the normal distribution is centred on the interpolated value of the \mist{} grid
with a fractional uncertainty of 3\%.

\tablenotemark{e}%
Approximated by adding a logistic function
$f(x) = -10^{100} / (1 + e^{-3000(x - 10)})$
to the log-probability of the model.

\emph{Note.}
$\uniformdist(a, b)$ is the uniform distribution over the interval $[a, b]$.
$\normaldist(\mu, \sigma)$ is
the normal distribution with mean $\mu$
and standard deviation $\sigma$.
\added[id=s]{The values and uncertainties quoted are 68\% credible intervals centred on medians.}

\emph{References.}
\gaia{} \edrthree{}: \citet{2021A&A...649A...1G}.
TIC\thinspace v8.2: \citet{tess_tic8,tess_tic82}.
Schlegel~1998: \citet{Schlegel:1998}.

\end{table*}

\begin{table*}
\caption{\toitwothousand{} planetary system parameters.}
\label{tab:planet}
\renewcommand{\arraystretch}{1.25}
\begin{tabular}{lr@{}l@{}cr@{}l@{}c}
\toprule
Planet
& \multicolumn{3}{c}{b}
& \multicolumn{3}{c}{c}
\\
Parameter \dotfill Unit:
& \multicolumn{2}{c}{Value}
& Prior
& \multicolumn{2}{c}{Value}
& Prior
\\
\midrule
\addlinespace[4pt]
Sampled \\
\addlinespace[1pt]
                                                        
$T_\text{c}$, time of conjunction  \dotfill   BJD     & \sysParamTzeroSubOne & \sysParamTzeroSubOneUnc & $2458855 + \uniformdist(0.224, 0.264)$ & \sysParamTzeroSubZero & \sysParamTzeroSubZeroUnc & $2459110 + \uniformdist(0.0609, 0.0709)$ \\
$P$, period \dotfill                            day     & \sysParamPeriodSubOne & \sysParamPeriodSubOneUnc & $3.09833 + \uniformdist(-2, 2)\times10^{-4}$ & \sysParamPeriodSubZero & \sysParamPeriodSubZeroUnc & $9.127055 + \uniformdist(-10^{-4}, 10^4)$ \\
$\sqrt{e}\cos\omega$ \dotfill                           & 0 & & fixed & \sysParamSqrtEccVecZeroSubZero & \sysParamSqrtEccVecZeroSubZeroUnc & $e < 1$\tablenotemark{a} \\
$\sqrt{e}\sin\omega$ \dotfill                           & 0 & & fixed & \sysParamSqrtEccVecZeroSubOne & \sysParamSqrtEccVecZeroSubOneUnc & $e < 1$\tablenotemark{a} \\
$ b \equiv a \cos(i) / R_\star $, impact parameter \dotfill & \sysParamBSubOne & \sysParamBSubOneUnc & $\uniformdist(0, 1 + R_\text{p}/R_\star)$ & \sysParamBSubZero & \sysParamBSubZeroUnc & $\uniformdist(0, 1 + R_\text{p}/R_\star)$ \\
$ R_{\text{p}} / R_\star $ \dotfill                   & \sysParamRpSubOne & \sysParamRpSubOneUnc & $\uniformdist(0, 0.5)$ & \sysParamRpSubZero & \sysParamRpSubZeroUnc & $\uniformdist(0, 0.5)$ \\
$ M_{\text{p}}/M_\star $ \dotfill                     & \sysParamMPlanetSubOne & \sysParamMPlanetSubOneUnc & $\uniformdist(0, 2.8605 \times 10^{-4})$\tablenotemark{b} & \sysParamMPlanetSubZero & \sysParamMPlanetSubZeroUnc & $\uniformdist(0, 8.5814 \times 10^{-4})$\tablenotemark{b} \\
\addlinespace[4pt]
Derived \\
\addlinespace[1pt]
$T_{14}$, total transit duration \dotfill       hour                    & \sysParamTdurSubOne & \sysParamTdurSubOneUnc & -- & \sysParamTdurSubZero & \sysParamTdurSubZeroUnc & -- \\
$e$, eccentricity \dotfill                                              & 0 & & fixed & \sysParamEccSubZero & \sysParamEccSubZeroUnc & -- \\
$\omega$, argument of periastron \dotfill       deg                     & \multicolumn2c{--} & -- & \sysParamOmegaFoldSubZero & \sysParamOmegaFoldSubZeroUnc &  -- \\
$a$, semimajor axis \dotfill                    au                      & \sysParamASubOne & \sysParamASubOneUnc & -- & \sysParamASubZero & \sysParamASubZeroUnc & -- \\
$ a / R_\star $ \dotfill                                                & \sysParamAorSubOne & \sysParamAorSubOneUnc & -- & \sysParamAorSubZero & \sysParamAorSubZeroUnc & -- \\
$i$, inclination \dotfill                       deg                     & \sysParamInclSubOne & \sysParamInclSubOneUnc & -- & \sysParamInclSubZero & \sysParamInclSubZeroUnc & -- \\
$K$, RV semiamplitude \dotfill                  \si{\m\per\s}           & \sysParamKSubOne & \sysParamKSubOneUnc & -- & \sysParamKSubZero & \sysParamKSubOneUnc & -- \\
$R_{\text{p}}$, radius \dotfill     \si{\radius\earth}      & \sysParamRPlanetEarthSubOne & \sysParamRPlanetEarthSubOneUnc & -- & \sysParamRPlanetEarthSubZero & \sysParamRPlanetEarthSubZeroUnc & -- \\
$R_{\text{p}}$, radius \dotfill     \si{\radius\jupiter}      & \sysParamRPlanetJupiterSubOne & \sysParamRPlanetJupiterSubOneUnc & -- & \sysParamRPlanetJupiterSubZero & \sysParamRPlanetJupiterSubZeroUnc & -- \\
$M_{\text{p}}$, mass \dotfill           \si{\mass\earth}      & \sysParamMPlanetEarthSubOne & \sysParamMPlanetEarthSubOneUnc & -- & \sysParamMPlanetEarthSubZero & \sysParamMPlanetEarthSubZeroUnc & -- \\
$M_{\text{p}}$, mass \dotfill           \si{\mass\jupiter}      & \sysParamMPlanetJupiterSubOne & \sysParamMPlanetJupiterSubOneUnc & -- & \sysParamMPlanetJupiterSubZero & \sysParamMPlanetJupiterSubZeroUnc & -- \\
$\rho_{\text{p}}$, density \dotfill \si{\gram\per\cm\cubed}           & \sysParamRhoPlanetSubOne & \sysParamRhoPlanetSubOneUnc & -- & \sysParamRhoPlanetSubZero & \sysParamRhoPlanetSubZeroUnc & -- \\
Stellar irradiation \dotfill \si{\erg\per\s\per\cm\squared}             & \sysParamIrradiationSubOne & \sysParamIrradiationSubOneUnc & -- & \sysParamIrradiationSubZero & \sysParamIrradiationSubZeroUnc & -- \\
\tempeq{}, equilibrium temperature\tablenotemark{c} \dotfill K            & \sysParamTempEqSubOne & \sysParamTempEqSubOneUnc & -- & \sysParamTempEqSubZero & \sysParamTempEqSubZeroUnc & -- \\
\bottomrule
\end{tabular}

\flushleft

\tablenotemark{a}%
The prior for the eccentricity vector
$(\sqrt{e}\cos\omega, \sqrt{e}\sin\omega)$
is uniform in the unit disc.

\tablenotemark{b}%
The mass upper limits for the two planets are
roughly \SI{100}{\mass\earth}
and \SI{300}{\mass\earth}
when combined with the mean of the prior for $M_\star$.

\tablenotemark{c}%
Assuming the planets are tidally locked with no heat circulation and have a Bond albedo uniformly drawn from the interval $[0, 0.7]$.
For \planetouter{}, we adopt the mean-flux averaged \tempeq{} following \citet{2022RNAAS...6...56Q},
although here the correction due to eccentricity is negligible compared to the posterior uncertainty.

\textit{Note.}
$\uniformdist(a, b)$ is the uniform distribution over the interval $[a, b]$.
$\normaldist(\mu, \sigma)$ is
the normal distribution with mean $\mu$
and standard deviation $\sigma$.
\added[id=s]{The values and uncertainties quoted are 68\% credible intervals centred on medians.}

\end{table*}

\begin{table}
\caption{Additional sampled joint model parameters.}
\label{tab:nuissance}
\scriptsize
\renewcommand{\arraystretch}{1.25}
\begin{tabular}{lr@{}lc}
\toprule
Parameter \dotfill Unit & \multicolumn2c{Value} & Prior \\
\midrule
\addlinespace[4pt]
Limb darkening \\
\addlinespace[1pt]
$u_1$, \textit{TESS} \dotfill& \sysParamUTessSubZero & \sysParamUTessSubZeroUnc & Kipping \\
$u_2$, \textit{TESS} \dotfill& \sysParamUTessSubOne & \sysParamUTessSubOneUnc & Kipping \\
$u_1$, \filterRc{} \dotfill& \sysParamURcSubZero & \sysParamURcSubZeroUnc & Kipping \\
$u_2$, \filterRc{} \dotfill& \sysParamURcSubOne & \sysParamURcSubOneUnc & Kipping \\
$u_1$, \filterzs{} \dotfill& \sysParamUZsSubZero & \sysParamUZsSubZeroUnc & Kipping \\
$u_2$, \filterzs{} \dotfill& \sysParamUZsSubOne & \sysParamUZsSubOneUnc & Kipping \\
\addlinespace[4pt]
Photometry \\
\addlinespace[1pt]
Offset, \tess{} year 1 \dotfill & \sysParamMeanFluxZero & \sysParamMeanFluxZeroUnc & $\uniformdist(-0.5, 0.5)$ \\
Offset, \tess{} year 3 \dotfill & \sysParamMeanFluxOne & \sysParamMeanFluxOneUnc & $\uniformdist(-0.5, 0.5)$ \\
Jitter, \tess{} year 1 \dotfill & \sysParamLcJitterSubZero & \sysParamLcJitterSubZeroUnc & $\uniformdist(0, 1)$ \\
Jitter, \tess{} year 3 \dotfill & \sysParamLcJitterSubOne & \sysParamLcJitterSubOneUnc & $\uniformdist(0, 1)$ \\
Jitter, \astep{} \filterRc{} \dotfill & \sysParamLcJitterSubTwo & \sysParamLcJitterSubTwoUnc & $\uniformdist(0, 1)$ \\
Jitter, \lcogt{} SSO \filterzs{} \dotfill & \sysParamLcJitterSubThree & \sysParamLcJitterSubThreeUnc & $\uniformdist(0, 1)$ \\
\addlinespace[4pt]
\multicolumn4l{Radial velocity} \\
\addlinespace[1pt]
$\gamma$, \chiron{} \dotfill \si{\m\per\s} & \sysParamRvGammaSubZero & \sysParamRvGammaSubZeroUnc & $\uniformdist(5659.0, 7659.0)$ \\
$\gamma$, \feros{} \dotfill \si{\m\per\s} & \sysParamRvGammaSubOne & \sysParamRvGammaSubOneUnc & $\uniformdist(7114.6, 9114.6)$ \\
$\gamma$, \harps{} \dotfill \si{\m\per\s} & \sysParamRvGammaSubTwo & \sysParamRvGammaSubTwoUnc & $\uniformdist(7113.5, 9113.5)$ \\
Jitter, \chiron{} \dotfill \si{\m\per\s} & \sysParamRvJitterSubZero & \sysParamRvJitterSubZeroUnc & $\sim \uniformdist(0, \infty)\,\normaldist(0, 15)$ \\
Jitter, \feros{} \dotfill \si{\m\per\s} & \sysParamRvJitterSubOne & \sysParamRvJitterSubOneUnc & $\sim \uniformdist(0, \infty)\,\normaldist(0, 30)$ \\
Jitter, \harps{} \dotfill \si{\m\per\s} & \sysParamRvJitterSubTwo & \sysParamRvJitterSubTwoUnc & $\sim \uniformdist(0, \infty)\,\normaldist(0, 15)$ \\
\addlinespace[4pt]
\multicolumn4l{Gaussian process for RV} \\
\addlinespace[1pt]
$\rho$, undamped period \dotfill day & \sysParamGpRho & \sysParamGpRhoUnc & $\uniformdist(15, 200)$ \\
$\tau$, damping timescale \dotfill day & \sysParamGpTau & \sysParamGpTauUnc & $\uniformdist(0, 200)$ \\
$\sigma$, standard deviation \dotfill \si{\meter\per\second} & \sysParamGpSigma & \sysParamGpSigmaUnc & $\uniformdist(0, 100)$ \\
\addlinespace[4pt]
\multicolumn4l{SED} \\
\addlinespace[1pt]
$T_\text{eff,SED}$, effective temperature \ldots K & \sysParamTeffSed & \sysParamTeffSedUnc & $\normaldist(\tempeff{}, 2.5\%)$ \\
$R_{\star,\text{SED}}$, stellar radius \dotfill \si{\radius\sun} & \sysParamRStarSed & \sysParamRStarSedUnc & $\uniformdist(0, 2 R_\star)$ \\
SED uncertainty scaling factor \dotfill & \sysParamSedUncScale & \sysParamSedUncScaleUnc & $\uniformdist(1, 4)$ \\
\bottomrule
\addlinespace[4pt]
\end{tabular}

\footnotesize

\textit{Note.}
The \citet{2013MNRAS.435.2152K} prior is a triangle sampling of the
space of physically plausible quadratic limb darkening parameters.
The RV parameters are reproduced in \autoref{tab:spec} for convenience.
$\uniformdist(a, b)$ is the uniform distribution over the interval $[a, b]$.
$\normaldist(\mu, \sigma)$ is
the normal distribution with mean $\mu$
and standard deviation $\sigma$.
\added[id=s]{For an explanation of $T_\text{eff,SED}$, $R_{\star,\text{SED}}$,
and the \enquote{SED uncertainty scaling factor}, see \autoref{sec:jointmodelsed}.
The values and uncertainties quoted are 68\% credible intervals centred on medians.}

\end{table}

We constructed a joint model of the photometric light curves, RV measurements,
metallicity, surface gravity,
and broadband photometric magnitudes of \toitwothousand{}
with the Python packages \exoplanetpy{}
\citep{2021JOSS....6.3285F,foreman_mackey_daniel_2021_5834934}
and \celeritetwo{} \citep{celerite1,celerite2}
as well as the \mist{} stellar evolutionary tracks
\citep{2016ApJS..222....8D,2016ApJ...823..102C}.
The model's parameters included the orbital elements of the two planets
(the inner planet~b's orbit is fixed to be circular%
\footnote{Using the median values from the posterior of the joint model,
the tidal circularization timescale for \planetinner{}
$\approx\SI{1}{\giga\year}$,
assuming a modified tidal quality factor $Q' = 5 \times 10^4$.
The estimate of $Q'$ has a wide range from $10^3$ to $10^6$,
according to \citet[Supplementary Figure 8]{2019NatAs...3..424M}.
Whilst we do not have definitive evidence that \planetinner{} has fully
circularized,
we also do not have sufficient RV data to measure its eccentricity reliably,
thus we fix the eccentricity to 0.}),
stellar parameters,
limb-darkening parameters,
Gaussian process (GP) parameters for modelling the long-term trend in RVs,
and other observational nuisance parameters.
The posterior distribution of these parameters
was then sampled via Hamiltonian Monte Carlo \citep[HMC;][]{1987PhLB..195..216D}
implemented by the \pymc{} probabilistic programming framework \citep{pymc3}.
\added[id=s]{The posteriors and priors of the joint model's stellar, planetary, and other parameters
are reported in \autoref{tab:stellar}, \autoref{tab:planet}, and \autoref{tab:nuissance}.}

\subsubsection{Light curve and RV models}

We modelled the photometric observations with
the quadratic law \texttt{LimbDarkLightCurve} class built into
the \exoplanetpy{} package.
To account for long exposure time's smearing effect on the shape of transit light curves,
we oversampled each \tess{} 30-min exposure by a factor of $15$
but did not oversample the 20-s exposures.
Among the ground-based photometry, we included
two sets of observations with the best S/N,
the \astep{} data in the \filterRc{} band
and the \lcogt-SSO data in the \filterzs{} band,
in the joint model.
These light curves were simultaneously detrended
using their respective \enquote{sky} time series
in Tables~\ref{tab:astep} and \ref{tab:lcogt}.
We used separate sets of limb darkening parameters for each of the
\textit{TESS}, \filterRc{}, \filterzs{} filters,
each subject to the uniform triangular sampling prior of
\citet{2013MNRAS.435.2152K}.

We also used \exoplanetpy{}'s built-in RV model.
For each instrument, we introduced a constant RV offset $\gamma$
and a jitter term representing systematic uncertainty
that is added in quadrature to the reported instrumental uncertainties.
We modelled long-term variations in the RV residuals
with a GP simple harmonic oscillator (SHO) kernel
implemented by the \celeritetwo{} \citep{celerite1,celerite2}
Python package.
The SHO kernel has a power spectral density of
\begin{equation}
    S(\omega) = \sqrt{\frac{2}{\pi}}
        \frac{S_0 \omega_0^4}{(\omega^2 - \omega_0^2)^2 + \omega_0^2 \omega^2 / Q^2},
\end{equation}
and we use a parameterization in terms of
the undamped period of the oscillator
$\rho = 2\pi / \omega_0$,
the damping timescale of the process
$\tau = 2 Q / \omega_0$,
and the standard deviation of the process
$\sigma = \sqrt{S_0 \omega_0 Q}$.
We require $\rho \geq \SI{15}{\day}$ so that the GP does not interfere with
the RV signals of the two planets at shorter periods.
This exclusion of shorter periods is justified
since we do not expect \toitwothousand{} to be rapidly rotating
as it is a middle-aged Sun-like star,
and there are no significant short-term periodic variations in the undetrended \tess{} light curves.
In an RV-only model that uses Gaussian priors
corresponding to the joint model posterior values in \autoref{tab:planet}
for the planets' period and time of conjunction,
introducing GP with the SHO kernel causes
the RV jitter of \harps{} to decrease from $11.6$ to \SI{3.1}{\m\per\s}.
Thus, the GP captures some of the long-term RV variations.

To ensure that the joint model's ability to detect the RV signal of planet~b
is not sensitive to the choice of GP kernel,
we also ran another model using the Matérn-3/2 kernel
\begin{equation}
    k(\tau) = \sigma^2 \left( 1 + \frac{\sqrt{3} \tau}{\rho} \right)
        \exp \! \left(-\frac{\sqrt{3} \tau}{\rho}\right),
\end{equation}
where $\tau$ is the pairwise distance
and $\sigma, \rho$ are arbitrary parameters.
The Matérn-3/2 kernel did not lead to a meaningful difference
in the sampled posterior distribution
compared to the SHO kernel.

\added[id=s]{%
Even though the variations in the RV residuals
are sufficiently captured by the GP model,
which is customarily used to model correlated noise due to stellar activity,
we cannot definitively establish that stellar activity is the cause of these variations.
An alternative explanation is that the variations
are due to undetected additional planets in the system,
a possibility that we explore in \autoref{sec:extra_planets}.
}

\subsubsection{MIST evolutionary track interpolation} \label{sec:jointmodelmist}

We used \exoplanetpy{}'s \texttt{RegularGridInterpolator}
to interpolate the \mist{} stellar evolutionary tracks \citep{2016ApJS..222....8D,2016ApJ...823..102C}
built into the \isochrones{} Python package \citep{2015ascl.soft03010M}.
We used the initial stellar mass $M_{\star, 0}$,
initial metallicity $\feh_0$,
and equivalent evolutionary point (EEP) as inputs to the interpolated grid,
which then output present values of
$M_\star$, $\feh{}_{\text{MIST}}$, $\tempeff{}_{,\text{MIST}}$, and $R_{\star,\text{MIST}}$.
Although we sampled EEP instead of age,
we multiplied the overall model likelihood by
$ \partial(\mathrm{EEP}) / \partial(\mathrm{age}) $
as was done by \citet{2019arXiv190709480E} for \exofast{}v2
to guarantee that the prior distribution is uniform in stellar age rather than EEP.
We restricted EEP to the main sequence because spectroscopic evidence indicated that the star was not evolved.
We also constrained the stellar age to be $< \SI{10}{Gyr}$,
because \gaia{} \edrthree{} kinematics suggest that
\toitwothousand{} is most likely in the Galactic thin disk.
To ensure that the probability density of the model is continuous across the stellar age cutoff,
an important consideration because HMC relies on partial derivatives of the model,
we implemented the maximum-age constraint by adding a logistic function
\begin{equation}
    f(x) = -\frac{10^{100}}{1 + e^{-3000(x - 10)}} ,
\end{equation}
where $x$ is the stellar age in Gyr,
to the log-probability of the model.
The coefficients of the logistic function are chosen to
be as large as possible in order to achieve a sharp cutoff,
but not so large that
its derivative exceeds what can be represented by IEEE double precision floating point numbers,
which would cause HMC to fail.

To better reflect realistic uncertainties in theoretical modelling,
we made the stellar radius $R_\star$,
stellar effective temperature \tempeff{},
and stellar metallicity \feh{}
independently sampled model parameters
subject to Gaussian priors with a fractional width of 3~per cent centred on their respective interpolated \mist{} grid values,
following how \citet{2019arXiv190709480E} sampled these parameters for \exofast{}v2.
However, $M_\star$ was taken directly from the interpolated grid value.
This model can be represented in pseudocode as
\begin{subequations}
\begin{align}
    (M_{\star, 0}, \feh{}_0, \mathrm{EEP}) &\mapsto
        (M_\star, R_{\star,\text{MIST}}, \feh{}_{\text{MIST}}, \tempeff{}_{,\text{MIST}}), \\
    R_\star &\sim \normaldist(R_{\star,\text{MIST}}, 0.03\, R_{\star,\text{MIST}}), \\
    \feh{} &\sim \normaldist(\feh{}_{\text{MIST}}, 0.03\, \feh{}_{\text{MIST}}), \\
    \tempeff{} &\sim \normaldist(\tempeff{}_{,\text{MIST}}, 0.03\, \tempeff{}_{,\text{MIST}}),
\end{align}
\end{subequations}
where the variables on the left are sampled, the variables on the right are deterministically computed,
the arrow $\mapsto$ represents interpolation of the \mist{} grid,
and the relation $\sim$ represents draws from a prior distribution,
which in this case is the normal distribution $\normaldist(\mu, \sigma)$
with mean $\mu$ and standard deviation $\sigma$.
In addition, we computed the likelihood of \feh{} and \tempeff{}
using the \harps{} spectroscopic values in \autoref{sec:stellarparams},
but the uncertainty of the \harps{} \tempeff{} was artificially inflated to \SI{100}{\kelvin}
to better represent the uncertainty of the atmospheric models used to derive it.

\subsubsection{SED model} \label{sec:jointmodelsed}

\added[id=s]{%
We interpolated the \mist{} bolometric correction, or BC, grid \citep{2016ApJ...823..102C}
to convert the bolometric magnitude computed from the joint model's stellar parameters
into broadband photometric magnitudes.
For direct comparison with observed apparent magnitudes,
we sampled parallax from a Gaussian prior based on the \gaia{} \edrthree{} value and uncertainty
corrected for a zero-point offset
dependent on magnitude, colour, and ecliptic latitude using the prescription of \citet{2021A&A...649A...4L}.
Independently from the SED analysis in \autoref{sec:sed},
we computed the likelihood of the interpolated broadband magnitudes in
the \gaia{} \edrthree{} \gaiaG{}, \gaiaRP{}, and \gaiaBP{} bands,
the \twomass{} \filterJ{}, \filterH{}, and \filterK{} bands,
as well as the \wise{} \filterWone{}--\filterWfour{} bands
using their observed values and uncertainties
but applied an uncertainty floor of $0.02$ to the \gaia{} magnitudes.
We also multiplied all photometric uncertainties by a multiplicative factor
constrained to be between $1$ and $4$
in order to avoid overweighting the SED constraints within the joint model.
This factor is named \enquote{SED uncertainty scaling factor} in \autoref{tab:nuissance}.
}

\added[id=s]{%
We wanted to ensure that the SED model does not constrain stellar parameters more precisely
than the systematic uncertainty floors estimated by \citet{2022ApJ...927...31T}.
We loosely followed the latest SED model construction of \exofast{}v2 described by \citet{2022arXiv220914301E}.
The key was to partially decouple the stellar parameters used by the SED model,
in particular the calculation of BC,
from those used for the light curve and RV models or the \mist{} evolutionary tracks.
To that end, we introduced additional sampled parameters
$T_\text{eff,SED}$ and $R_{\star,\text{SED}}$,
which are reported in \autoref{tab:nuissance}.
The SED radius $R_{\star,\text{SED}}$ was then combined with the $M_\star$
from interpolating the \mist{} track to calculate $\log g_\text{SED}$.
The \mist{}-interpolated \feh{} was used as is,
and the extinction $A_V$ was sampled from a uniform prior with a lower limit of 0 and
an upper limit given by \citet{Schlegel:1998},
which we notate as $\uniformdist(0, 0.70)$.
Altogether, four of these parameters formed the inputs to the interpolated BC grid as
\begin{subequations}
\begin{align}
    A_V &\sim \uniformdist(0, 0.70), \\
    (M_\star, R_{\star,\text{SED}}) &\mapsto \log g_\text{SED}, \\
    (T_\text{eff,SED}, \log g_\text{SED}, \feh, A_V) &\mapsto \mathrm{BC},
\end{align}
\end{subequations}
using the notation of the pseudocode in \autoref{sec:jointmodelmist}.
}

\added[id=s]{%
To take account of the systematic uncertainty floors estimated for Solar-type stars
by \citet{2022ApJ...927...31T},
the parameters $T_\text{eff,SED}$ and $R_{\star,\text{SED}}$
were partially coupled to the \tempeff{} and $R_\star$
adopted by the light curve and RV models.
The parameter $T_\text{eff,SED}$ was coupled to the adopted \tempeff{} by a Gaussian prior
with a 2.5 per cent width.
Implicitly constraining $R_{\star,\text{SED}}$
was a 2-per-cent-wide Gaussian likelihood coupling the bolometric flux $F_\text{bol,SED}$
computed from $T_\text{eff,SED}$ and $R_{\star,\text{SED}}$
to $F_\text{bol}$ from the adopted \tempeff{} and $R_\star$.
Using the notation of the pseudocode in \autoref{sec:jointmodelmist},
\begin{subequations}
\begin{align}
    T_\text{eff,SED} &\sim \normaldist(\tempeff, 0.025 \tempeff{}), \\
    R_{\star,\text{SED}} &\sim \normaldist(0, 2 R_\star), \\
    (T_\text{eff,SED}, R_{\star,\text{SED}}, \varpi) &\mapsto F_\text{bol,SED}, \\
    (T_\text{eff}, R_\star, \varpi) &\mapsto F_\text{bol}, \\
    F_\text{bol,SED}  &\propto \normaldist(F_\text{bol}, 0.02 F_\text{bol}),
\end{align}
\end{subequations}
where $\varpi$ is the parallax and the relation $\propto$ indicates likelihood.
Converting the $F_\text{bol,SED}$ to apparent bolometric magnitude
and subtracting the BC gave the predicted broadband photometric magnitudes
that were compared to observations.
}

\subsubsection{Sampling the posterior distribution}

We ran \pymc{}'s HMC No-U-Turn Sampler \citep[NUTS;][]{JMLR:v15:hoffman14a}
for 4096 tuning steps and 4096 draws on 32 independent chains.
The resulting samples converged well,
with the convergence statistic of \citet{1992StaSc...7..457G}
$\hat{R} \approx 1$
and effective sample sizes ranging from
\added[id=s]{\numrange{8300}{95000}}
across all parameters and chains
as calculated by the \arviz{} Python package
\citep{2019JOSS....4.1143K}.
The median and the middle 68 per cent credible interval of the posterior distribution,
as well as the prior distribution of each fitted parameter,
are reported in Tables \ref{tab:stellar}, \ref{tab:planet}, and \ref{tab:nuissance}.

\begin{figure*}
    \includegraphics[width=\textwidth]{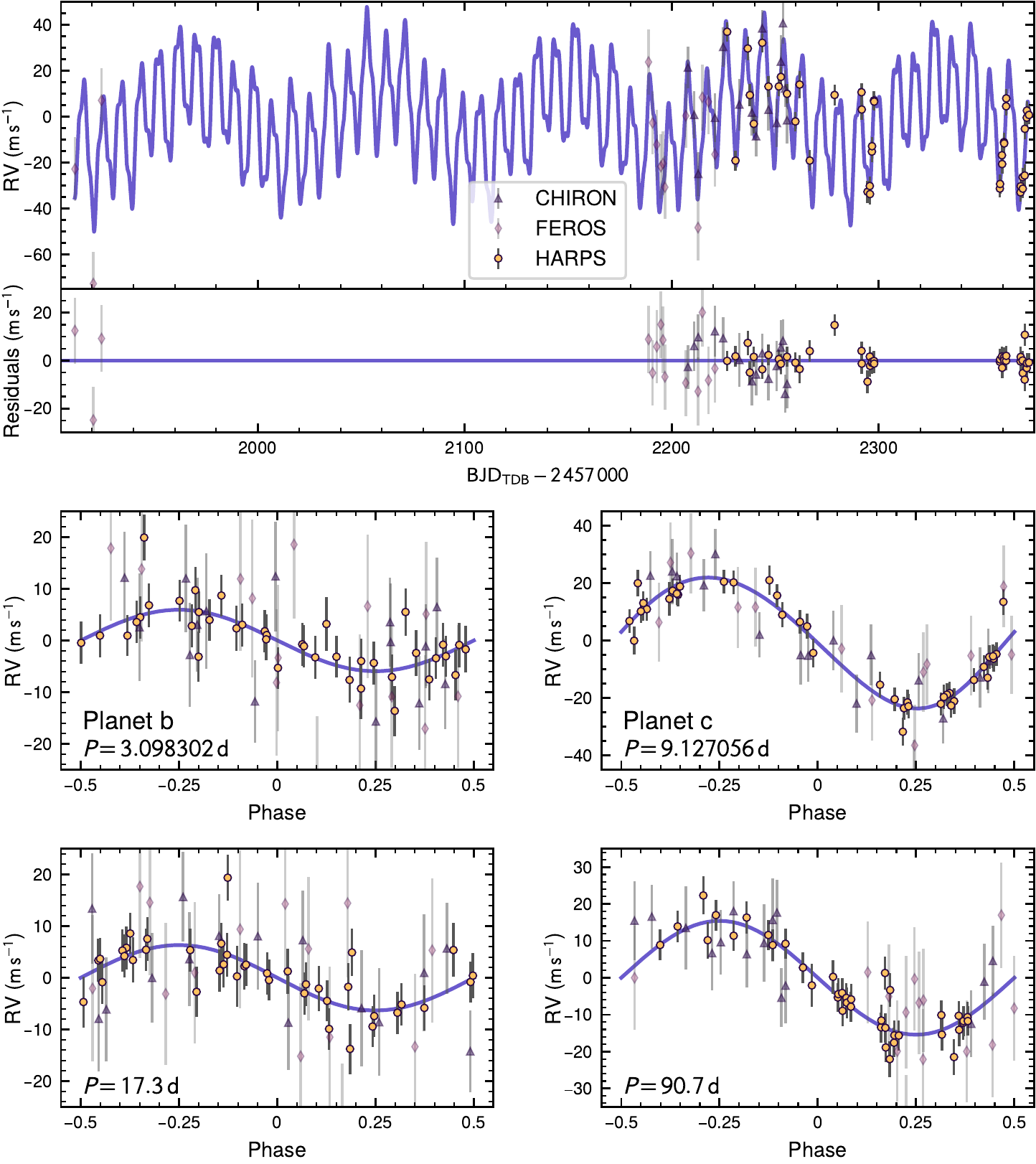}
    \caption{\added[id=s]{%
    Alternative four-planet model of the RV measurements of \toitwothousand{}.
    The shape and colour of each mark indicate which spectrograph it corresponds to,
    and its error bar is the quadrature sum of uncertainty and an instrumental jitter term.
    Solid purple lines represent the median-posterior RV model.
    Top: The median-posterior RV model and the residuals of the RV measurements with respect to the model.
    Middle: Phase-folded RV variations due to planets~b and c only.
    Bottom: Phase-folded RV variations due to the hypothetical non-transiting planets only.}}
    \label{fig:extra_planets}
\end{figure*}

\subsection{Alternative RV model with additional planets} \label{sec:extra_planets}

\added[id=s]{%
To explore alternative interpretations for the RV residuals
not explained by the two transiting planet candidates,
we modelled their variations as additional non-transiting planets in the system.
We ran two RV-only models: a three-planet model with the third planet around a 90-day orbital period,
and a four-planet model with an additional fourth planet around a 17-day period.
These periods corresponded to the next two most significant peaks in the Lomb--Scargle periodogram of the RVs
after the signal due to the hot saturn at 9.13~d was removed.
For these RV-only models, the periods and times of conjunction of the two inner planets
were drawn from Gaussian prior distributions
with means and standard deviations corresponding to
those of the posterior of the joint model in \autoref{sec:globalmodel}.
The eccentricity $e$ and argument of periapsis $\omega$
of the 9.13-day hot saturn were free parameters,
with a uniform prior in the unit disc of $(\sqrt{e}\cos\omega, \sqrt{e}\sin\omega)$,
but the other three planets' orbits were fixed to be circular.
}

\added[id=s]{%
The resulting median-posterior four-planet RV model is presented in \autoref{fig:extra_planets}.
As measured by the four-planet model,
the proposed two outer planets have periods
\SI[parse-numbers=false]{17.29^{+0.09}_{-0.12}}{\day}
and \SI[parse-numbers=false]{90.7^{+1.6}_{-1.5}}{\day}
and RV semiamplitudes
\SI{5.3 \pm 1.1}{\m\per\s}
and \SI{15.4 \pm 1.2}{\m\per\s},
corresponding to minimum masses ($M_\text{p} \sin i$)
\SI[parse-numbers=false]{26.9 \pm 4.8}{\mass\earth}
and \SI[parse-numbers=false]{114.1^{+10.2}_{-9.8}}{\mass\earth}.
We do not show the three-planet model here because it has significantly higher instrumental jitters
compared to the four-planet model.
For \chiron{}, \feros{}, and \harps{}, respectively,
the three-planet model has jitters
\SI[parse-numbers=false]{4.2^{+3.6}_{-2.8}}{\m\per\s},
\SI[parse-numbers=false]{14.6^{+4.4}_{-3.3}}{\m\per\s},
and \SI[parse-numbers=false]{5.93^{+0.94}_{-0.79}}{\m\per\s},
while the four-planet model has
\SI[parse-numbers=false]{4.0^{+3.4}_{-2.7}}{\m\per\s},
\SI[parse-numbers=false]{12.3^{+3.9}_{-3.0}}{\m\per\s},
and \SI[parse-numbers=false]{3.60^{+0.80}_{-0.69}}{\m\per\s}.
}

Nevertheless, \added[id=t]{the evidence for the two additional signals in our RV data is not conclusive}
(\autoref{sec:rv_analysis}).
\added[id=s]{%
The longest contiguous RV time baseline covers just a little more than one period of the 90-day planet (from approximately \bjdtdb{} 2459189 to 2459298),
so it is difficult to assess if the signal of the proposed 90-day planet is truly periodic.
Moreover, due to the sparsity of data,
we cannot rule out that non-transiting planets
of a different combination of orbital periods
are the true sources of the 90-day and 17-day signals.
}
\added[id=t]{We conclude that our present RV data is insufficient
for confidently claiming the detection of these two additional planets in the RVs.}

\added[id=s]{%
Thus, in the absence of overwhelming evidence for the two additional RV-only planets,
we chose to adopt planetary and stellar parameters from the the joint model with GP.
Compared to the four-planet RV-only model, the two-planet GP-based joint model
prefers a slightly lower mass for the 3.1-day mini-neptune by about 23~per cent ($1.3\sigma$).
However,
this difference would not change the qualitative characterization of the inner planet as a mini-neptune
that we discuss in \autoref{sec:massradius}.
}

\section{Confirmation of planetary candidates} \label{sec:confirm}

\subsection{Characteristics of TOI-2000 c} \label{sec:analyzec}

\planetouter{} is a hot saturn with a radius of
\sysParamRPlanetEarthSubZero\sysParamRPlanetEarthSubZeroUnc\,\si{\radius\earth}
and a mass of
\sysParamMPlanetEarthSubZero\sysParamMPlanetEarthSubZeroUnc\,\si{\mass\earth}
in a \sysParamPeriodSubZero\sysParamPeriodSubZeroUnc-day orbit.
We detected an eccentricity of
\sysParamEccSubZero\sysParamEccSubZeroUnc,
and the posterior distribution slightly favours a non-zero eccentricity at the 2--$3\sigma$ level.
The mass and eccentricity were measured from a combination of 41 \harps{} observations, 14 \feros{} observations and 15 \chiron{} observations.
Each RV instrument independently detected a strong signal
that was consistent with a companion of planetary mass
at the period corresponding to the transit signal.
In addition, ground-based observations showed that the transit signal of \planetouter{} was on target and consistent with the depth measured by \tess{}.
The Gemini/Zorro speckle imaging did not reveal any nearby stellar companions down to $\Delta\mathrm{mag} = 5$
at an angular separation more than \ang{;;0.1} (projected separation of \SI{17.4}{\au}) away from the star.
Therefore, we confirm that \planetouter{} is a planet with high confidence.

\subsection{Characteristics of TOI-2000 b}

\planetinner{} is a mini-neptune with a radius of
\sysParamRPlanetEarthSubOne\sysParamRPlanetEarthSubOneUnc\,\si{\radius\earth}
and a mass of
\sysParamMPlanetEarthSubOne\sysParamMPlanetEarthSubOneUnc\,\si{\mass\earth}
in a \sysParamPeriodSubOne\sysParamPeriodSubOneUnc-day orbit,
which corresponds to an RV semiamplitude of
\sysParamKSubOne\sysParamKSubOneUnc\,\si{\meter\per\second}.
In an RV-and-GP-only model that does \emph{not} require the RV semiamplitude to be strictly positive,
the $4 \sigma$ lower limit is \SI{0.39}{\meter\per\second}.
This simplified model uses Gaussian priors
corresponding to the joint model posterior values in \autoref{tab:planet}
for the planets' orbital period and time of conjunction
and constrains the undamped period of the GP kernel to $> 15\,\text{d}$, more than either planet's period.
This result is independent of the choice of either the SHO or the Matérn-3/2 kernel for GP.

Ordinarily, an independent RV detection suffices to confirm transiting planet candidates.
In the case of \planetinner{}, however,
because of its relatively small transit depth ($\approx 470 \, \text{ppm}$)
and low measured RV semiamplitude (\sysParamKSubOne\sysParamKSubOneUnc\,\si{\m\per\s}),
we apply extra scrutiny to possible false positive scenarios.

\subsection{False-positive analysis of TOI-2000 b} \label{sec:analyzeb}

There are two main classes of false positive transiting planet candidates:
instrumental artefacts and astrophysical signals that mimic transiting planets.
We can rule out the possibility that \planetinner{} is due to an instrumental artefact because
the transit detection is robust across different observational campaigns and data reduction software.
Transits occurred periodically and at a consistent depth during both the first and third years of \tess{} observations
at 30-min or 20-s cadences.
Both the \qlp{} and the SPOC pipeline independently detected the transits of \planetinner{} as a threshold-crossing event during their multi-year transit planet searches.
Using an approach independent of the standard SPOC pipeline and \qlp{},
we also decorrelated the \tess{} light curves from the spacecraft pointing quaternion time series (\autoref{sec:obs_tess}),
thereby removing variations caused by spacecraft pointing jitter, a common source of instrumental systematics.
Thus, it is highly unlikely that the transit signal was an instrumental artefact.

Having ruled out instrumental artefacts,
there are four possible scenarios for astrophysical false positives.
We will eliminate them one by one in the following subsubsections
through statistical validation.

\subsubsection{Is TOI-2000 an eclipsing binary (EB)?}
We consider the scenario that the transit signal of \planetinner{}
is due to a grazing EB.
Our RV analysis shows that \planetinner{} is well below stellar mass if it indeed orbits its host star.
In an RV-only model that does not fit for long-term trends with GP and uses Gaussian priors
corresponding to the joint model posterior values in \autoref{tab:planet}
for the planets' period and time of conjunction,
the $5\sigma$ upper limit of \planetinner{}'s RV semiamplitude is \SI{13.1}{\meter\per\second},
corresponding to a mass of \SI{31.4}{\mass\earth},
less than \SI{0.1}{\mass\jupiter}.
Thus, we can confidently rule out the EB scenario.

\subsubsection{Is light from a physically associated companion blended with TOI-2000?}
In this scenario, an EB or transiting planetary system that is gravitationally bound to \toitwothousand{}
is the true source of the transit signal of \planetinner{}.
We use the \molusc{} tool from \citet{2021AJ....162..128W} to rule out potential bound companions with a Monte Carlo method.
The \molusc{} tool starts by randomly generating two million companions of stars or brown dwarfs
according to physically motivated prior distributions of mass and orbital parameters.
The expected observational effect of each companion
is then compared with
\gaia{} imaging,
the scatter of \gaia{} astrometric measurements
(parameterized by the renormalized unit weight error, or RUWE, metric),
RVs, and the contrast curve from Gemini/Zorro speckle imaging.
The surviving companions, which are not ruled out by the observational constraints,
are the samples for the posterior distribution.
Using the RVs and \geminisouth{} contrast curve,
\molusc{} finds that about 23 per cent of generated companions survive
regardless of whether the generated companion is restricted to impact parameter $< 1$,
and that objects with mass $> \SI{0.1}{\mass\sun}$
are almost completely ruled out at periods $< \SI{1000}{\day}$, equivalent to orbital separations $< \SI{2}{\au}$.
Thus, we can rule out most, but perhaps not all the possible physically associated EBs or transiting planetary systems.

To incorporate information from the shape of the transit light curve,
we turn to \triceratops{} by \citet{2021AJ....161...24G}.
The \triceratops{} tool uses
the \tess{} light curve and the Gemini/Zorro contrast curve
to calculate the marginal likelihoods of various scenarios
that could potentially cause the transit-like signal that we attribute to \planetinner{}.
The marginal likelihood of each scenario is then scaled by appropriately chosen priors
to produce the Bayesian posterior probability that the transit-like signal is produced by that scenario.
Considering all the false positive scenarios featuring an unresolved bound companion%
\footnote{Using the shorthand of \citet{2021AJ....161...24G},
the scenarios considered are PEB, PEBx2P, STP, SEB, and SEBx2P.},
\triceratops{} finds the probability to be
$0.0073 \pm 0.0002$.

\subsubsection{Is light from a resolved and unassociated background system blended with TOI-2000?}
In this scenario, a chance alignment of the foreground star \toitwothousand{} with a binary system in the Galactic background
caused the signal to be wrongly identified with the foreground.
A nearby EB that is no more than $8.3$~magnitudes fainter than \toitwothousand{}
is needed to produce the $\approx 470\,\text{ppm}$ transit depth.
A ground-based photometric observation on \ut{} 2021 February 6 from \lcogt{} CTIO
during the transit window of \planetinner{}
confidently rules out that the transit signal appears on nearly all nearby resolved stars
of sufficient brightness,
with the sole exception of
TIC\,845089585 ($\Delta \mathrm{mag} = 8.71$ in \emph{TESS} band, \ang{;;5.32} separation at a position angle of \ang{3.96}).
Nevertheless,
\triceratops{} rules out that TIC\,845089585 was the source of the signal.
The nearby false-positive probability%
\footnote{The NFPP includes scenarios NTP, NEB, and NEBx2P in the parlance of \citet{2021AJ....161...24G}.}
arising from TIC\,845089585
is $(3.02 \pm 0.06) \times 10^{-12}$, virtually zero.

There is also the possibility that the transit signal of \planetinner{} was contaminated by EBs
falling on the same \tess{} CCD as \toitwothousand{}
but outside of its immediate vicinity.
Whilst this type of contamination was common during the \kepler{} mission
\citep[e.g. due to crosstalk or charge transfer inefficiency; see][]{coughlin2014},
it is unlikely to apply to \tess{} observations of \toitwothousand{}.
Across multiple \tess{} sectors,
the field of view rotated relative to the orientation of the CCDs,
so any contaminating source would not be able to affect \toitwothousand{} consistently.
Furthermore,
the transit depth is consistent between photometry using a 1-pixel aperture versus larger apertures,
providing strong evidence that the signal was indeed localized on the CCD.
For these reasons,
we only consider contaminating sources in the immediate vicinity of \toitwothousand{} for the false-positive analysis in this section.

\subsubsection{Is light from an unresolved and unassociated background system blended with TOI-2000?}
The analysis above rules out on-target EBs, bound companions, and nearby EBs
but not unresolved background EBs arising from chance alignment.
As the proper motion of \toitwothousand{} is too small for archival Digitized Sky Survey images to reveal the background under its current position,
we turn to \triceratops{} once more
to quantify how likely the remaining scenarios are.
To generate a population of simulated unresolved background stars
in a $0.1 \, \text{deg}^2$ region of the sky,
\triceratops{} uses the TRILEGAL galactic model \citep{2005A&A...436..895G}
and considers all simulated stars with \tess{} magnitudes fainter
than that of the target star
and brighter than $21$ \citep{2021AJ....161...24G}.
For false-positive scenarios involving these simulated background stars%
\footnote{The scenarios here are
DEB, DEBx2P, BTP, BEB, and BEBx2P.},
\triceratops{} finds a probability of $0.0020 \pm 0.0001$.

\subsubsection{TOI-2000\thinspace b is a confirmed planet}

Altogether, \triceratops{} gives a false-positive probability (FPP) of
$0.0094 \pm 0.0002$.
Because the inner planet has been detected by the \tess{} SPOC pipeline,
this FPP can be further divided by the
multiplicity boost of $\approx 20$
calculated using the TOI catalogue from the \tess{} prime mission
\citep{2021ApJS..254...39G},
which puts the final FPP ($\approx 0.0005$) at well below any reasonable
threshold for validated planets \citep[e.g. $< 0.015$;][]{2021AJ....161...24G}.
This highly confident statistical validation,
in addition to the $> 4\sigma$ RV detection,
is overwhelming evidence in favour of a small inner planet orbiting \toitwothousand{}.

\section{Discussion} \label{sec:discussion}

\begin{figure}
    \includegraphics[width=\columnwidth]{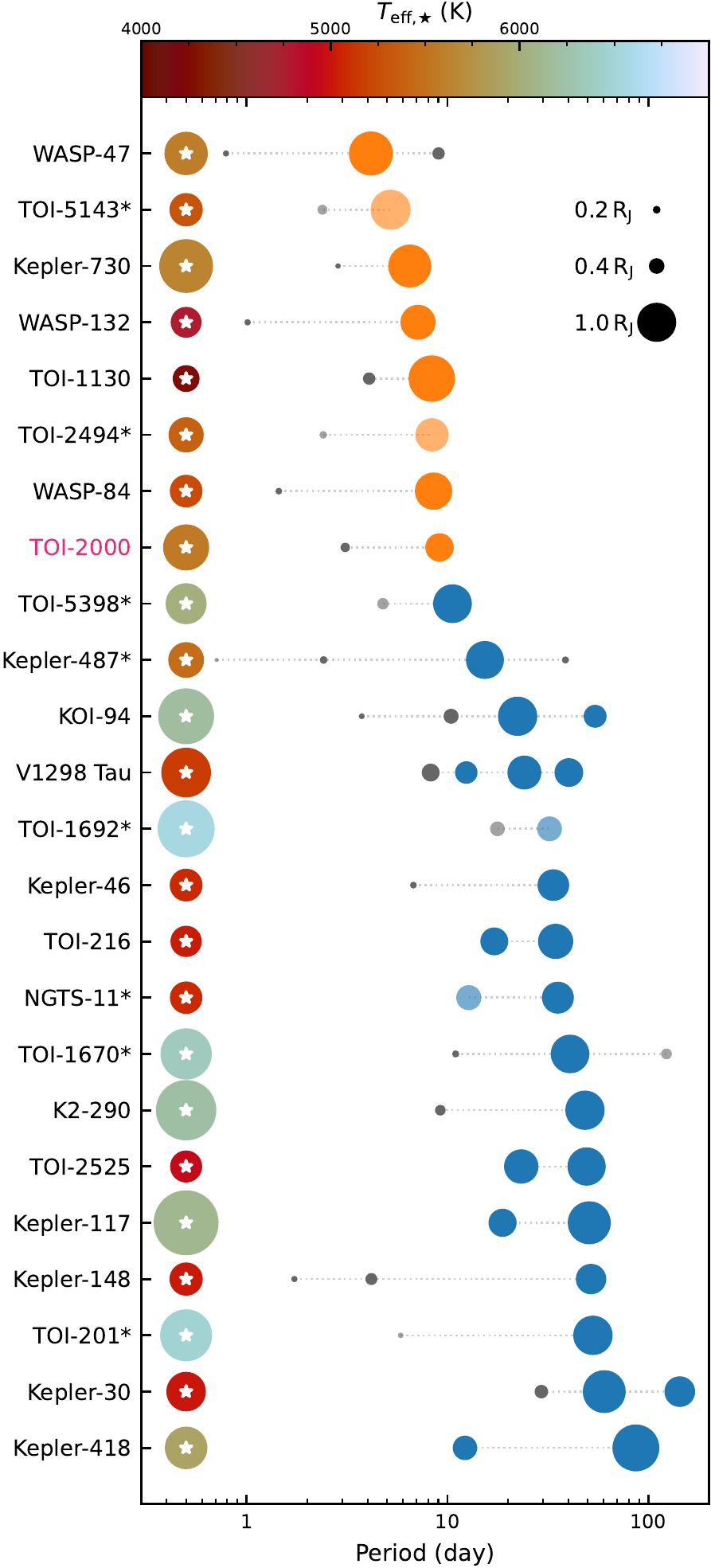}
    \caption{Transiting planetary systems hosting small planets orbiting interior to giant planets ($P < 100\,\text{d}$).
    Non-transiting planets in these systems are omitted.
    The leftmost circle in each row represent the host star,
    with the mark's linear size proportional to the stellar radius
    and fill colour indicating the effective temperature
    \citep[from TIC\thinspace 8.1;][]{tess_tic8}.
    The solid circles represent the transiting planets in the system,
    with their linear sizes proportional to planetary radii
    and fill colours indicating whether the planet is a
    giant planet (orange, $P < 10\,\text{d}$; blue, $P \geq 10\,\text{d}$)
    or a small planet (grey, $R < \SI{6}{\radius\earth}$).
    The relative mark sizes among either the stars or the planets are to scale,
    but not between a star and its planets.
    The slightly translucent marks are planet candidates.
    \added[id=s]{Systems with an asterisk appended to their names contain planet candidates.}
    The systems are sorted in ascending order of the period of the largest planet from top to bottom.
    \autoref{tab:giant_multi} contains the numbers and references underlying this figure.
    }
    \label{fig:family}
\end{figure}

\begin{figure*}
    \centering
    \includegraphics[width=\textwidth]{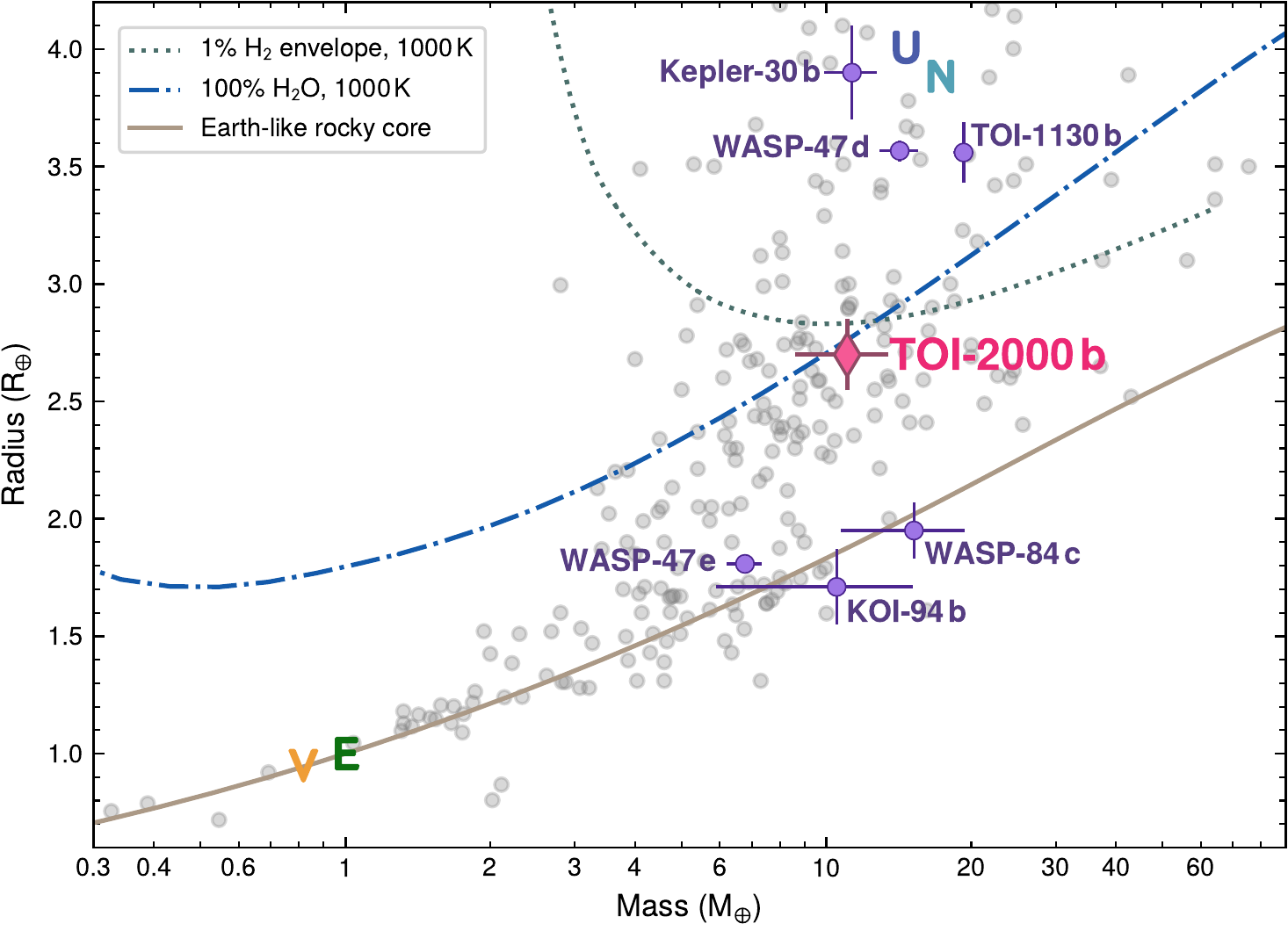}
    \caption{Mass--radius relationship of small exoplanets ($< \SI{4}{\radius\earth}$).
    Only planets with mass determination better than 33\% have been included.
    The values used are the \enquote{default parameter set} of the Planetary Systems Table
    from the NASA Exoplanet Archive \citep[][accessed on 2022 August 19]{nasa_exoplanet_archive,nasa_exoplanet_archive_ps}.
    The mini-Neptune \planetinner{},
    together with \replaced[id=r]{six}{four} small planets
    \planet{\waspfortyseven}{d} and e \citep{2022AJ....163..197B},
    \toieleventhirtyinner{} \citep{2023arXiv230515565K},
    \waspeightyfourinner{} \citep{2023arXiv230509177M},
    \planet{Kepler-30}{b} \citep{2012Natur.487..449S},
    and \planet{KOI-94}{b} \citep[also known as Kepler-89;][]{2013ApJ...768...14W},
    which are also found in systems with giant planets,
    are highlighted, with error bars representing quoted uncertainties.
    The single letters denote the Solar System planets Venus, Earth, Uranus, and Neptune.
    Three theoretical mass--radius curves by
    \citet[available online at \url{https://lweb.cfa.harvard.edu/~lzeng/planetmodels.html}]{2019PNAS..116.9723Z}
    are plotted for reference:
    solid brown for an Earth-like rocky core (32.5\% Fe and 67.5\% $\text{MgSiO}_3$),
    dotted-dash blue for an 100\% water world at \SI{1000}{\kelvin},
    and dotted teal for a planet with 1\% hydrogen envelope and 99\% Earth-like rocky core at \SI{1000}{\kelvin}.%
    }
    \label{fig:massradius}
\end{figure*}

\subsection{Diversity among inner companions to transiting gas giants} \label{sec:massradius}

The \toitwothousand{} system
hosts the smallest transiting hot gas giant planet
($P < \SI{10}{\day}, M > \SI{60}{\mass\earth}$)
known to have an inner companion.
The hot gas giant \planetouter{} has mass and radius similar to those of Saturn
(\autoref{sec:analyzec})
and has roughly the same mean density as \wasponethirtytwohj{}
\citep[\SI{0.84}{\gram\per\cm\cubed}, ][]{2017MNRAS.465.3693H},
in contrast to the larger
\waspfortysevenhj{},
\keplerseventhirtyhj{},
and \toieleventhirtyhj{}%
\footnote{The true radius of \toieleventhirtyhj{} is uncertain because its transits are grazing.}%
,
which all have radii $\approx \SI{1.1}{\radius\jupiter}$.
The companion \planetinner{} is a mini-neptune
(\autoref{sec:analyzeb})
whose size sits between
the three super-earth-sized companions
(\keplerseventhirtyinner{},
\waspfortyseveninner{},
and \wasponethirtytwoinner{})
and the Neptune-sized \toieleventhirtyinner{}.

The mean density of \planetinner{} stands out
among the handful of gas giant inner companions
that have measured masses.
Among the 23 confirmed or candidate transiting planetary systems
that contain small planets ($< \SI{4}{\radius\earth}$)
orbiting interior to gas giants of any period (\autoref{fig:family}),
only \replaced[id=r]{six}{five} small planets other than \planetinner{} have measured mass
(\autoref{fig:massradius}).
The densities of \replaced[id=r]{three}{two} of these planets,
\waspfortyseveninner{}\replaced[id=r]{,}{and} \keplereightynineb{}, \added[id=r]{and \waspeightyfourinner{},}
are consistent with them being exposed rocky cores.
The other three planets, \keplerthirtyb{}, \waspfortysevenneptune{}, and \toieleventhirtyinner{},
have masses and radii comparable to those of Neptune and Uranus.
Whilst \planetinner{} is not dense enough to be an exposed rocky core,
its atmospheric mass fraction is likely considerably smaller than those of the three Neptune-sized planets,
further demonstrating that these rare inner companions to giant planets
are just as diverse in density as other small planets.

\added[id=s]{%
The qualitative results in this subsection would not change if we had adopted
the alternative four-planet model (\autoref{sec:extra_planets}, \autoref{fig:extra_planets})
instead of the two planets plus GP model (\autoref{fig:rv}) for the RVs.
Respectively for \planetinner{} and \planetouter{},
the four-planet model gives RV semiamplitudes
\SI{5.9 \pm 1.0}{\m\per\s}
and \SI{22.8 \pm 1.0}{\m\per\s},
corresponding to planet masses
\SI[parse-numbers=false]{14.3^{+2.6}_{-2.5}}{\mass\earth}
and \SI[parse-numbers=false]{78.3^{+4.7}_{-4.6}}{\mass\earth}.
Compared to these values,
the posterior values from the joint model (\autoref{sec:globalmodel})
is smaller by $1.3\sigma$ for the mass of \planetinner{}
and larger by $0.7\sigma$ for the mass of \planetouter{}.
Regardless of whether this difference is statistically significant for \planetinner{},
it is far short of what is needed to make \planetinner{} a rocky core instead.
The alternative location of \planetinner{} on the mass--radius diagram
is still intermediate between the rocky and the Neptune-sized planets.
}

\subsection{Expected transit timing variations} \label{sec:ttv}

We expect TTVs to be present in the \tess{} light curves of \toitwothousand{},
as the periods of \planetouter{} and \planetinner{} are within 1.8 per cent of an exact 3:1 ratio
and the orbit of \planetouter{} is probably slightly eccentric (\autoref{sec:analyzec}).
Drawing from an ensemble of orbital parameters and planetary masses,
we estimate the peak-to-peak TTV amplitude of \planetinner{} to be 3--30\thinspace min
with a superperiod $\approx 168$\thinspace d
using \textsc{TTVFast} \citep{2014ApJ...787..132D}.
The eccentricity vector and mass of \planetouter{}
and the mass of \planetinner{} are drawn from the joint modelling posterior distribution (\autoref{sec:globalmodel}),
and the eccentricity of \planetinner{} is drawn from a Rayleigh distribution with a width of $0.06$.
The TTV amplitude \planetouter{} is similarly estimated to be $< 5\,\text{min}$.
This wide range of expected TTV amplitudes is attributable to uncertainties in the eccentricities and TTV phases of the two planets.
However, the extremely shallow transit depth ($\approx 470\, \text{ppm}$) of \planetinner{}
makes it challenging to directly measure TTV from its individual transits in the \tess{} data.%
\footnote{Unfortunately, \toitwothousand{} is not in the \emph{CHEOPS} viewing zone.}

In order to compare the simulated TTV effect to observation despite lacking a firm TTV detection,
we instead measure the difference in the average time of conjunction $T_\text{c}$ for each year of \tess{} observations
using the same set of \textsc{TTVFast} calculations,
similar to the method employed by \citet{2016Natur.533..509M}.
The resulting difference of 0.5--7~minutes is comparable to
the joint model's uncertainty of 3~minutes in the $T_\text{c}$ of \planetinner{}.
Whilst we are unable to prove or rule out the existence of TTVs in current \tess{} data,
\tess{} will observe \toitwothousand{} again during its Extended Mission 2
in sectors 63--65 (\ut{} 2023 March 10 -- June 2),
which will undoubtedly improve the S/N for TTV detection.

\subsection{Opportunities for atmospheric characterization}

The \toitwothousand{} system currently provides the second best opportunity
for measuring the atmospheric compositions of a hot gas giant and small inner planet together,
which potentially provide us with insight into where they formed within their protoplanetary disc.
The \jwst{} transmission spectrum metric
\citep[TSM; ][]{2018PASP..130k4401K}
for \planetinner{} and c are 29 and 68, respectively.
Although \toitwothousand{} is brighter in the \filterV{} band,
its TSMs trail those of the \toieleventhirty{} system
because that system has a smaller host star brighter in the \filterJ{} band and bigger planets.
The host star \toieleventhirty{} is a K dwarf rather than a G dwarf like \toitwothousand{},
and its inner planet \toieleventhirtyinner{} is approximately 40 per cent larger than \planetinner{}.
However, that \toieleventhirtyinner{} shows TTV
with peak-to-peak amplitude of at least two hours \citep{2023arXiv230515565K}
means that an accurate ephemeris based on $N$-body numerical integration
must be available,
unlike \planetinner{},
which does not conclusively show TTV at the present level of photometric precision
(\autoref{sec:ttv}).

Measuring the atmospheric metallicity of both planets could help us understand their structure and origin.
Using CEPAM \citep{1995A&AS..109..109G,2006A&A...453L..21G}
and a non-grey atmosphere \citep{2015A&A...574A..35P},
we model the evolution of both planets in the system
assuming a simple structure consisting of a central rocky core
surrounded by a H--He envelope of Solar composition.
Under this model, the core mass of \planetouter{} is between \added[id=t]{$36$ and \SI{46}{\mass\earth}},
about twice as large as Saturn’s total mass of heavy elements
\citep[$16.5$ to \SI{21}{\mass\earth};][]{2021NatAs...5.1103M}.
The presumed H--He envelope of \planetinner{} must be smaller than $\SI{0.1}{\mass\earth}$
(1~per cent of the mass of the planet),
in line with \autoref{fig:massradius},
which indicates that the radius of \planetinner{} is \replaced[id=t]{up}{2} to 10~per cent smaller
than the radius predicted by the theoretical mass--radius curve of \citet{2019PNAS..116.9723Z}
for planets with 1~per cent $\text{H}_2$ envelope.
Overall, this analysis implies that both planets likely contain a proportion
of heavy elements that is significantly larger than that of planets with similar mass in the Solar System.

By comparing the atmospheric abundances of two planets around the same host star,
we can infer whether they formed at disc locations with similar compositions.
The atmospheres of short-period planets forming \emph{in situ}
should contain more refractory metals
(Fe, Cr, Ti, VO, Na, K, and P),
which are more abundant closer to the host star,
than volatile elements (O, C, and N).
Even though it is unlikely that measured atmospheric abundances alone
could pinpoint where a planet formed in its disc,
such as with the method suggested by \citet{oberg},
we can still distinguish scenarios where the gas giant and the small planet formed
together far from the host star
(i.e. beyond the snow line)
from others where the gas giant formed first
and then swept material near its orbital resonances
on its migration path to form the small planet.
The former scenario would result in two planets of similar atmospheric composition,
whereas the latter scenario would result in the inner planet having a higher ratio of volatile to refractory elements than the outer planet.

\subsection{Interpolating MIST tracks under Hamiltonian Monte Carlo}

In addition to the scientific results, this paper also introduces code
that efficiently samples the posterior of planetary and stellar parameters with HMC
while incorporating constraints from the MIST stellar evolutionary tracks.
Unlike traditional Markov chain Monte Carlo (MCMC) methods,
HMC also uses the gradient of the model in proposing chain step movements.
By including information from the gradient,
a technique inspired by Hamiltonian mechanics,
successive samples under HMC are much less correlated, thus achieving a larger effective sample size
with the same number of \added[id=s]{sampler steps}.
\added[id=s]{%
Sampling the joint model with 4096 tuning steps and 4096 sampling steps across 32 parallel chains
took 36.5~hours on a workstation equipped with an AMD Ryzen 5950X 16-core (32-thread) CPU.
A preliminary run using fewer steps and binning the 20-s \tess{} light curves to 2~min
could be completed in just a few hours.
In contrast, a traditional MCMC sampler would require many times more sampling steps and weeks of computation time
to achieve a similar effective sample size and level of convergence.
}

However, because the gradient of the model can only be generated practically through automatic differentiation,
such as via the \textsc{Aesara} backend of \pymc{},
it can be challenging to incorporate existing code,
such as the \isochrones{} package \citep{2015ascl.soft03010M},
that has not been specially adapted to \pymc{}.
This complication is why no previous paper in the literature
that uses the \exoplanetpy{} and \pymc{} packages for exoplanet parameter fitting
has incorporated the constraint from MIST tracks.
By writing custom code to interpolate MIST tracks in a way compatible with \pymc{},
we are able to significantly speed up the run time of posterior sampling,
which in turn allowed us to rapidly iterate and improve on the design of our joint model.

\section{Conclusions}

The \toitwothousand{} system is the latest example of rare planetary systems that host a hot gas giant
with a inner companion.
By jointly modelling \tess{} and ground-based transit light curves,
precise RVs from \chiron{}, \feros{}, and \harps{},
and broadband photometric magnitudes,
we have confirmed both \planetinner{} and \planetouter{} by direct mass measurement.
The inner \planetinner{} is a mini-neptune with a mass of
\sysParamMPlanetEarthSubOne\sysParamMPlanetEarthSubOneUnc\,\si{\mass\earth}
and a radius of
\sysParamRPlanetEarthSubOne\sysParamRPlanetEarthSubOneUnc\,\si{\radius\earth}
on a 3.09833-day orbit,
whereas the outer \planetouter{} is a hot saturn with a mass of
\sysParamMPlanetJupiterSubZero\sysParamMPlanetJupiterSubZeroUnc\,\si{\mass\jupiter}
and a radius of
\sysParamRPlanetJupiterSubZero\sysParamRPlanetJupiterSubZeroUnc\,\si{\radius\jupiter}
on a 9.127055-day orbit.
\replaced[id=t]%
{Radial velocity residuals hint at additional non-transiting planets in the system
with possible orbital periods of \SIlist{17.3;90}{\day},
but more observations are needed for a conclusive detection.}%
{Future radial velocity measurements may reveal additional non-transiting planets in the system.}
Because \planetinner{} survived to the present day around a mature main-sequence G dwarf,
it is unlikely that \planetouter{} formed via high-eccentricity migration (HEM).
Although no theory presently accounts for the formation of all hot gas giants,
finding more companions like \planetinner{}
and calculating their intrinsic occurrence rate
will help us understand quantitatively the relative frequency between
HEM and other pathways, such as disc migration and \emph{in situ} formation.
Future atmospheric characterization of both planets
can help answer whether they formed together far from their host star
or \planetinner{} formed out of materials
swept by \planetouter{} along its migration path.

\section*{Acknowledgements}

\added[id=s]{%
We thank the anonymous reviewer for detailed feedback that improved this paper.
We also thank Jason D. Eastman for in-depth discussions of the best way to account for
systematic uncertainties in the \mist{} evolutionary tracks and bolometric correction grid
in the joint model.}

Funding for the \tess{} mission is provided by the National Aeronautics and Space Administration's (NASA) Science Mission Directorate.
We acknowledge the use of public \tess{} data from pipelines at the \tess{} Science Office and at the \tess{} Science Processing Operations Center.
This paper includes data collected by the \tess{} mission that are publicly available from the Mikulski Archive for Space Telescopes (MAST).
This research has made use of the Exoplanet Follow-up Observation Program (ExoFOP) website and the NASA Exoplanet Archive, which are operated by the California Institute of Technology, under contract with NASA under the Exoplanet Exploration Program.
Resources supporting this work were provided by the NASA High-End Computing (HEC) Program through the NASA Advanced Supercomputing (NAS) Division at Ames Research Center for the production of the SPOC data products.

Some of the observations in the paper made use of the High-Resolution Imaging instrument Zorro obtained under Gemini LLP Proposal Number: GN/S-2021A-LP-105. Zorro was funded by the NASA Exoplanet Exploration Program and built at the NASA Ames Research Center by Steve B. Howell, Nic Scott, Elliott P. Horch, and Emmett Quigley. Zorro was mounted on the Gemini South telescope of the international Gemini Observatory, a programme of NSF’s OIR Lab, which is managed by the Association of Universities for Research in Astronomy (AURA) under a cooperative agreement with the National Science Foundation. on behalf of the Gemini partnership: the National Science Foundation (United States), National Research Council (Canada), Agencia Nacional de Investigación y Desarrollo (ANID, Chile), Ministerio de Ciencia, Tecnología e Innovación (Argentina), Ministério da Ciência, Tecnologia, Inovações e Comunicações (Brazil), and Korea Astronomy and Space Science Institute (Republic of Korea).

This work makes use of observations from the LCOGT network. Part of the LCOGT telescope time was granted by NOIRLab through the Mid-Scale Innovations Program (MSIP). MSIP is funded by NSF.

This work makes use of observations from the ASTEP telescope. ASTEP benefited from the support of the French and Italian polar agencies IPEV and PNRA in the framework of
the Concordia station programme.

\added[id=t]{This publication makes use of The Data \& Analysis Center for Exoplanets (DACE), which is a facility based at the University of Geneva (CH) dedicated to extrasolar planets data visualisation, exchange and analysis. DACE is a platform of the Swiss National Centre of Competence in Research (NCCR) PlanetS, federating the Swiss expertise in Exoplanet research. The DACE platform is available at
\url{https://dace.unige.ch}.}

C.X.H. and G.Z. acknowledge the support of the Australian Research Council
Discovery Early Career Researcher Award (ARC DECRA)
programmes DE200101840 and DE210101893, respectively.

D.J.A. is supported by UK Research \& Innovation (UKRI) through the Science and Technology Facilities Council (STFC) (ST/R00384X/1) and the Engineering and Physical Sciences Research Council (EPSRC) (EP/X027562/1).

R.B., A.J., and M.H. acknowledge support from ANID Millennium Science Initiative ICN12\_009. A.J.\ acknowledges additional support from FONDECYT project 1210718. R.B.\ acknowledges support from FONDECYT Project 11200751.

\added[id=t]{We acknowledge the support from Fundação para a Ciência e a Tecnologia (FCT) through national funds and from FEDER through COMPETE2020 by the following grants:
UIDB/04434/2020 \& UIDP/04434/2020.
E.D.M. acknowledges the support from FCT through Stimulus FCT contract 2021.01294.CEECIND.}
S.G.S. acknowledges the support from FCT through Investigador FCT contract nr.CEECIND/00826/2018 and POPH/FSE (EC).
\added[id=t]{V.A. acknowledges the support from FCT through the 2022.06962.PTDC grant.}

T.G. and S.H. acknowledge support from the Programme National de Planétologie.

T.T. acknowledges support by the DFG Research Unit FOR 2544 `Blue Planets around Red Stars' project No. KU 3625/2-1.
T.T. further acknowledges support by the BNSF programme `VIHREN-2021' project No. \foreignlanguage{bulgarian}{КП-06-ДВ/5}.

This paper makes use of \textsc{AstroImageJ} \citep{Collins:2017} for photometric reduction.
This paper makes use of the Python packages
\arviz{} \citep{2019JOSS....4.1143K},
\textsc{astropy} \citep{astropy:2013,astropy:2018},
\celeritetwo{}  \citep{celerite1,celerite2},
\exoplanetpy{} \citep{foreman_mackey_daniel_2021_5834934,2021JOSS....6.3285F},
\textsc{matplotlib} \citep{Hunter:2007,thomas_a_caswell_2021_5194481},
\textsc{numpy} \citep{harris2020array},
\textsc{pandas} \citep{mckinney-proc-scipy-2010,jeff_reback_2021_5013202},
\textsc{scipy} \citep{2020SciPy-NMeth},
and
\textsc{TTVFast} \citep{2014ApJ...787..132D},
as well as their dependencies.

\section*{Data Availability}
All photometric and RV measurements used in our joint model analysis
have been included in this paper's tables,
\added[id=t]{which are available in machine-readable format as online Supplementary Materials}.
The \tess{} data products from which we derived our light curves
are publicly available online from the Mikulski Archive for Space Telescopes
(\href{https://archive.stsci.edu/missions-and-data/tess}{MAST}).
Additional information of the ground-based observations used in this paper
is available from
\href{https://exofop.ipac.caltech.edu/tess/target.php?id=371188886}{ExoFOP-TESS}.
The MCMC samples of the posterior distribution of the joint model's parameters are available from Zenodo
\citep{toi2000_mcmc}
in the NetCDF/HDF5 format,
which can be most conveniently read using the \arviz{} Python package.
The Python Jupyter notebooks that performed the MCMC fitting and generated the figures,
\added[id=t]{together with a copy of the data tables in this paper,
are available in a \href{https://github.com/vulpicastor/toi2000-code}{GitHub repository},
which is archived on Zenodo \citep{toi2000_code}}.

\bibliography{toi2000,family_portrait}
\bibliographystyle{mnras}

\appendix

\section{Multiplanet systems with transiting hot or warm gas giants}

We collect the numbers and references underlying \autoref{fig:family}
in \autoref{tab:giant_multi}.
We prefer sources that give RV-derived masses,
and when there are multiple available for the same system,
we prefer the one with the best precision.
\added[id=s]{%
For certain \kepler{} systems where no single source gives parameters for all planets,
we have used values from the \kepler{} Objects of Interest (KOI) table, Data Release~25 \citep{2018ApJS..235...38T}.
The values for \tess{} Objects of Interest \citep[TOI;][]{2021ApJS..254...39G}
candidates are from the \href{https://exofop.ipac.caltech.edu/tess/view_toi.php}{ExoFOP-TESS website},
accessed on 2022 September 26.
The \enquote{flags} column indicates whether there is literature reporting RV (R),
transit (T),
or TTV (V) detection.
For the disposition column,
planets that only appear in the TOI or KOI catalogues
are considered \enquote{candidates}.
We consider planets detectable via RV or TTV \enquote{confirmed}
(except Kepler-418\thinspace b, which was confirmed with multicolor differential photometry),
and others statistically \enquote{validated}.
}

\added[id=r]{%
While this paper was under review, \citet{2023arXiv230509177M} announced the discovery
of a rocky super-earth interior to the 8.5-d hot jupiter \waspeightyfourhj{}.
We have updated \autoref{fig:family}, \autoref{fig:massradius}, and \autoref{tab:giant_multi} with their results,
as well as minimally modified the main text in \autoref{sec:massradius} where appropriate.
}

\begin{table*}
    \centering
    \caption{Multiplanet systems with transiting hot or warm gas giants ($P < 100\,\text{d}$).
    \comment[id=t]{Parameters updated the with latest numbers in literature.}
    }
    \label{tab:giant_multi}
    \renewcommand{\arraystretch}{1.05}
    \begin{tabular}{llr@{}c@{}lccccl}
    \toprule
        Host star
        & Planet
        & \multicolumn{3}{c}{Flags}
        & Orbital period
        & Radius
        & Mass
        & Disposition
        & Reference
        \\
        & & & & & (d) & (\si{\radius\earth}) & (\si{\mass\earth}) & & \\
    \midrule
WASP-47 & WASP-47 e & R & T &   & $0.7895933\pm 0.0000044$ & $1.808\pm 0.026$ & $6.77\pm 0.57$ & confirmed & \citet{2022AJ....163..197B} \\
WASP-47 & WASP-47 b & R & T & V & $4.15914920\pm 0.00000060$ & $12.64\pm 0.15$ & $363.6\pm 7.3$ & confirmed & \citet{2022AJ....163..197B} \\
WASP-47 & WASP-47 d & R & T & V & $9.030550\pm 0.000080$ & $3.567\pm 0.045$ & $14.2\pm 1.3$ & confirmed & \citet{2022AJ....163..197B} \\
WASP-47 & WASP-47 c & R &   &   & $588.8\pm 2.0$ & -- & $\geq 398.9\pm 9.1$\tablenotemark{a} & confirmed & \citet{2022AJ....163..197B} \\
TOI-5143 & TOI-5143.02 &   & T &   & $2.38517\pm 0.00020$ & $2.9\pm 2.9$ & -- & candidate & \citet{2021ApJS..254...39G} \\
TOI-5143 & TOI-5143.01 &   & T &   & $5.20923\pm 0.00040$ & $11.5\pm 1.9$ & -- & candidate & \citet{2021ApJS..254...39G} \\
Kepler-730 & Kepler-730 c &   & T &   & $2.85188338$ & $1.57\pm 0.13$ & -- & validated & \citet{2019ApJ...870L..17C} \\
Kepler-730 & Kepler-730 b &   & T &   & $6.491682808$ & $12.33_{-0.56}^{+0.53}$ & -- & validated & \citet{2019ApJ...870L..17C} \\
WASP-132 & WASP-132 c &   & T &   & $1.0115340\pm 0.0000050$ & $1.85\pm 0.10$ & $< 37.4$ & validated & \citet{2022AJ....164...13H} \\
WASP-132 & WASP-132 b & R & T &   & $7.1335140\pm 0.0000040$ & $10.05\pm 0.34$ & $130.3\pm 9.5$\tablenotemark{b} & confirmed & \citet{2022AJ....164...13H} \\
TOI-1130 & TOI-1130 b & R & T & V & $4.07445\pm 0.00046$ & $3.56\pm 0.13$ & $19.28\pm 0.97$ & confirmed & \citet{2023arXiv230515565K} \\
TOI-1130 & TOI-1130 c & R & T & V & $8.350231\pm 0.000098$ & $13.32_{-1.41}^{+1.55}$ & $325.59\pm 5.59$ & confirmed & \citet{2023arXiv230515565K} \\
TOI-2494 & TOI-2494.02 &   & T &   & $2.408765\pm 0.000021$ & $2.24\pm 0.21$ & -- & candidate & \citet{2021ApJS..254...39G} \\
TOI-2494 & TOI-2494.01 &   & T &   & $8.376132\pm 0.000012$ & $9.54\pm 0.59$ & -- & candidate & \citet{2021ApJS..254...39G} \\
WASP-84 & WASP-84 c & R & T &   & $1.4468849_{-0.0000016}^{+0.0000022}$ & $1.96 \pm 0.12$ & $15.2^{+4.5}_{-4.2}$ & confirmed & \citet{2023arXiv230509177M} \\
WASP-84 & WASP-84 b & R & T &   & $8.52349648 \pm 0.00000060 $ & $10.72 \pm 0.27 $ & $ 220 \pm 18 $ & confirmed & \citet{2023arXiv230509177M} \\
TOI-2000 & TOI-2000 b & R & T &   & $3.098330_{-0.000019}^{+0.000021}$ & $2.70\pm 0.15$ & $11.0\pm 2.4$ & confirmed & This work \\
TOI-2000 & TOI-2000 c & R & T &   & $9.1270550_{-0.0000072}^{+0.0000073}$ & $8.14_{-0.30}^{+0.31}$ & $81.7_{-4.6}^{+4.7}$ & confirmed & This work \\
TOI-5398 & TOI-5398.02 &   & T &   & $4.77290\pm 0.00091$ & $3.30\pm 0.17$ & -- & candidate & \citet{2021ApJS..254...39G} \\
TOI-5398 & TOI-5398 b &   & T &   & $10.59092\pm 0.00068$ & $11.12\pm 0.44$ & -- & validated\tablenotemark{c} & \citet{2021ApJS..254...39G} \\
Kepler-487 & KOI-191.03 &   & T &   & $0.7086211\pm 0.0000014$ & $1.200_{-0.090}^{+0.120}$ & -- & candidate & \citet{2018ApJS..235...38T} \\
Kepler-487 & Kepler-487 d &   & T &   & $2.4184068\pm 0.0000025$ & $2.25_{-0.17}^{+0.23}$ & -- & validated & \citet{2018ApJS..235...38T} \\
Kepler-487 & Kepler-487 b &   & T &   & $15.3587678\pm 0.0000021$ & $10.89_{-0.82}^{+1.11}$ & -- & validated & \citet{2018ApJS..235...38T} \\
Kepler-487 & Kepler-487 c &   & T &   & $38.65208\pm 0.00026$ & $2.07_{-0.16}^{+0.21}$ & -- & validated & \citet{2018ApJS..235...38T} \\
KOI-94\tablenotemark{d} & KOI-94 b & R & T &   & $3.743208\pm 0.000015$ & $1.71\pm 0.16$ & $10.5\pm 4.6$ & confirmed & \citet{2013ApJ...768...14W} \\
KOI-94\tablenotemark{d} & KOI-94 c & R & T & V & $10.423648\pm 0.000016$ & $4.32\pm 0.41$ & $15.6_{-15.6}^{+5.7}$ & confirmed & \citet{2013ApJ...768...14W} \\
KOI-94\tablenotemark{d} & KOI-94 d & R & T & V & $22.3429890\pm 0.0000067$ & $11.3\pm 1.1$ & $106\pm 11$ & confirmed & \citet{2013ApJ...768...14W} \\
KOI-94\tablenotemark{d} & KOI-94 e & R & T & V & $54.32031\pm 0.00012$ & $6.56\pm 0.62$ & $35_{-28}^{+18}$ & confirmed & \citet{2013ApJ...768...14W} \\
V1298 Tau & V1298 Tau c &   & T & V\tablenotemark{e} & $8.24892\pm 0.00083$ & $5.16\pm 0.38$ & $< 76.3$ & confirmed & \citet{2022NatAs...6..232S} \\
V1298 Tau & V1298 Tau d &   & T & V\tablenotemark{e} & $12.4058\pm 0.0018$ & $6.43\pm 0.46$ & $< 98.5$ & confirmed & \citet{2022NatAs...6..232S} \\
V1298 Tau & V1298 Tau b & R & T &   & $24.1399\pm 0.0015$ & $9.73\pm 0.63$ & $203\pm 60$ & confirmed & \citet{2022NatAs...6..232S} \\
V1298 Tau & V1298 Tau e & R & T &   & $40.2\pm 1.0$ & $8.24\pm 0.81$ & $369\pm 95$ & confirmed & \citet{2022NatAs...6..232S} \\
TOI-1692 & TOI-1692.01 &   & T &   & $17.72888\pm 0.00016$ & $4.25\pm 0.26$ & -- & candidate & \citet{2021ApJS..254...39G} \\
TOI-1692 & TOI-1692.02 &   & T &   & $32.20835\pm 0.00017$ & $7.12\pm 0.37$ & -- & candidate & \citet{2021ApJS..254...39G} \\
Kepler-46 & Kepler-46 d &   & T &   & $6.766529\pm 0.000016$ & $1.970_{-0.120}^{+0.090}$ & -- & validated & \citet{2018ApJS..235...38T} \\
Kepler-46 & Kepler-46 b &   & T & V & $33.6480_{-0.0050}^{+0.0040}$ & $9.08_{-0.40}^{+0.39}$ & $281_{-109}^{+119}$ & confirmed & \citet{2017AJ....153..198S} \\
Kepler-46 & Kepler-46 c &   &   &   & $57.325_{-0.098}^{+0.116}$ & -- & $115.1\pm 5.1$ & confirmed & \citet{2017AJ....153..198S} \\
TOI-216 & TOI-216 b & R & T & V & $17.09680\pm 0.00070$ & $8.0_{-2.0}^{+3.0}$ & $18.75\pm 0.64$ & confirmed & \citet{2021AJ....161..161D} \\
TOI-216 & TOI-216 c & R & T & V & $34.55160\pm 0.00030$ & $10.10\pm 0.20$ & $178.0\pm 6.4$ & confirmed & \citet{2021AJ....161..161D} \\
NGTS-11 & TOI-1847.02 &   & T &   & $12.76458\pm 0.00014$ & $7.2_{-1.6}^{+4.2}$ & -- & candidate & \citet{2021ApJS..254...39G} \\
NGTS-11 & NGTS-11 b & R & T &   & $35.45533\pm 0.00019$ & $9.16_{-0.36}^{+0.31}$ & $109_{-23}^{+29}$ & confirmed & \citet{2020ApJ...898L..11G} \\
TOI-1670 & TOI-1670 b &   & T &   & $10.98462_{-0.00051}^{+0.00046}$ & $2.06_{-0.15}^{+0.19}$ & $< 41.3$ & validated & \citet{2022AJ....163..225T} \\
TOI-1670 & TOI-1670 c & R & T &   & $40.74976_{-0.00021}^{+0.00022}$\tablenotemark{f} & $11.06\pm 0.28$ & $200_{-25}^{+29}$ & confirmed & \citet{2022AJ....163..225T} \\
TOI-1670 & TOI-1670.03 &   & T &   & $123.0619\pm 0.0016$ & $3.10\pm 0.61$ & -- & candidate & \citet{2021ApJS..254...39G} \\
K2-290 & K2-290 b &   & T &   & $9.21165_{-0.00034}^{+0.00033}$ & $3.06\pm 0.16$ & $< 21.1$ & confirmed & \citet{2019MNRAS.484.3522H} \\
K2-290 & K2-290 c & R & T &   & $48.36685_{-0.00040}^{+0.00041}$ & $11.28\pm 0.56$ & $246\pm 15$ & confirmed & \citet{2019MNRAS.484.3522H} \\
TOI-2525 & TOI-2525 b & R & T & V & $23.288_{-0.002}^{+0.001} $ & $9.86\pm0.22$ & $28.0^{+1.6}_{-1.3}$ & confirmed & \citet{2023AJ....165..179T} \\
TOI-2525 & TOI-2525 c & R & T & V & $49.260\pm 0.001$ & $10.98\pm0.22$ & $225\pm 11$ & confirmed & \citet{2023AJ....165..179T} \\
Kepler-117 & Kepler-117 b &   & T & V & $18.7959228\pm 0.0000075$ & $8.06\pm 0.27$ & $30\pm 10$ & confirmed & \citet{2015AnA...573A.124B} \\
Kepler-117 & Kepler-117 c & R & T & V & $50.790391\pm 0.000014$ & $12.34\pm 0.39$ & $585\pm 57$ & confirmed & \citet{2015AnA...573A.124B} \\
Kepler-148 & Kepler-148 b &   & T &   & $1.7293671\pm 0.0000026$ & $1.820_{-0.090}^{+0.150}$ & -- & validated & \citet{2018ApJS..235...38T} \\
Kepler-148 & Kepler-148 c &   & T &   & $4.1800481\pm 0.0000027$ & $3.44_{-0.17}^{+0.27}$ & -- & validated & \citet{2018ApJS..235...38T} \\
Kepler-148 & Kepler-148 d &   & T &   & $51.846890\pm 0.000029$ & $8.73_{-0.42}^{+0.70}$ & -- & validated & \citet{2018ApJS..235...38T} \\
TOI-201 & TOI-201.02 &   & T &   & $5.849229\pm 0.000022$ & $1.64\pm 0.14$ & $< 37.5$\tablenotemark{g} & candidate & \citet{2021ApJS..254...39G} \\
TOI-201 & TOI-201 b & R & T &   & $52.97800\pm 0.00040$ & $11.30_{-0.17}^{+0.13}$ & $133.5_{-9.5}^{+15.9}$ & confirmed & \citet{2021AJ....161..235H} \\
Kepler-30 & Kepler-30 b &   & T & V & $29.3343\pm 0.0081$ & $3.90\pm 0.20$ & $11.3\pm 1.4$ & confirmed & \citet{2012Natur.487..449S} \\
Kepler-30 & Kepler-30 c &   & T & V & $60.32310\pm 0.00024$ & $12.30\pm 0.40$ & $640\pm 50$ & confirmed & \citet{2012Natur.487..449S} \\
Kepler-30 & Kepler-30 d &   & T & V & $143.3439\pm 0.0086$ & $8.80\pm 0.50$ & $23.1\pm 2.7$ & confirmed & \citet{2012Natur.487..449S} \\
Kepler-418 & Kepler-418 c &   & T &   & $12.218260\pm 0.000010$ & $7.01\pm 0.93$ & $< 203$ & validated & \citet{2014AnA...567A..14T} \\
Kepler-418 & Kepler-418 b &   & T &   & $86.678560\pm 0.000070$ & $13.5\pm 1.8$ & $< 350$ & confirmed\tablenotemark{h} & \citet{2014AnA...567A..14T} \\
    \bottomrule
    \end{tabular}

    \flushleft
    \emph{Notes.}
    \tablenotemark{a}Minimum mass ($M_\text{p} \sin i$).
    \tablenotemark{b}Mass from \citet{2017MNRAS.465.3693H}.
    \tablenotemark{c}Validated by \citet{2022MNRAS.516.4432M}.
    \tablenotemark{d}Also known as Kepler-89.
    \tablenotemark{e}TTV detection by \citet{2022ApJ...925L...2F}.
    \tablenotemark{f}Upper uncertainty of period corrected from the original paper (Q.H. Tran 2022, personal communication).
    \tablenotemark{g}Mass upper limit from \citet{2021AJ....161..235H}.
    \tablenotemark{h}Confirmed with multicolour photometry.
\end{table*}

\section{Diagnostics for HARPS spectra}

\added[id=s]{%
We present the diagnostics and stellar activity indicators computed by the \harps{} Data Reduction Software in \autoref{tab:rv_harps}.
}
\added[id=r]{%
Periodograms of select stellar activity indicators are shown in \autoref{fig:rv_activity}.
}

\begin{table*}
    \rotatebox{90}{%
    \begin{minipage}{\textheight}
    \caption{\added[id=s]{Diagnostics for the HARPS spectra of \toitwothousand{}.}}
    \label{tab:rv_harps}
    \footnotesize
    \begin{tabular}{@{}S@{}S@{}S@{}SSS@{}S@{}S@{}S@{}S@{}S@{}SSSS}
    \toprule
    {\bjdtdb{}}
        & {RV}
        & {$\sigma_\text{RV}$}
        & FWHM
        & Contrast
        & {Bisector span}
        & {$\sigma_\text{Bis. span}$}
        & {$S_\text{MW}$}
        & {$\sigma_{S_\text{MW}}$}
        & $R_\text{HK}$
        & {$\sigma_{R_\text{HK}}$}
        & S/N
        & S/N
        & S/N
        & S/N
    \\
    {${} - \num{2400000}$}
        & {(\si{\kilo\meter\per\second})}
        & {(\si{\kilo\meter\per\second})}
        & {(\si{\kilo\meter\per\second})}
        &
        & {(\si{\kilo\meter\per\second})}
        & {(\si{\kilo\meter\per\second})}
        &
        &
        &
        &
        & {(\ion{Ca}{ii} HK)}
        & {(Ord. 10)}
        & {(Ord. 50)}
        & {(Ord. 60)}
    \\
    \midrule
59226.762805	& 8.15581	& 0.00210	& 7.29519	& 58.089	& -0.02003	& 0.00198	& 0.146577	& 0.007525	& -5.1119	& 0.0401	& 7.30	& 11.10	& 40.00	& 39.60 \\
59230.798279	& 8.09973	& 0.00246	& 7.29097	& 58.062	& -0.02907	& 0.00237	& 0.131114	& 0.009402	& -5.2034	& 0.0619	& 6.10	& 9.50	& 35.80	& 34.80 \\
59236.721230	& 8.14855	& 0.00374	& 7.31032	& 57.997	& -0.01522	& 0.00368	& 0.083744	& 0.015215	& -5.7532	& 0.3553	& 4.00	& 6.40	& 25.80	& 25.20 \\
59237.738282	& 8.12825	& 0.00342	& 7.30489	& 57.965	& -0.02853	& 0.00336	& 0.146261	& 0.013803	& -5.1136	& 0.0739	& 4.40	& 6.90	& 27.90	& 27.10 \\
59239.628780	& 8.11574	& 0.00211	& 7.31819	& 58.010	& -0.02555	& 0.00200	& 0.130054	& 0.006915	& -5.2104	& 0.0463	& 7.80	& 11.70	& 40.80	& 39.30 \\
59243.727714	& 8.15102	& 0.00186	& 7.30821	& 57.958	& -0.01419	& 0.00174	& 0.169925	& 0.005445	& -5.0024	& 0.0226	& 9.40	& 14.00	& 44.40	& 41.80 \\
59246.723274	& 8.13201	& 0.00250	& 7.29066	& 58.163	& -0.02039	& 0.00240	& 0.176227	& 0.009203	& -4.9771	& 0.0360	& 6.20	& 9.60	& 35.10	& 34.10 \\
59251.726865	& 8.13202	& 0.00194	& 7.30339	& 58.085	& -0.02793	& 0.00181	& 0.134105	& 0.006252	& -5.1841	& 0.0394	& 8.30	& 12.90	& 45.00	& 42.40 \\
59252.729261	& 8.13616	& 0.00237	& 7.29288	& 58.053	& -0.02954	& 0.00228	& 0.117973	& 0.008328	& -5.2999	& 0.0685	& 6.60	& 10.40	& 37.50	& 35.70 \\
59255.684397	& 8.12881	& 0.00205	& 7.29914	& 58.012	& -0.02344	& 0.00193	& 0.129155	& 0.006589	& -5.2165	& 0.0447	& 8.00	& 12.50	& 42.90	& 40.20 \\
59259.722980	& 8.11677	& 0.00214	& 7.30380	& 58.033	& -0.03367	& 0.00202	& 0.114993	& 0.007491	& -5.3251	& 0.0653	& 7.20	& 11.40	& 42.00	& 39.70 \\
59261.780234	& 8.13285	& 0.00467	& 7.30647	& 58.082	& -0.01148	& 0.00461	& 0.156819	& 0.019056	& -5.0605	& 0.0903	& 3.20	& 5.20	& 22.50	& 21.50 \\
59266.671034	& 8.09974	& 0.00277	& 7.31129	& 58.202	& -0.02296	& 0.00268	& 0.129927	& 0.010651	& -5.2113	& 0.0714	& 5.40	& 8.60	& 33.50	& 32.00 \\
59278.660823	& 8.12831	& 0.00267	& 7.30873	& 58.031	& -0.01664	& 0.00258	& 0.178656	& 0.011154	& -4.9677	& 0.0427	& 5.00	& 8.80	& 35.40	& 34.40 \\
59282.759174	& 8.11321	& 0.03150	& 7.37968	& 56.618	& -0.04154	& 0.03149	& 0.263719	& 0.147719	& -4.7248	& 0.3231	& 0.30	& 0.50	& 6.10	& 7.20 \\
59291.662808	& 8.12950	& 0.00172	& 7.29472	& 57.997	& -0.02874	& 0.00155	& 0.174546	& 0.005551	& -4.9837	& 0.0220	& 9.10	& 15.00	& 52.30	& 50.30 \\
59291.781561	& 8.12197	& 0.00246	& 7.29181	& 57.839	& -0.02839	& 0.00234	& 0.172245	& 0.011396	& -4.9929	& 0.0462	& 5.00	& 8.80	& 39.20	& 39.20 \\
59294.581135	& 8.08614	& 0.00329	& 7.28208	& 58.045	& -0.02301	& 0.00320	& 0.117349	& 0.014166	& -5.3050	& 0.1178	& 4.30	& 7.60	& 29.70	& 29.00 \\
59295.668144	& 8.08866	& 0.00184	& 7.29308	& 58.157	& -0.02035	& 0.00168	& 0.138994	& 0.006438	& -5.1544	& 0.0379	& 8.10	& 13.70	& 49.80	& 48.20 \\
59295.771282	& 8.08515	& 0.00260	& 7.29257	& 58.247	& -0.03065	& 0.00249	& 0.134204	& 0.012714	& -5.1835	& 0.0800	& 4.70	& 8.40	& 37.60	& 37.30 \\
59296.686053	& 8.10365	& 0.00274	& 7.29130	& 58.155	& -0.02568	& 0.00265	& 0.097573	& 0.013265	& -5.5118	& 0.1777	& 4.40	& 8.10	& 35.70	& 34.80 \\
59296.754306	& 8.10600	& 0.00193	& 7.28994	& 58.070	& -0.01686	& 0.00177	& 0.119957	& 0.009468	& -5.2838	& 0.0750	& 5.70	& 10.60	& 50.70	& 50.90 \\
59297.682295	& 8.12591	& 0.00189	& 7.29785	& 58.078	& -0.02833	& 0.00173	& 0.134521	& 0.007471	& -5.1815	& 0.0468	& 7.20	& 12.40	& 49.40	& 48.40 \\
59297.765977	& 8.12542	& 0.00173	& 7.29147	& 58.187	& -0.02310	& 0.00155	& 0.129571	& 0.007975	& -5.2137	& 0.0538	& 6.70	& 12.00	& 56.40	& 56.50 \\
59358.496526	& 8.08755	& 0.00166	& 7.30293	& 57.897	& -0.02440	& 0.00148	& 0.153473	& 0.006230	& -5.0766	& 0.0306	& 8.10	& 14.10	& 55.30	& 57.00 \\
59358.601790	& 8.08954	& 0.00238	& 7.30258	& 57.777	& -0.01141	& 0.00226	& 0.145167	& 0.012376	& -5.1195	& 0.0672	& 4.60	& 8.60	& 40.20	& 42.40 \\
59359.495659	& 8.10205	& 0.00175	& 7.31994	& 57.749	& -0.01947	& 0.00157	& 0.157244	& 0.006331	& -5.0585	& 0.0299	& 8.00	& 13.80	& 51.90	& 53.10 \\
59359.592451	& 8.09822	& 0.00216	& 7.30569	& 57.640	& -0.01853	& 0.00202	& 0.140468	& 0.010724	& -5.1458	& 0.0618	& 5.20	& 9.40	& 44.20	& 46.50 \\
59360.476170	& 8.10776	& 0.00183	& 7.31540	& 57.668	& -0.01792	& 0.00166	& 0.160370	& 0.006937	& -5.0440	& 0.0316	& 7.50	& 13.00	& 50.00	& 51.70 \\
59360.565046	& 8.10720	& 0.00208	& 7.30539	& 57.447	& -0.02062	& 0.00194	& 0.168561	& 0.009647	& -5.0081	& 0.0405	& 5.70	& 10.20	& 45.30	& 47.60 \\
59361.516541	& 8.12338	& 0.00187	& 7.31504	& 57.678	& -0.01767	& 0.00171	& 0.151552	& 0.007901	& -5.0862	& 0.0397	& 6.70	& 11.80	& 49.60	& 52.00 \\
59361.572662	& 8.12681	& 0.00216	& 7.29524	& 57.542	& -0.01947	& 0.00203	& 0.153522	& 0.010778	& -5.0764	& 0.0530	& 5.30	& 9.50	& 44.00	& 46.80 \\
59368.480158	& 8.08589	& 0.00191	& 7.30645	& 58.034	& -0.00711	& 0.00174	& 0.144500	& 0.007885	& -5.1231	& 0.0432	& 6.70	& 11.70	& 47.40	& 50.00 \\
59368.549786	& 8.08856	& 0.00214	& 7.29602	& 58.077	& -0.00804	& 0.00200	& 0.144560	& 0.010787	& -5.1228	& 0.0590	& 5.30	& 9.40	& 43.50	& 47.20 \\
59369.477343	& 8.09255	& 0.00170	& 7.30277	& 58.057	& -0.01401	& 0.00152	& 0.146717	& 0.006618	& -5.1112	& 0.0352	& 7.80	& 13.20	& 53.00	& 55.80 \\
59369.566148	& 8.08750	& 0.00178	& 7.30921	& 58.067	& -0.01480	& 0.00161	& 0.151638	& 0.009366	& -5.0857	& 0.0470	& 5.80	& 10.40	& 52.80	& 57.60 \\
59370.486575	& 8.09324	& 0.00326	& 7.29495	& 58.131	& -0.02144	& 0.00318	& 0.158053	& 0.019231	& -5.0547	& 0.0899	& 3.10	& 6.00	& 30.20	& 32.40 \\
59370.571550	& 8.11356	& 0.00287	& 7.23011	& 53.176	& 0.00617	& 0.00277	& 0.020057	& 0.010755	& 0.0000	& 0.0000	& 3.00	& 5.20	& 38.70	& 44.80 \\
59371.478262	& 8.11841	& 0.00170	& 7.28721	& 58.138	& -0.00334	& 0.00151	& 0.148970	& 0.007174	& -5.0993	& 0.0372	& 7.20	& 12.60	& 54.00	& 57.10 \\
59371.550679	& 8.12149	& 0.00193	& 7.29430	& 58.078	& -0.00891	& 0.00177	& 0.149964	& 0.009868	& -5.0942	& 0.0505	& 5.40	& 10.00	& 48.50	& 52.20 \\
59372.571612	& 8.11961	& 0.00196	& 7.29552	& 58.137	& -0.01785	& 0.00179	& 0.142847	& 0.010176	& -5.1323	& 0.0569	& 5.40	& 9.90	& 48.20	& 51.70 \\
    \bottomrule
    \end{tabular}
    \end{minipage}
    }
\end{table*}

\clearpage
\newpage

{ \itshape \footnotesize \noindent
$^{1}$Department of Astronomy, University of Wisconsin--Madison, 475 N Charter St, Madison, WI 53706, USA \\
$^{2}$Department of Physics and Kavli Institute for Astrophysics and Space Research, Massachusetts Institute of Technology, 77 Massachusetts Ave, Cambridge, MA 02139, USA \\
$^{3}$Centre for Astrophysics, University of Southern Queensland, West Street, Toowoomba, QLD 4350 Australia \\
$^{4}$Department of Physics, University of Warwick, Gibbet Hill Road, Coventry CV4 7AL, UK \\
$^{5}$Centre for Exoplanets and Habitability, University of Warwick, Gibbet Hill Road, Coventry CV4 7AL, UK \\
$^{6}$Facultad de Ingeniería y Ciencias, Universidad Adolfo Ibáñez, Av.\ Diagonal las Torres 2640, Peñalolén, Santiago, Chile \\
$^{7}$Millennium Institute for Astrophysics, Chile \\
$^{8}$Data Observatory Foundation, Chile \\
$^{9}$Department of Astronomy, University of California at Berkeley, Berkeley, CA 94720, USA \\
$^{10}$Department of Physics and Astronomy, The University of North Carolina at Chapel Hill, Chapel Hill, NC 27599-3255, USA \\
$^{11}$Center for Astrophysics \textbar\ Harvard \& Smithsonian, 60 Garden Street, Cambridge, MA 02138, USA \\
$^{12}$European Southern Observatory, Karl-Schwarzschild-Straße 2, 85748 Garching bei München, Germany \\
$^{13}$Max-Planck-Institut für Astronomie, Königstuhl 17, Heidelberg 69117, Germany \\
$^{14}$Millenium Institute for Astrophysics, Chile \\
$^{15}$Department of Physics, Engineering and Astronomy, Stephen F. Austin State University, 1936 North St, Nacogdoches, TX 75962, USA \\
$^{16}$NASA Ames Research Center, Moffett Field, CA 94035, USA \\
$^{17}$Instituto de Astrofísica, Pontificia Universidad Católica de Chile, Avda. Vicuña Mackenna 4860, Macul, Santiago, Chile \\
$^{18}$Instituto de Astrofisica e Ciencias do Espaco, Universidade do Porto, CAUP, Rua das Estrelas, P-4150-762 Porto, Portugal \\
$^{19}$Université Côte d'Azur, Observatoire de la Côte d'Azur, CNRS, Laboratoire Lagrange, Bd de l'Observatoire, CS 34229, 06304 Nice CEDEX 4, France \\
$^{20}$Department of Astronomy, Sofia University ``St Kliment Ohridski'', 5 James Bourchier Blvd, BG-1164 Sofia, Bulgaria \\
$^{21}$Caltech/IPAC, Mail Code 100-22, Pasadena, CA 91125, USA \\
$^{22}$Perth Exoplanet Survey Telescope, Perth, Australia \\
$^{23}$Tsinghua International School, Beijing 100084, China \\
$^{24}$Villa '39 Observatory, Landers, CA 92285, USA \\
$^{25}$Hazelwood Observatory, Australia \\
$^{26}$NASA Exoplanet Science Institute, Caltech/IPAC, Pasadena, CA 91125, USA \\
$^{27}$Kotizarovci Observatory, Sarsoni 90, 51216 Viskovo, Croatia \\
$^{28}$Centro de Astrobiología (CSIC-INTA), ESAC Campus, Camino Bajo del Castillo s/n, 28692 Villanueva de la Cañada, Madrid, Spain \\
$^{29}$Department of Physics and Astronomy, Vanderbilt University, Nashville, TN 37235, USA \\
$^{30}$Department of Astrophysical Sciences, Princeton University, 4 Ivy Lane, Princeton, NJ 08544, USA \\
$^{31}$Department of Earth, Atmospheric and Planetary Sciences, Massachusetts Institute of Technology, Cambridge, MA 02139, USA \\
$^{32}$Department of Aeronautics and Astronautics, Massachusetts Institute of Technology, Cambridge, MA 02139, USA \\
$^{33}$NASA Goddard Space Flight Center, 8800 Greenbelt Road, Greenbelt, MD 20771, USA \\
$^{34}$University of Maryland, Baltimore County, 1000 Hilltop Circle, Baltimore, MD 21250, USA \\
$^{35}$Department of Astronomy, California Institute of Technology, Pasadena, CA 91125, USA \\
$^{36}$Bryant Space Science Center, Department of Astronomy, University of Florida, Gainesville, FL 32611, USA
}

\bsp	%
\label{lastpage}

\end{document}